\documentclass[longauth]{aa_gaia}
\usepackage{graphicx}
\usepackage{txfonts}
\usepackage[draft]{hyperref}
\usepackage{longtable}

\begin{document} 

\title{The {\it Gaia} mission}

\author{
{\it Gaia} Collaboration\relax
\and T.        ~Prusti                        \inst{\ref{inst:0001}}\relax
\and J.H.J.    ~de Bruijne                    \inst{\ref{inst:0001}}\relax
\and A.G.A.    ~Brown                         \inst{\ref{inst:0003}}\relax
\and A.        ~Vallenari                     \inst{\ref{inst:0004}}\relax
\and C.        ~Babusiaux                     \inst{\ref{inst:0005}}\relax
\and C.A.L.    ~Bailer-Jones                  \inst{\ref{inst:0006}}\relax
\and U.        ~Bastian                       \inst{\ref{inst:0007}}\relax
\and M.        ~Biermann                      \inst{\ref{inst:0007}}\relax
\and D.W.      ~Evans                         \inst{\ref{inst:0009}}\relax
\and L.        ~Eyer                          \inst{\ref{inst:0010}}\relax
\and F.        ~Jansen                        \inst{\ref{inst:0011}}\relax
\and C.        ~Jordi                         \inst{\ref{inst:0012}}\relax
\and S.A.      ~Klioner                       \inst{\ref{inst:0013}}\relax
\and U.        ~Lammers                       \inst{\ref{inst:0014}}\relax
\and L.        ~Lindegren                     \inst{\ref{inst:0015}}\relax
\and X.        ~Luri                          \inst{\ref{inst:0012}}\relax
\and F.        ~Mignard                       \inst{\ref{inst:0017}}\relax
\and D.J.      ~Milligan                      \inst{\ref{inst:0018}}\relax
\and C.        ~Panem                         \inst{\ref{inst:0019}}\relax
\and V.        ~Poinsignon                    \inst{\ref{inst:0020}}\relax
\and D.        ~Pourbaix                      \inst{\ref{inst:0021},\ref{inst:0022}}\relax
\and S.        ~Randich                       \inst{\ref{inst:0023}}\relax
\and G.        ~Sarri                         \inst{\ref{inst:0024}}\relax
\and P.        ~Sartoretti                    \inst{\ref{inst:0005}}\relax
\and H.I.      ~Siddiqui                      \inst{\ref{inst:0026}}\relax
\and C.        ~Soubiran                      \inst{\ref{inst:0027}}\relax
\and V.        ~Valette                       \inst{\ref{inst:0019}}\relax
\and F.        ~van Leeuwen                   \inst{\ref{inst:0009}}\relax
\and N.A.      ~Walton                        \inst{\ref{inst:0009}}\relax
\and C.        ~Aerts                         \inst{\ref{inst:0031},\ref{inst:0032}}\relax
\and F.        ~Arenou                        \inst{\ref{inst:0005}}\relax
\and M.        ~Cropper                       \inst{\ref{inst:0034}}\relax
\and R.        ~Drimmel                       \inst{\ref{inst:0035}}\relax
\and E.        ~H{\o}g                        \inst{\ref{inst:0036}}\relax
\and D.        ~Katz                          \inst{\ref{inst:0005}}\relax
\and M.G.      ~Lattanzi                      \inst{\ref{inst:0035}}\relax
\and W.        ~O'Mullane                     \inst{\ref{inst:0014}}\relax
\and E.K.      ~Grebel                        \inst{\ref{inst:0007}}\relax
\and A.D.      ~Holland                       \inst{\ref{inst:0041}}\relax
\and C.        ~Huc                           \inst{\ref{inst:0019}}\relax
\and X.        ~Passot                        \inst{\ref{inst:0019}}\relax
\and L.        ~Bramante                      \inst{\ref{inst:0044}}\relax
\and C.        ~Cacciari                      \inst{\ref{inst:0045}}\relax
\and J.        ~Casta\~{n}eda                 \inst{\ref{inst:0012}}\relax
\and L.        ~Chaoul                        \inst{\ref{inst:0019}}\relax
\and N.        ~Cheek                         \inst{\ref{inst:0048}}\relax
\and F.        ~De Angeli                     \inst{\ref{inst:0009}}\relax
\and C.        ~Fabricius                     \inst{\ref{inst:0012}}\relax
\and R.        ~Guerra                        \inst{\ref{inst:0014}}\relax
\and J.        ~Hern\'{a}ndez                 \inst{\ref{inst:0014}}\relax
\and A.        ~Jean-Antoine-Piccolo          \inst{\ref{inst:0019}}\relax
\and E.        ~Masana                        \inst{\ref{inst:0012}}\relax
\and R.        ~Messineo                      \inst{\ref{inst:0044}}\relax
\and N.        ~Mowlavi                       \inst{\ref{inst:0010}}\relax
\and K.        ~Nienartowicz                  \inst{\ref{inst:0057}}\relax
\and D.        ~Ord\'{o}\~{n}ez-Blanco        \inst{\ref{inst:0057}}\relax
\and P.        ~Panuzzo                       \inst{\ref{inst:0005}}\relax
\and J.        ~Portell                       \inst{\ref{inst:0012}}\relax
\and P.J.      ~Richards                      \inst{\ref{inst:0061}}\relax
\and M.        ~Riello                        \inst{\ref{inst:0009}}\relax
\and G.M.      ~Seabroke                      \inst{\ref{inst:0034}}\relax
\and P.        ~Tanga                         \inst{\ref{inst:0017}}\relax
\and F.        ~Th\'{e}venin                  \inst{\ref{inst:0017}}\relax
\and J.        ~Torra                         \inst{\ref{inst:0012}}\relax
\and S.G.      ~Els                           \inst{\ref{inst:0067},\ref{inst:0007}}\relax
\and G.        ~Gracia-Abril                  \inst{\ref{inst:0067},\ref{inst:0012}}\relax
\and G.        ~Comoretto                     \inst{\ref{inst:0026}}\relax
\and M.        ~Garcia-Reinaldos              \inst{\ref{inst:0014}}\relax
\and T.        ~Lock                          \inst{\ref{inst:0014}}\relax
\and E.        ~Mercier                       \inst{\ref{inst:0067},\ref{inst:0007}}\relax
\and M.        ~Altmann                       \inst{\ref{inst:0007},\ref{inst:0077}}\relax
\and R.        ~Andrae                        \inst{\ref{inst:0006}}\relax
\and T.L.      ~Astraatmadja                  \inst{\ref{inst:0006}}\relax
\and I.        ~Bellas-Velidis                \inst{\ref{inst:0080}}\relax
\and K.        ~Benson                        \inst{\ref{inst:0034}}\relax
\and J.        ~Berthier                      \inst{\ref{inst:0082}}\relax
\and R.        ~Blomme                        \inst{\ref{inst:0083}}\relax
\and G.        ~Busso                         \inst{\ref{inst:0009}}\relax
\and B.        ~Carry                         \inst{\ref{inst:0017},\ref{inst:0082}}\relax
\and A.        ~Cellino                       \inst{\ref{inst:0035}}\relax
\and G.        ~Clementini                    \inst{\ref{inst:0045}}\relax
\and S.        ~Cowell                        \inst{\ref{inst:0009}}\relax
\and O.        ~Creevey                       \inst{\ref{inst:0017},\ref{inst:0091}}\relax
\and J.        ~Cuypers                       \inst{\ref{inst:0083}}\relax
\and M.        ~Davidson                      \inst{\ref{inst:0093}}\relax
\and J.        ~De Ridder                     \inst{\ref{inst:0031}}\relax
\and A.        ~de Torres                     \inst{\ref{inst:0095}}\relax
\and L.        ~Delchambre                    \inst{\ref{inst:0096}}\relax
\and A.        ~Dell'Oro                      \inst{\ref{inst:0023}}\relax
\and C.        ~Ducourant                     \inst{\ref{inst:0027}}\relax
\and Y.        ~Fr\'{e}mat                    \inst{\ref{inst:0083}}\relax
\and M.        ~Garc\'{i}a-Torres             \inst{\ref{inst:0100}}\relax
\and E.        ~Gosset                        \inst{\ref{inst:0096},\ref{inst:0022}}\relax
\and J.-L.     ~Halbwachs                     \inst{\ref{inst:0103}}\relax
\and N.C.      ~Hambly                        \inst{\ref{inst:0093}}\relax
\and D.L.      ~Harrison                      \inst{\ref{inst:0009},\ref{inst:0106}}\relax
\and M.        ~Hauser                        \inst{\ref{inst:0007}}\relax
\and D.        ~Hestroffer                    \inst{\ref{inst:0082}}\relax
\and S.T.      ~Hodgkin                       \inst{\ref{inst:0009}}\relax
\and H.E.      ~Huckle                        \inst{\ref{inst:0034}}\relax
\and A.        ~Hutton                        \inst{\ref{inst:0111}}\relax
\and G.        ~Jasniewicz                    \inst{\ref{inst:0112}}\relax
\and S.        ~Jordan                        \inst{\ref{inst:0007}}\relax
\and M.        ~Kontizas                      \inst{\ref{inst:0114}}\relax
\and A.J.      ~Korn                          \inst{\ref{inst:0115}}\relax
\and A.C.      ~Lanzafame                     \inst{\ref{inst:0116},\ref{inst:0117}}\relax
\and M.        ~Manteiga                      \inst{\ref{inst:0118}}\relax
\and A.        ~Moitinho                      \inst{\ref{inst:0119}}\relax
\and K.        ~Muinonen                      \inst{\ref{inst:0120},\ref{inst:0121}}\relax
\and J.        ~Osinde                        \inst{\ref{inst:0122}}\relax
\and E.        ~Pancino                       \inst{\ref{inst:0023},\ref{inst:0124}}\relax
\and T.        ~Pauwels                       \inst{\ref{inst:0083}}\relax
\and J.-M.     ~Petit                         \inst{\ref{inst:0126}}\relax
\and A.        ~Recio-Blanco                  \inst{\ref{inst:0017}}\relax
\and A.C.      ~Robin                         \inst{\ref{inst:0126}}\relax
\and L.M.      ~Sarro                         \inst{\ref{inst:0129}}\relax
\and C.        ~Siopis                        \inst{\ref{inst:0021}}\relax
\and M.        ~Smith                         \inst{\ref{inst:0034}}\relax
\and K.W.      ~Smith                         \inst{\ref{inst:0006}}\relax
\and A.        ~Sozzetti                      \inst{\ref{inst:0035}}\relax
\and W.        ~Thuillot                      \inst{\ref{inst:0082}}\relax
\and W.        ~van Reeven                    \inst{\ref{inst:0111}}\relax
\and Y.        ~Viala                         \inst{\ref{inst:0005}}\relax
\and U.        ~Abbas                         \inst{\ref{inst:0035}}\relax
\and A.        ~Abreu Aramburu                \inst{\ref{inst:0138}}\relax
\and S.        ~Accart                        \inst{\ref{inst:0139}}\relax
\and J.J.      ~Aguado                        \inst{\ref{inst:0129}}\relax
\and P.M.      ~Allan                         \inst{\ref{inst:0061}}\relax
\and W.        ~Allasia                       \inst{\ref{inst:0142}}\relax
\and G.        ~Altavilla                     \inst{\ref{inst:0045}}\relax
\and M.A.      ~\'{A}lvarez                   \inst{\ref{inst:0118}}\relax
\and J.        ~Alves                         \inst{\ref{inst:0145}}\relax
\and R.I.      ~Anderson                      \inst{\ref{inst:0146},\ref{inst:0010}}\relax
\and A.H.      ~Andrei                        \inst{\ref{inst:0148},\ref{inst:0149},\ref{inst:0077}}\relax
\and E.        ~Anglada Varela                \inst{\ref{inst:0122},\ref{inst:0048}}\relax
\and E.        ~Antiche                       \inst{\ref{inst:0012}}\relax
\and T.        ~Antoja                        \inst{\ref{inst:0001}}\relax
\and S.        ~Ant\'{o}n                     \inst{\ref{inst:0155},\ref{inst:0156}}\relax
\and B.        ~Arcay                         \inst{\ref{inst:0118}}\relax
\and A.        ~Atzei                         \inst{\ref{inst:0024}}\relax
\and L.        ~Ayache                        \inst{\ref{inst:0159}}\relax
\and N.        ~Bach                          \inst{\ref{inst:0111}}\relax
\and S.G.      ~Baker                         \inst{\ref{inst:0034}}\relax
\and L.        ~Balaguer-N\'{u}\~{n}ez        \inst{\ref{inst:0012}}\relax
\and C.        ~Barache                       \inst{\ref{inst:0077}}\relax
\and C.        ~Barata                        \inst{\ref{inst:0119}}\relax
\and A.        ~Barbier                       \inst{\ref{inst:0139}}\relax
\and F.        ~Barblan                       \inst{\ref{inst:0010}}\relax
\and M.        ~Baroni                        \inst{\ref{inst:0024}}\relax
\and D.        ~Barrado y Navascu\'{e}s       \inst{\ref{inst:0168}}\relax
\and M.        ~Barros                        \inst{\ref{inst:0119}}\relax
\and M.A.      ~Barstow                       \inst{\ref{inst:0170}}\relax
\and U.        ~Becciani                      \inst{\ref{inst:0117}}\relax
\and M.        ~Bellazzini                    \inst{\ref{inst:0045}}\relax
\and G.        ~Bellei                        \inst{\ref{inst:0173}}\relax
\and A.        ~Bello Garc\'{i}a              \inst{\ref{inst:0174}}\relax
\and V.        ~Belokurov                     \inst{\ref{inst:0009}}\relax
\and P.        ~Bendjoya                      \inst{\ref{inst:0017}}\relax
\and A.        ~Berihuete                     \inst{\ref{inst:0177}}\relax
\and L.        ~Bianchi                       \inst{\ref{inst:0142}}\relax
\and O.        ~Bienaym\'{e}                  \inst{\ref{inst:0103}}\relax
\and F.        ~Billebaud                     \inst{\ref{inst:0027}}\relax
\and N.        ~Blagorodnova                  \inst{\ref{inst:0009}}\relax
\and S.        ~Blanco-Cuaresma               \inst{\ref{inst:0010},\ref{inst:0027}}\relax
\and T.        ~Boch                          \inst{\ref{inst:0103}}\relax
\and A.        ~Bombrun                       \inst{\ref{inst:0095}}\relax
\and R.        ~Borrachero                    \inst{\ref{inst:0012}}\relax
\and S.        ~Bouquillon                    \inst{\ref{inst:0077}}\relax
\and G.        ~Bourda                        \inst{\ref{inst:0027}}\relax
\and H.        ~Bouy                          \inst{\ref{inst:0168}}\relax
\and A.        ~Bragaglia                     \inst{\ref{inst:0045}}\relax
\and M.A.      ~Breddels                      \inst{\ref{inst:0191}}\relax
\and N.        ~Brouillet                     \inst{\ref{inst:0027}}\relax
\and T.        ~Br\"{ u}semeister             \inst{\ref{inst:0007}}\relax
\and B.        ~Bucciarelli                   \inst{\ref{inst:0035}}\relax
\and F.        ~Budnik                        \inst{\ref{inst:0018}}\relax
\and P.        ~Burgess                       \inst{\ref{inst:0009}}\relax
\and R.        ~Burgon                        \inst{\ref{inst:0041}}\relax
\and A.        ~Burlacu                       \inst{\ref{inst:0019}}\relax
\and D.        ~Busonero                      \inst{\ref{inst:0035}}\relax
\and R.        ~Buzzi                         \inst{\ref{inst:0035}}\relax
\and E.        ~Caffau                        \inst{\ref{inst:0005}}\relax
\and J.        ~Cambras                       \inst{\ref{inst:0202}}\relax
\and H.        ~Campbell                      \inst{\ref{inst:0009}}\relax
\and R.        ~Cancelliere                   \inst{\ref{inst:0204}}\relax
\and T.        ~Cantat-Gaudin                 \inst{\ref{inst:0004}}\relax
\and T.        ~Carlucci                      \inst{\ref{inst:0077}}\relax
\and J.M.      ~Carrasco                      \inst{\ref{inst:0012}}\relax
\and M.        ~Castellani                    \inst{\ref{inst:0208}}\relax
\and P.        ~Charlot                       \inst{\ref{inst:0027}}\relax
\and J.        ~Charnas                       \inst{\ref{inst:0057}}\relax
\and P.        ~Charvet                       \inst{\ref{inst:0020}}\relax
\and F.        ~Chassat                       \inst{\ref{inst:0020}}\relax
\and A.        ~Chiavassa                     \inst{\ref{inst:0017}}\relax
\and M.        ~Clotet                        \inst{\ref{inst:0012}}\relax
\and G.        ~Cocozza                       \inst{\ref{inst:0045}}\relax
\and R.S.      ~Collins                       \inst{\ref{inst:0093}}\relax
\and P.        ~Collins                       \inst{\ref{inst:0018}}\relax
\and G.        ~Costigan                      \inst{\ref{inst:0003}}\relax
\and F.        ~Crifo                         \inst{\ref{inst:0005}}\relax
\and N.J.G.    ~Cross                         \inst{\ref{inst:0093}}\relax
\and M.        ~Crosta                        \inst{\ref{inst:0035}}\relax
\and C.        ~Crowley                       \inst{\ref{inst:0095}}\relax
\and C.        ~Dafonte                       \inst{\ref{inst:0118}}\relax
\and Y.        ~Damerdji                      \inst{\ref{inst:0096},\ref{inst:0225}}\relax
\and A.        ~Dapergolas                    \inst{\ref{inst:0080}}\relax
\and P.        ~David                         \inst{\ref{inst:0082}}\relax
\and M.        ~David                         \inst{\ref{inst:0228}}\relax
\and P.        ~De Cat                        \inst{\ref{inst:0083}}\relax
\and F.        ~de Felice                     \inst{\ref{inst:0230}}\relax
\and P.        ~de Laverny                    \inst{\ref{inst:0017}}\relax
\and F.        ~De Luise                      \inst{\ref{inst:0232}}\relax
\and R.        ~De March                      \inst{\ref{inst:0044}}\relax
\and D.        ~de Martino                    \inst{\ref{inst:0234}}\relax
\and R.        ~de Souza                      \inst{\ref{inst:0235}}\relax
\and J.        ~Debosscher                    \inst{\ref{inst:0031}}\relax
\and E.        ~del Pozo                      \inst{\ref{inst:0111}}\relax
\and M.        ~Delbo                         \inst{\ref{inst:0017}}\relax
\and A.        ~Delgado                       \inst{\ref{inst:0009}}\relax
\and H.E.      ~Delgado                       \inst{\ref{inst:0129}}\relax
\and F.        ~di Marco                      \inst{\ref{inst:0241}}\relax
\and P.        ~Di Matteo                     \inst{\ref{inst:0005}}\relax
\and S.        ~Diakite                       \inst{\ref{inst:0126}}\relax
\and E.        ~Distefano                     \inst{\ref{inst:0117}}\relax
\and C.        ~Dolding                       \inst{\ref{inst:0034}}\relax
\and S.        ~Dos Anjos                     \inst{\ref{inst:0235}}\relax
\and P.        ~Drazinos                      \inst{\ref{inst:0114}}\relax
\and J.        ~Dur\'{a}n                     \inst{\ref{inst:0122}}\relax
\and Y.        ~Dzigan                        \inst{\ref{inst:0249},\ref{inst:0250}}\relax
\and E.        ~Ecale                         \inst{\ref{inst:0020}}\relax
\and B.        ~Edvardsson                    \inst{\ref{inst:0115}}\relax
\and H.        ~Enke                          \inst{\ref{inst:0253}}\relax
\and M.        ~Erdmann                       \inst{\ref{inst:0024}}\relax
\and D.        ~Escolar                       \inst{\ref{inst:0024}}\relax
\and M.        ~Espina                        \inst{\ref{inst:0018}}\relax
\and N.W.      ~Evans                         \inst{\ref{inst:0009}}\relax
\and G.        ~Eynard Bontemps               \inst{\ref{inst:0139}}\relax
\and C.        ~Fabre                         \inst{\ref{inst:0259}}\relax
\and M.        ~Fabrizio                      \inst{\ref{inst:0124},\ref{inst:0232}}\relax
\and S.        ~Faigler                       \inst{\ref{inst:0262}}\relax
\and A.J.      ~Falc\~{a}o                    \inst{\ref{inst:0263}}\relax
\and M.        ~Farr\`{a}s Casas              \inst{\ref{inst:0012}}\relax
\and F.        ~Faye                          \inst{\ref{inst:0020}}\relax
\and L.        ~Federici                      \inst{\ref{inst:0045}}\relax
\and G.        ~Fedorets                      \inst{\ref{inst:0120}}\relax
\and J.        ~Fern\'{a}ndez-Hern\'{a}ndez   \inst{\ref{inst:0048}}\relax
\and P.        ~Fernique                      \inst{\ref{inst:0103}}\relax
\and A.        ~Fienga                        \inst{\ref{inst:0270}}\relax
\and F.        ~Figueras                      \inst{\ref{inst:0012}}\relax
\and F.        ~Filippi                       \inst{\ref{inst:0044}}\relax
\and K.        ~Findeisen                     \inst{\ref{inst:0005}}\relax
\and A.        ~Fonti                         \inst{\ref{inst:0044}}\relax
\and M.        ~Fouesneau                     \inst{\ref{inst:0006}}\relax
\and E.        ~Fraile                        \inst{\ref{inst:0276}}\relax
\and M.        ~Fraser                        \inst{\ref{inst:0009}}\relax
\and J.        ~Fuchs                         \inst{\ref{inst:0278}}\relax
\and R.        ~Furnell                       \inst{\ref{inst:0024}}\relax
\and M.        ~Gai                           \inst{\ref{inst:0035}}\relax
\and S.        ~Galleti                       \inst{\ref{inst:0045}}\relax
\and L.        ~Galluccio                     \inst{\ref{inst:0017}}\relax
\and D.        ~Garabato                      \inst{\ref{inst:0118}}\relax
\and F.        ~Garc\'{i}a-Sedano             \inst{\ref{inst:0129}}\relax
\and P.        ~Gar\'{e}                      \inst{\ref{inst:0024}}\relax
\and A.        ~Garofalo                      \inst{\ref{inst:0045}}\relax
\and N.        ~Garralda                      \inst{\ref{inst:0012}}\relax
\and P.        ~Gavras                        \inst{\ref{inst:0005},\ref{inst:0080},\ref{inst:0114}}\relax
\and J.        ~Gerssen                       \inst{\ref{inst:0253}}\relax
\and R.        ~Geyer                         \inst{\ref{inst:0013}}\relax
\and G.        ~Gilmore                       \inst{\ref{inst:0009}}\relax
\and S.        ~Girona                        \inst{\ref{inst:0294}}\relax
\and G.        ~Giuffrida                     \inst{\ref{inst:0124}}\relax
\and M.        ~Gomes                         \inst{\ref{inst:0119}}\relax
\and A.        ~Gonz\'{a}lez-Marcos           \inst{\ref{inst:0297}}\relax
\and J.        ~Gonz\'{a}lez-N\'{u}\~{n}ez    \inst{\ref{inst:0048},\ref{inst:0299}}\relax
\and J.J.      ~Gonz\'{a}lez-Vidal            \inst{\ref{inst:0012}}\relax
\and M.        ~Granvik                       \inst{\ref{inst:0120}}\relax
\and A.        ~Guerrier                      \inst{\ref{inst:0139}}\relax
\and P.        ~Guillout                      \inst{\ref{inst:0103}}\relax
\and J.        ~Guiraud                       \inst{\ref{inst:0019}}\relax
\and A.        ~G\'{u}rpide                   \inst{\ref{inst:0012}}\relax
\and R.        ~Guti\'{e}rrez-S\'{a}nchez     \inst{\ref{inst:0026}}\relax
\and L.P.      ~Guy                           \inst{\ref{inst:0057}}\relax
\and R.        ~Haigron                       \inst{\ref{inst:0005}}\relax
\and D.        ~Hatzidimitriou                \inst{\ref{inst:0114}}\relax
\and M.        ~Haywood                       \inst{\ref{inst:0005}}\relax
\and U.        ~Heiter                        \inst{\ref{inst:0115}}\relax
\and A.        ~Helmi                         \inst{\ref{inst:0191}}\relax
\and D.        ~Hobbs                         \inst{\ref{inst:0015}}\relax
\and W.        ~Hofmann                       \inst{\ref{inst:0007}}\relax
\and B.        ~Holl                          \inst{\ref{inst:0010}}\relax
\and G.        ~Holland                       \inst{\ref{inst:0009}}\relax
\and J.A.S.    ~Hunt                          \inst{\ref{inst:0034}}\relax
\and A.        ~Hypki                         \inst{\ref{inst:0003}}\relax
\and V.        ~Icardi                        \inst{\ref{inst:0044}}\relax
\and M.        ~Irwin                         \inst{\ref{inst:0009}}\relax
\and G.        ~Jevardat de Fombelle          \inst{\ref{inst:0057}}\relax
\and P.        ~Jofr\'{e}                     \inst{\ref{inst:0009},\ref{inst:0027}}\relax
\and P.G.      ~Jonker                        \inst{\ref{inst:0324},\ref{inst:0032}}\relax
\and A.        ~Jorissen                      \inst{\ref{inst:0021}}\relax
\and F.        ~Julbe                         \inst{\ref{inst:0012}}\relax
\and A.        ~Karampelas                    \inst{\ref{inst:0114},\ref{inst:0080}}\relax
\and A.        ~Kochoska                      \inst{\ref{inst:0330}}\relax
\and R.        ~Kohley                        \inst{\ref{inst:0014}}\relax
\and K.        ~Kolenberg                     \inst{\ref{inst:0332},\ref{inst:0031},\ref{inst:0334}}\relax
\and E.        ~Kontizas                      \inst{\ref{inst:0080}}\relax
\and S.E.      ~Koposov                       \inst{\ref{inst:0009}}\relax
\and G.        ~Kordopatis                    \inst{\ref{inst:0253},\ref{inst:0017}}\relax
\and P.        ~Koubsky                       \inst{\ref{inst:0278}}\relax
\and A.        ~Kowalczyk                     \inst{\ref{inst:0018}}\relax
\and A.        ~Krone-Martins                 \inst{\ref{inst:0119}}\relax
\and M.        ~Kudryashova                   \inst{\ref{inst:0082}}\relax
\and I.        ~Kull                          \inst{\ref{inst:0262}}\relax
\and R.K.      ~Bachchan                      \inst{\ref{inst:0015}}\relax
\and F.        ~Lacoste-Seris                 \inst{\ref{inst:0139}}\relax
\and A.F.      ~Lanza                         \inst{\ref{inst:0117}}\relax
\and J.-B.     ~Lavigne                       \inst{\ref{inst:0139}}\relax
\and C.        ~Le Poncin-Lafitte             \inst{\ref{inst:0077}}\relax
\and Y.        ~Lebreton                      \inst{\ref{inst:0005},\ref{inst:0350}}\relax
\and T.        ~Lebzelter                     \inst{\ref{inst:0145}}\relax
\and S.        ~Leccia                        \inst{\ref{inst:0234}}\relax
\and N.        ~Leclerc                       \inst{\ref{inst:0005}}\relax
\and I.        ~Lecoeur-Taibi                 \inst{\ref{inst:0057}}\relax
\and V.        ~Lemaitre                      \inst{\ref{inst:0139}}\relax
\and H.        ~Lenhardt                      \inst{\ref{inst:0007}}\relax
\and F.        ~Leroux                        \inst{\ref{inst:0139}}\relax
\and S.        ~Liao                          \inst{\ref{inst:0035},\ref{inst:0359}}\relax
\and E.        ~Licata                        \inst{\ref{inst:0142}}\relax
\and H.E.P.    ~Lindstr{\o}m                  \inst{\ref{inst:0036},\ref{inst:0362}}\relax
\and T.A.      ~Lister                        \inst{\ref{inst:0363}}\relax
\and E.        ~Livanou                       \inst{\ref{inst:0114}}\relax
\and A.        ~Lobel                         \inst{\ref{inst:0083}}\relax
\and W.        ~L\"{ o}ffler                  \inst{\ref{inst:0007}}\relax
\and M.        ~L\'{o}pez                     \inst{\ref{inst:0168}}\relax
\and A.        ~Lopez-Lozano                  \inst{\ref{inst:0368}}\relax
\and D.        ~Lorenz                        \inst{\ref{inst:0145}}\relax
\and T.        ~Loureiro                      \inst{\ref{inst:0018}}\relax
\and I.        ~MacDonald                     \inst{\ref{inst:0093}}\relax
\and T.        ~Magalh\~{a}es Fernandes       \inst{\ref{inst:0263}}\relax
\and S.        ~Managau                       \inst{\ref{inst:0139}}\relax
\and R.G.      ~Mann                          \inst{\ref{inst:0093}}\relax
\and G.        ~Mantelet                      \inst{\ref{inst:0007}}\relax
\and O.        ~Marchal                       \inst{\ref{inst:0005}}\relax
\and J.M.      ~Marchant                      \inst{\ref{inst:0377}}\relax
\and M.        ~Marconi                       \inst{\ref{inst:0234}}\relax
\and J.        ~Marie                         \inst{\ref{inst:0379}}\relax
\and S.        ~Marinoni                      \inst{\ref{inst:0208},\ref{inst:0124}}\relax
\and P.M.      ~Marrese                       \inst{\ref{inst:0208},\ref{inst:0124}}\relax
\and G.        ~Marschalk\'{o}                \inst{\ref{inst:0384},\ref{inst:0385}}\relax
\and D.J.      ~Marshall                      \inst{\ref{inst:0386}}\relax
\and J.M.      ~Mart\'{i}n-Fleitas            \inst{\ref{inst:0111}}\relax
\and M.        ~Martino                       \inst{\ref{inst:0044}}\relax
\and N.        ~Mary                          \inst{\ref{inst:0139}}\relax
\and G.        ~Matijevi\v{c}                 \inst{\ref{inst:0253}}\relax
\and T.        ~Mazeh                         \inst{\ref{inst:0262}}\relax
\and P.J.      ~McMillan                      \inst{\ref{inst:0015}}\relax
\and S.        ~Messina                       \inst{\ref{inst:0117}}\relax
\and A.        ~Mestre                        \inst{\ref{inst:0394}}\relax
\and D.        ~Michalik                      \inst{\ref{inst:0015}}\relax
\and N.R.      ~Millar                        \inst{\ref{inst:0009}}\relax
\and B.M.H.    ~Miranda                       \inst{\ref{inst:0119}}\relax
\and D.        ~Molina                        \inst{\ref{inst:0012}}\relax
\and R.        ~Molinaro                      \inst{\ref{inst:0234}}\relax
\and M.        ~Molinaro                      \inst{\ref{inst:0400}}\relax
\and L.        ~Moln\'{a}r                    \inst{\ref{inst:0384}}\relax
\and M.        ~Moniez                        \inst{\ref{inst:0402}}\relax
\and P.        ~Montegriffo                   \inst{\ref{inst:0045}}\relax
\and D.        ~Monteiro                      \inst{\ref{inst:0024}}\relax
\and R.        ~Mor                           \inst{\ref{inst:0012}}\relax
\and A.        ~Mora                          \inst{\ref{inst:0111}}\relax
\and R.        ~Morbidelli                    \inst{\ref{inst:0035}}\relax
\and T.        ~Morel                         \inst{\ref{inst:0096}}\relax
\and S.        ~Morgenthaler                  \inst{\ref{inst:0409}}\relax
\and T.        ~Morley                        \inst{\ref{inst:0241}}\relax
\and D.        ~Morris                        \inst{\ref{inst:0093}}\relax
\and A.F.      ~Mulone                        \inst{\ref{inst:0044}}\relax
\and T.        ~Muraveva                      \inst{\ref{inst:0045}}\relax
\and I.        ~Musella                       \inst{\ref{inst:0234}}\relax
\and J.        ~Narbonne                      \inst{\ref{inst:0139}}\relax
\and G.        ~Nelemans                      \inst{\ref{inst:0032},\ref{inst:0031}}\relax
\and L.        ~Nicastro                      \inst{\ref{inst:0418}}\relax
\and L.        ~Noval                         \inst{\ref{inst:0139}}\relax
\and C.        ~Ord\'{e}novic                 \inst{\ref{inst:0017}}\relax
\and J.        ~Ordieres-Mer\'{e}             \inst{\ref{inst:0421}}\relax
\and P.        ~Osborne                       \inst{\ref{inst:0009}}\relax
\and C.        ~Pagani                        \inst{\ref{inst:0170}}\relax
\and I.        ~Pagano                        \inst{\ref{inst:0117}}\relax
\and F.        ~Pailler                       \inst{\ref{inst:0019}}\relax
\and H.        ~Palacin                       \inst{\ref{inst:0139}}\relax
\and L.        ~Palaversa                     \inst{\ref{inst:0010}}\relax
\and P.        ~Parsons                       \inst{\ref{inst:0026}}\relax
\and T.        ~Paulsen                       \inst{\ref{inst:0024}}\relax
\and M.        ~Pecoraro                      \inst{\ref{inst:0142}}\relax
\and R.        ~Pedrosa                       \inst{\ref{inst:0431}}\relax
\and H.        ~Pentik\"{ a}inen              \inst{\ref{inst:0120}}\relax
\and J.        ~Pereira                       \inst{\ref{inst:0024}}\relax
\and B.        ~Pichon                        \inst{\ref{inst:0017}}\relax
\and A.M.      ~Piersimoni                    \inst{\ref{inst:0232}}\relax
\and F.-X.     ~Pineau                        \inst{\ref{inst:0103}}\relax
\and E.        ~Plachy                        \inst{\ref{inst:0384}}\relax
\and G.        ~Plum                          \inst{\ref{inst:0005}}\relax
\and E.        ~Poujoulet                     \inst{\ref{inst:0439}}\relax
\and A.        ~Pr\v{s}a                      \inst{\ref{inst:0440}}\relax
\and L.        ~Pulone                        \inst{\ref{inst:0208}}\relax
\and S.        ~Ragaini                       \inst{\ref{inst:0045}}\relax
\and S.        ~Rago                          \inst{\ref{inst:0035}}\relax
\and N.        ~Rambaux                       \inst{\ref{inst:0082}}\relax
\and M.        ~Ramos-Lerate                  \inst{\ref{inst:0445}}\relax
\and P.        ~Ranalli                       \inst{\ref{inst:0015}}\relax
\and G.        ~Rauw                          \inst{\ref{inst:0096}}\relax
\and A.        ~Read                          \inst{\ref{inst:0170}}\relax
\and S.        ~Regibo                        \inst{\ref{inst:0031}}\relax
\and F.        ~Renk                          \inst{\ref{inst:0018}}\relax
\and C.        ~Reyl\'{e}                     \inst{\ref{inst:0126}}\relax
\and R.A.      ~Ribeiro                       \inst{\ref{inst:0263}}\relax
\and L.        ~Rimoldini                     \inst{\ref{inst:0057}}\relax
\and V.        ~Ripepi                        \inst{\ref{inst:0234}}\relax
\and A.        ~Riva                          \inst{\ref{inst:0035}}\relax
\and G.        ~Rixon                         \inst{\ref{inst:0009}}\relax
\and M.        ~Roelens                       \inst{\ref{inst:0010}}\relax
\and M.        ~Romero-G\'{o}mez              \inst{\ref{inst:0012}}\relax
\and N.        ~Rowell                        \inst{\ref{inst:0093}}\relax
\and F.        ~Royer                         \inst{\ref{inst:0005}}\relax
\and A.        ~Rudolph                       \inst{\ref{inst:0018}}\relax
\and L.        ~Ruiz-Dern                     \inst{\ref{inst:0005}}\relax
\and G.        ~Sadowski                      \inst{\ref{inst:0021}}\relax
\and T.        ~Sagrist\`{a} Sell\'{e}s       \inst{\ref{inst:0007}}\relax
\and J.        ~Sahlmann                      \inst{\ref{inst:0014}}\relax
\and J.        ~Salgado                       \inst{\ref{inst:0122}}\relax
\and E.        ~Salguero                      \inst{\ref{inst:0122}}\relax
\and M.        ~Sarasso                       \inst{\ref{inst:0035}}\relax
\and H.        ~Savietto                      \inst{\ref{inst:0469}}\relax
\and A.        ~Schnorhk                      \inst{\ref{inst:0024}}\relax
\and M.        ~Schultheis                    \inst{\ref{inst:0017}}\relax
\and E.        ~Sciacca                       \inst{\ref{inst:0117}}\relax
\and M.        ~Segol                         \inst{\ref{inst:0473}}\relax
\and J.C.      ~Segovia                       \inst{\ref{inst:0048}}\relax
\and D.        ~Segransan                     \inst{\ref{inst:0010}}\relax
\and E.        ~Serpell                       \inst{\ref{inst:0241}}\relax
\and I-C.      ~Shih                          \inst{\ref{inst:0005}}\relax
\and R.        ~Smareglia                     \inst{\ref{inst:0400}}\relax
\and R.L.      ~Smart                         \inst{\ref{inst:0035}}\relax
\and C.        ~Smith                         \inst{\ref{inst:0480}}\relax
\and E.        ~Solano                        \inst{\ref{inst:0168},\ref{inst:0482}}\relax
\and F.        ~Solitro                       \inst{\ref{inst:0044}}\relax
\and R.        ~Sordo                         \inst{\ref{inst:0004}}\relax
\and S.        ~Soria Nieto                   \inst{\ref{inst:0012}}\relax
\and J.        ~Souchay                       \inst{\ref{inst:0077}}\relax
\and A.        ~Spagna                        \inst{\ref{inst:0035}}\relax
\and F.        ~Spoto                         \inst{\ref{inst:0017}}\relax
\and U.        ~Stampa                        \inst{\ref{inst:0007}}\relax
\and I.A.      ~Steele                        \inst{\ref{inst:0377}}\relax
\and H.        ~Steidelm\"{ u}ller            \inst{\ref{inst:0013}}\relax
\and C.A.      ~Stephenson                    \inst{\ref{inst:0026}}\relax
\and H.        ~Stoev                         \inst{\ref{inst:0493}}\relax
\and F.F.      ~Suess                         \inst{\ref{inst:0009}}\relax
\and M.        ~S\"{ u}veges                  \inst{\ref{inst:0057}}\relax
\and J.        ~Surdej                        \inst{\ref{inst:0096}}\relax
\and L.        ~Szabados                      \inst{\ref{inst:0384}}\relax
\and E.        ~Szegedi-Elek                  \inst{\ref{inst:0384}}\relax
\and D.        ~Tapiador                      \inst{\ref{inst:0499},\ref{inst:0500}}\relax
\and F.        ~Taris                         \inst{\ref{inst:0077}}\relax
\and G.        ~Tauran                        \inst{\ref{inst:0139}}\relax
\and M.B.      ~Taylor                        \inst{\ref{inst:0503}}\relax
\and R.        ~Teixeira                      \inst{\ref{inst:0235}}\relax
\and D.        ~Terrett                       \inst{\ref{inst:0061}}\relax
\and B.        ~Tingley                       \inst{\ref{inst:0506}}\relax
\and S.C.      ~Trager                        \inst{\ref{inst:0191}}\relax
\and C.        ~Turon                         \inst{\ref{inst:0005}}\relax
\and A.        ~Ulla                          \inst{\ref{inst:0509}}\relax
\and E.        ~Utrilla                       \inst{\ref{inst:0111}}\relax
\and G.        ~Valentini                     \inst{\ref{inst:0232}}\relax
\and A.        ~van Elteren                   \inst{\ref{inst:0003}}\relax
\and E.        ~Van Hemelryck                 \inst{\ref{inst:0083}}\relax
\and M.        ~van Leeuwen                   \inst{\ref{inst:0009}}\relax
\and M.        ~Varadi                        \inst{\ref{inst:0010},\ref{inst:0384}}\relax
\and A.        ~Vecchiato                     \inst{\ref{inst:0035}}\relax
\and J.        ~Veljanoski                    \inst{\ref{inst:0191}}\relax
\and T.        ~Via                           \inst{\ref{inst:0202}}\relax
\and D.        ~Vicente                       \inst{\ref{inst:0294}}\relax
\and S.        ~Vogt                          \inst{\ref{inst:0521}}\relax
\and H.        ~Voss                          \inst{\ref{inst:0012}}\relax
\and V.        ~Votruba                       \inst{\ref{inst:0278}}\relax
\and S.        ~Voutsinas                     \inst{\ref{inst:0093}}\relax
\and G.        ~Walmsley                      \inst{\ref{inst:0019}}\relax
\and M.        ~Weiler                        \inst{\ref{inst:0012}}\relax
\and K.        ~Weingrill                     \inst{\ref{inst:0253}}\relax
\and D.        ~Werner                        \inst{\ref{inst:0018}}\relax
\and T.        ~Wevers                        \inst{\ref{inst:0032}}\relax
\and G.        ~Whitehead                     \inst{\ref{inst:0018}}\relax
\and \L{}.     ~Wyrzykowski                   \inst{\ref{inst:0009},\ref{inst:0532}}\relax
\and A.        ~Yoldas                        \inst{\ref{inst:0009}}\relax
\and M.        ~\v{Z}erjal                    \inst{\ref{inst:0330}}\relax
\and S.        ~Zucker                        \inst{\ref{inst:0249}}\relax
\and C.        ~Zurbach                       \inst{\ref{inst:0112}}\relax
\and T.        ~Zwitter                       \inst{\ref{inst:0330}}\relax
\and A.        ~Alecu                         \inst{\ref{inst:0009}}\relax
\and M.        ~Allen                         \inst{\ref{inst:0001}}\relax
\and C.        ~Allende Prieto                \inst{\ref{inst:0034},\ref{inst:0541},\ref{inst:0542}}\relax
\and A.        ~Amorim                        \inst{\ref{inst:0119}}\relax
\and G.        ~Anglada-Escud\'{e}            \inst{\ref{inst:0012}}\relax
\and V.        ~Arsenijevic                   \inst{\ref{inst:0119}}\relax
\and S.        ~Azaz                          \inst{\ref{inst:0001}}\relax
\and P.        ~Balm                          \inst{\ref{inst:0026}}\relax
\and M.        ~Beck                          \inst{\ref{inst:0057}}\relax
\and H.-H.     ~Bernstein$^\dagger$           \inst{\ref{inst:0007}}\relax
\and L.        ~Bigot                         \inst{\ref{inst:0017}}\relax
\and A.        ~Bijaoui                       \inst{\ref{inst:0017}}\relax
\and C.        ~Blasco                        \inst{\ref{inst:0552}}\relax
\and M.        ~Bonfigli                      \inst{\ref{inst:0232}}\relax
\and G.        ~Bono                          \inst{\ref{inst:0208}}\relax
\and S.        ~Boudreault                    \inst{\ref{inst:0034},\ref{inst:0556}}\relax
\and A.        ~Bressan                       \inst{\ref{inst:0557}}\relax
\and S.        ~Brown                         \inst{\ref{inst:0009}}\relax
\and P.-M.     ~Brunet                        \inst{\ref{inst:0019}}\relax
\and P.        ~Bunclark$^\dagger$            \inst{\ref{inst:0009}}\relax
\and R.        ~Buonanno                      \inst{\ref{inst:0208}}\relax
\and A.G.      ~Butkevich                     \inst{\ref{inst:0013}}\relax
\and C.        ~Carret                        \inst{\ref{inst:0431}}\relax
\and C.        ~Carrion                       \inst{\ref{inst:0129}}\relax
\and L.        ~Chemin                        \inst{\ref{inst:0027},\ref{inst:0566}}\relax
\and F.        ~Ch\'{e}reau                   \inst{\ref{inst:0005}}\relax
\and L.        ~Corcione                      \inst{\ref{inst:0035}}\relax
\and E.        ~Darmigny                      \inst{\ref{inst:0019}}\relax
\and K.S.      ~de Boer                       \inst{\ref{inst:0570}}\relax
\and P.        ~de Teodoro                    \inst{\ref{inst:0048}}\relax
\and P.T.      ~de Zeeuw                      \inst{\ref{inst:0003},\ref{inst:0573}}\relax
\and C.        ~Delle Luche                   \inst{\ref{inst:0005},\ref{inst:0139}}\relax
\and C.D.      ~Domingues                     \inst{\ref{inst:0576}}\relax
\and P.        ~Dubath                        \inst{\ref{inst:0057}}\relax
\and F.        ~Fodor                         \inst{\ref{inst:0019}}\relax
\and B.        ~Fr\'{e}zouls                  \inst{\ref{inst:0019}}\relax
\and A.        ~Fries                         \inst{\ref{inst:0012}}\relax
\and D.        ~Fustes                        \inst{\ref{inst:0118}}\relax
\and D.        ~Fyfe                          \inst{\ref{inst:0170}}\relax
\and E.        ~Gallardo                      \inst{\ref{inst:0012}}\relax
\and J.        ~Gallegos                      \inst{\ref{inst:0048}}\relax
\and D.        ~Gardiol                       \inst{\ref{inst:0035}}\relax
\and M.        ~Gebran                        \inst{\ref{inst:0012},\ref{inst:0587}}\relax
\and A.        ~Gomboc                        \inst{\ref{inst:0330},\ref{inst:0589}}\relax
\and A.        ~G\'{o}mez                     \inst{\ref{inst:0005}}\relax
\and E.        ~Grux                          \inst{\ref{inst:0126}}\relax
\and A.        ~Gueguen                       \inst{\ref{inst:0005},\ref{inst:0593}}\relax
\and A.        ~Heyrovsky                     \inst{\ref{inst:0093}}\relax
\and J.        ~Hoar                          \inst{\ref{inst:0014}}\relax
\and G.        ~Iannicola                     \inst{\ref{inst:0208}}\relax
\and Y.        ~Isasi Parache                 \inst{\ref{inst:0012}}\relax
\and A.-M.     ~Janotto                       \inst{\ref{inst:0019}}\relax
\and E.        ~Joliet                        \inst{\ref{inst:0095},\ref{inst:0600}}\relax
\and A.        ~Jonckheere                    \inst{\ref{inst:0083}}\relax
\and R.        ~Keil                          \inst{\ref{inst:0602},\ref{inst:0603}}\relax
\and D.-W.     ~Kim                           \inst{\ref{inst:0006}}\relax
\and P.        ~Klagyivik                     \inst{\ref{inst:0384}}\relax
\and J.        ~Klar                          \inst{\ref{inst:0253}}\relax
\and J.        ~Knude                         \inst{\ref{inst:0036}}\relax
\and O.        ~Kochukhov                     \inst{\ref{inst:0115}}\relax
\and I.        ~Kolka                         \inst{\ref{inst:0609}}\relax
\and J.        ~Kos                           \inst{\ref{inst:0330},\ref{inst:0611}}\relax
\and A.        ~Kutka                         \inst{\ref{inst:0278},\ref{inst:0613}}\relax
\and V.        ~Lainey                        \inst{\ref{inst:0082}}\relax
\and D.        ~LeBouquin                     \inst{\ref{inst:0139}}\relax
\and C.        ~Liu                           \inst{\ref{inst:0006},\ref{inst:0617}}\relax
\and D.        ~Loreggia                      \inst{\ref{inst:0035}}\relax
\and V.V.      ~Makarov                       \inst{\ref{inst:0619}}\relax
\and M.G.      ~Marseille                     \inst{\ref{inst:0139}}\relax
\and C.        ~Martayan                      \inst{\ref{inst:0083},\ref{inst:0622}}\relax
\and O.        ~Martinez-Rubi                 \inst{\ref{inst:0012}}\relax
\and B.        ~Massart                       \inst{\ref{inst:0017},\ref{inst:0139},\ref{inst:0020}}\relax
\and F.        ~Meynadier                     \inst{\ref{inst:0005},\ref{inst:0077}}\relax
\and S.        ~Mignot                        \inst{\ref{inst:0005}}\relax
\and U.        ~Munari                        \inst{\ref{inst:0004}}\relax
\and A.-T.     ~Nguyen                        \inst{\ref{inst:0019}}\relax
\and T.        ~Nordlander                    \inst{\ref{inst:0115}}\relax
\and P.        ~Ocvirk                        \inst{\ref{inst:0253},\ref{inst:0103}}\relax
\and K.S.      ~O'Flaherty                    \inst{\ref{inst:0635}}\relax
\and A.        ~Olias Sanz                    \inst{\ref{inst:0636}}\relax
\and P.        ~Ortiz                         \inst{\ref{inst:0170}}\relax
\and J.        ~Osorio                        \inst{\ref{inst:0155}}\relax
\and D.        ~Oszkiewicz                    \inst{\ref{inst:0120},\ref{inst:0640}}\relax
\and A.        ~Ouzounis                      \inst{\ref{inst:0093}}\relax
\and M.        ~Palmer                        \inst{\ref{inst:0012}}\relax
\and P.        ~Park                          \inst{\ref{inst:0010}}\relax
\and E.        ~Pasquato                      \inst{\ref{inst:0021}}\relax
\and C.        ~Peltzer                       \inst{\ref{inst:0009}}\relax
\and J.        ~Peralta                       \inst{\ref{inst:0012}}\relax
\and F.        ~P\'{e}turaud                  \inst{\ref{inst:0005}}\relax
\and T.        ~Pieniluoma                    \inst{\ref{inst:0120}}\relax
\and E.        ~Pigozzi                       \inst{\ref{inst:0044}}\relax
\and J.        ~Poels$^\dagger$               \inst{\ref{inst:0096}}\relax
\and G.        ~Prat                          \inst{\ref{inst:0651}}\relax
\and T.        ~Prod'homme                    \inst{\ref{inst:0003},\ref{inst:0024}}\relax
\and F.        ~Raison                        \inst{\ref{inst:0654},\ref{inst:0593}}\relax
\and J.M.      ~Rebordao                      \inst{\ref{inst:0576}}\relax
\and D.        ~Risquez                       \inst{\ref{inst:0003}}\relax
\and B.        ~Rocca-Volmerange              \inst{\ref{inst:0658}}\relax
\and S.        ~Rosen                         \inst{\ref{inst:0034},\ref{inst:0170}}\relax
\and M.I.      ~Ruiz-Fuertes                  \inst{\ref{inst:0057}}\relax
\and F.        ~Russo                         \inst{\ref{inst:0035}}\relax
\and S.        ~Sembay                        \inst{\ref{inst:0170}}\relax
\and I.        ~Serraller Vizcaino            \inst{\ref{inst:0664}}\relax
\and A.        ~Short                         \inst{\ref{inst:0001}}\relax
\and A.        ~Siebert                       \inst{\ref{inst:0103},\ref{inst:0253}}\relax
\and H.        ~Silva                         \inst{\ref{inst:0263}}\relax
\and D.        ~Sinachopoulos                 \inst{\ref{inst:0080}}\relax
\and E.        ~Slezak                        \inst{\ref{inst:0017}}\relax
\and M.        ~Soffel                        \inst{\ref{inst:0013}}\relax
\and D.        ~Sosnowska                     \inst{\ref{inst:0010}}\relax
\and V.        ~Strai\v{z}ys                  \inst{\ref{inst:0673}}\relax
\and M.        ~ter Linden                    \inst{\ref{inst:0095},\ref{inst:0675}}\relax
\and D.        ~Terrell                       \inst{\ref{inst:0676}}\relax
\and S.        ~Theil                         \inst{\ref{inst:0677}}\relax
\and C.        ~Tiede                         \inst{\ref{inst:0006},\ref{inst:0679}}\relax
\and L.        ~Troisi                        \inst{\ref{inst:0124},\ref{inst:0681}}\relax
\and P.        ~Tsalmantza                    \inst{\ref{inst:0006}}\relax
\and D.        ~Tur                           \inst{\ref{inst:0202}}\relax
\and M.        ~Vaccari                       \inst{\ref{inst:0684},\ref{inst:0685}}\relax
\and F.        ~Vachier                       \inst{\ref{inst:0082}}\relax
\and P.        ~Valles                        \inst{\ref{inst:0012}}\relax
\and W.        ~Van Hamme                     \inst{\ref{inst:0688}}\relax
\and L.        ~Veltz                         \inst{\ref{inst:0253},\ref{inst:0091}}\relax
\and J.        ~Virtanen                      \inst{\ref{inst:0120},\ref{inst:0121}}\relax
\and J.-M.     ~Wallut                        \inst{\ref{inst:0019}}\relax
\and R.        ~Wichmann                      \inst{\ref{inst:0694}}\relax
\and M.I.      ~Wilkinson                     \inst{\ref{inst:0009},\ref{inst:0170}}\relax
\and H.        ~Ziaeepour                     \inst{\ref{inst:0126}}\relax
\and S.        ~Zschocke                      \inst{\ref{inst:0013}}\relax
}

\institute{
     Scientific Support Office, Directorate of Science, European Space Research and Technology Centre (ESA/ESTEC), Keplerlaan 1, 2201AZ, Noordwijk, The Netherlands\relax                                    \label{inst:0001}
\and Leiden Observatory, Leiden University, Niels Bohrweg 2, 2333 CA Leiden, The Netherlands\relax                                                                                                           \label{inst:0003}
\and INAF - Osservatorio astronomico di Padova, Vicolo Osservatorio 5, 35122 Padova, Italy\relax                                                                                                             \label{inst:0004}
\and GEPI, Observatoire de Paris, PSL Research University, CNRS, Univ. Paris Diderot, Sorbonne Paris Cit{\'e}, 5 Place Jules Janssen, 92190 Meudon, France\relax                                             \label{inst:0005}
\and Max Planck Institute for Astronomy, K\"{ o}nigstuhl 17, 69117 Heidelberg, Germany\relax                                                                                                                 \label{inst:0006}
\and Astronomisches Rechen-Institut, Zentrum f\"{ u}r Astronomie der Universit\"{ a}t Heidelberg, M\"{ o}nchhofstr. 12-14, D-69120 Heidelberg, Germany\relax                                                 \label{inst:0007}
\and Institute of Astronomy, University of Cambridge, Madingley Road, Cambridge CB3 0HA, United Kingdom\relax                                                                                                \label{inst:0009}
\and Department of Astronomy, University of Geneva, Chemin des Maillettes 51, CH-1290 Versoix, Switzerland\relax                                                                                             \label{inst:0010}
\and Mission Operations Division, Operations Department, Directorate of Science, European Space Research and Technology Centre (ESA/ESTEC), Keplerlaan 1, 2201 AZ, Noordwijk, The Netherlands\relax          \label{inst:0011}
\and Institut de Ci\`{e}ncies del Cosmos, Universitat  de  Barcelona  (IEEC-UB), Mart\'{i}  Franqu\`{e}s  1, E-08028 Barcelona, Spain\relax                                                                  \label{inst:0012}
\and Lohrmann Observatory, Technische Universit\"{ a}t Dresden, Mommsenstra{\ss}e 13, 01062 Dresden, Germany\relax                                                                                           \label{inst:0013}
\and European Space Astronomy Centre (ESA/ESAC), Camino bajo del Castillo, s/n, Urbanizacion Villafranca del Castillo, Villanueva de la Ca\~{n}ada, E-28692 Madrid, Spain\relax                              \label{inst:0014}
\and Lund Observatory, Department of Astronomy and Theoretical Physics, Lund University, Box 43, SE-22100 Lund, Sweden\relax                                                                                 \label{inst:0015}
\and Laboratoire Lagrange, Universit\'{e} Nice Sophia-Antipolis, Observatoire de la C\^{o}te d'Azur, CNRS, CS 34229, F-06304 Nice Cedex, France\relax                                                        \label{inst:0017}
\and European Space Operations Centre (ESA/ESOC), Robert Bosch Stra{\ss}e 5, 64293 Darmstadt, Germany\relax                                                                                                  \label{inst:0018}
\and CNES Centre Spatial de Toulouse, 18 avenue Edouard Belin, 31401 Toulouse Cedex 9, France\relax                                                                                                          \label{inst:0019}
\and Airbus Defence and Space SAS, 31 Rue des Cosmonautes, 31402 Toulouse Cedex 4, France\relax                                                                                                              \label{inst:0020}
\and Institut d'Astronomie et d'Astrophysique, Universit\'{e} Libre de Bruxelles CP 226, Boulevard du Triomphe, 1050 Brussels, Belgium\relax                                                                 \label{inst:0021}
\and F.R.S.-FNRS, Rue d'Egmont 5, 1000 Brussels, Belgium\relax                                                                                                                                               \label{inst:0022}
\and INAF - Osservatorio Astrofisico di Arcetri, Largo Enrico Fermi 5, I-50125 Firenze, Italy\relax                                                                                                          \label{inst:0023}
\and Directorate of Science, European Space Research and Technology Centre (ESA/ESTEC), Keplerlaan 1, 2201AZ, Noordwijk, The Netherlands\relax                                                               \label{inst:0024}
\and Telespazio Vega UK Ltd for ESA/ESAC, Camino bajo del Castillo, s/n, Urbanizacion Villafranca del Castillo, Villanueva de la Ca\~{n}ada, E-28692 Madrid, Spain\relax                                     \label{inst:0026}
\and Laboratoire d'astrophysique de Bordeaux, Universit\'{e} de Bordeaux, CNRS, B18N, all{\'e}e Geoffroy Saint-Hilaire, 33615 Pessac, France\relax                                                           \label{inst:0027}
\and Instituut voor Sterrenkunde, KU Leuven, Celestijnenlaan 200D, 3001 Leuven, Belgium\relax                                                                                                                \label{inst:0031}
\and Department of Astrophysics/IMAPP, Radboud University Nijmegen, P.O.Box 9010, 6500 GL Nijmegen, The Netherlands\relax                                                                                    \label{inst:0032}
\and Mullard Space Science Laboratory, University College London, Holmbury St Mary, Dorking, Surrey RH5 6NT, United Kingdom\relax                                                                            \label{inst:0034}
\and INAF - Osservatorio Astrofisico di Torino, via Osservatorio 20, 10025 Pino Torinese (TO), Italy\relax                                                                                                   \label{inst:0035}
\and Niels Bohr Institute, University of Copenhagen, Juliane Maries Vej 30, 2100 Copenhagen {\O}, Denmark\relax                                                                                              \label{inst:0036}
\and Centre for Electronic Imaging, Department of Physical Sciences, The Open University, Walton Hall MK7 6AA Milton Keynes, United Kingdom\relax                                                            \label{inst:0041}
\and ALTEC S.p.a, Corso Marche, 79,10146 Torino, Italy\relax                                                                                                                                                 \label{inst:0044}
\and INAF - Osservatorio Astronomico di Bologna, via Ranzani 1, 40127 Bologna,  Italy\relax                                                                                                                  \label{inst:0045}
\and Serco Gesti\'{o}n de Negocios for ESA/ESAC, Camino bajo del Castillo, s/n, Urbanizacion Villafranca del Castillo, Villanueva de la Ca\~{n}ada, E-28692 Madrid, Spain\relax                              \label{inst:0048}
\and Department of Astronomy, University of Geneva, Chemin d'Ecogia 16, CH-1290 Versoix, Switzerland\relax                                                                                                   \label{inst:0057}
\and STFC, Rutherford Appleton Laboratory, Harwell, Didcot, OX11 0QX, United Kingdom\relax                                                                                                                   \label{inst:0061}
\and Gaia DPAC Project Office, ESAC, Camino bajo del Castillo, s/n, Urbanizacion Villafranca del Castillo, Villanueva de la Ca\~{n}ada, E-28692 Madrid, Spain\relax                                          \label{inst:0067}
\and SYRTE, Observatoire de Paris, PSL Research University, CNRS, Sorbonne Universit{\'e}s, UPMC Univ. Paris 06, LNE, 61 avenue de l'Observatoire, 75014 Paris, France\relax                                 \label{inst:0077}
\and National Observatory of Athens, I. Metaxa and Vas. Pavlou, Palaia Penteli, 15236 Athens, Greece\relax                                                                                                   \label{inst:0080}
\and IMCCE, Observatoire de Paris, PSL Research University, CNRS, Sorbonne Universit{\'e}s, UPMC Univ. Paris 06, Univ. Lille, 77 av. Denfert-Rochereau, 75014 Paris, France\relax                            \label{inst:0082}
\and Royal Observatory of Belgium, Ringlaan 3, 1180 Brussels, Belgium\relax                                                                                                                                  \label{inst:0083}
\and Institut d'Astrophysique Spatiale, Universit\'{e} Paris XI, UMR 8617, CNRS, B\^{a}timent 121, 91405, Orsay Cedex, France\relax                                                                          \label{inst:0091}
\and Institute for Astronomy, Royal Observatory, University of Edinburgh, Blackford Hill, Edinburgh EH9 3HJ, United Kingdom\relax                                                                            \label{inst:0093}
\and HE Space Operations BV for ESA/ESAC, Camino bajo del Castillo, s/n, Urbanizacion Villafranca del Castillo, Villanueva de la Ca\~{n}ada, E-28692 Madrid, Spain\relax                                     \label{inst:0095}
\and Institut d'Astrophysique et de G\'{e}ophysique, Universit\'{e} de Li\`{e}ge, 19c, All\'{e}e du 6 Ao\^{u}t, B-4000 Li\`{e}ge, Belgium\relax                                                              \label{inst:0096}
\and \'{A}rea de Lenguajes y Sistemas Inform\'{a}ticos, Universidad Pablo de Olavide, Ctra. de Utrera, km 1. 41013, Sevilla, Spain\relax                                                                     \label{inst:0100}
\and Observatoire Astronomique de Strasbourg, Universit\'{e} de Strasbourg, CNRS, UMR 7550, 11 rue de l'Universit\'{e}, 67000 Strasbourg, France\relax                                                       \label{inst:0103}
\and Kavli Institute for Cosmology, University of Cambridge, Madingley Road, Cambride CB3 0HA, United Kingdom\relax                                                                                          \label{inst:0106}
\and Aurora Technology for ESA/ESAC, Camino bajo del Castillo, s/n, Urbanizacion Villafranca del Castillo, Villanueva de la Ca\~{n}ada, E-28692 Madrid, Spain\relax                                          \label{inst:0111}
\and Laboratoire Univers et Particules de Montpellier, Universit\'{e} Montpellier, Place Eug\`{e}ne Bataillon, CC72, 34095 Montpellier Cedex 05, France\relax                                                \label{inst:0112}
\and Department of Astrophysics, Astronomy and Mechanics, National and Kapodistrian University of Athens, Panepistimiopolis, Zografos, 15783 Athens, Greece\relax                                            \label{inst:0114}
\and Department of Physics and Astronomy, Division of Astronomy and Space Physics, Uppsala University, Box 516, 75120 Uppsala, Sweden\relax                                                                  \label{inst:0115}
\and Universit\`{a} di Catania, Dipartimento di Fisica e Astronomia, Sezione Astrofisica, Via S. Sofia 78, I-95123 Catania, Italy\relax                                                                      \label{inst:0116}
\and INAF - Osservatorio Astrofisico di Catania, via S. Sofia 78, 95123 Catania, Italy\relax                                                                                                                 \label{inst:0117}
\and Universidade da Coru\~{n}a, Facultade de Inform\'{a}tica, Campus de Elvi\~{n}a S/N, 15071, A Coru\~{n}a, Spain\relax                                                                                    \label{inst:0118}
\and CENTRA, Universidade de Lisboa, FCUL, Campo Grande, Edif. C8, 1749-016 Lisboa, Portugal\relax                                                                                                           \label{inst:0119}
\and University of Helsinki, Department of Physics, P.O. Box 64, FI-00014 University of Helsinki, Finland\relax                                                                                              \label{inst:0120}
\and Finnish Geospatial Research Institute FGI, Geodeetinrinne 2, FI-02430 Masala, Finland\relax                                                                                                             \label{inst:0121}
\and Isdefe for ESA/ESAC, Camino bajo del Castillo, s/n, Urbanizacion Villafranca del Castillo, Villanueva de la Ca\~{n}ada, E-28692 Madrid, Spain\relax                                                     \label{inst:0122}
\and ASI Science Data Center, via del Politecnico SNC, 00133 Roma, Italy\relax                                                                                                                               \label{inst:0124}
\and Institut UTINAM UMR6213, CNRS, OSU THETA Franche-Comt\'{e} Bourgogne, Universit\'{e} Bourgogne Franche-Comt\'{e}, F-25000 Besan\c{c}on, France\relax                                                    \label{inst:0126}
\and Dpto. de Inteligencia Artificial, UNED, c/ Juan del Rosal 16, 28040 Madrid, Spain\relax                                                                                                                 \label{inst:0129}
\and Elecnor Deimos Space for ESA/ESAC, Camino bajo del Castillo, s/n, Urbanizacion Villafranca del Castillo, Villanueva de la Ca\~{n}ada, E-28692 Madrid, Spain\relax                                       \label{inst:0138}
\and Thales Services for CNES Centre Spatial de Toulouse, 18 avenue Edouard Belin, 31401 Toulouse Cedex 9, France\relax                                                                                      \label{inst:0139}
\and EURIX S.r.l., via Carcano 26, 10153, Torino, Italy\relax                                                                                                                                                \label{inst:0142}
\and University of Vienna, Department of Astrophysics, T\"{ u}rkenschanzstra{\ss}e 17, A1180 Vienna, Austria\relax                                                                                           \label{inst:0145}
\and Department of Physics and Astronomy, The Johns Hopkins University, 3400 N Charles St, Baltimore, MD 21218, USA\relax                                                                                    \label{inst:0146}
\and ON/MCTI-BR, Rua Gal. Jos\'{e} Cristino 77, Rio de Janeiro, CEP 20921-400, RJ,  Brazil\relax                                                                                                             \label{inst:0148}
\and OV/UFRJ-BR, Ladeira Pedro Ant\^{o}nio 43, Rio de Janeiro, CEP 20080-090, RJ, Brazil\relax                                                                                                               \label{inst:0149}
\and Faculdade Ciencias, Universidade do Porto, Departamento Matematica Aplicada, Rua do Campo Alegre, 687 4169-007 Porto, Portugal\relax                                                                    \label{inst:0155}
\and Instituto de Astrof\'{\i}sica e Ci\^encias do Espa\,co, Universidade de Lisboa Faculdade de Ci\^encias, Campo Grande, PT1749-016 Lisboa, Portugal\relax                                                 \label{inst:0156}
\and Vitrociset Belgium for ESA/ESTEC, Keplerlaan 1, 2201AZ, Noordwijk, The Netherlands\relax                                                                                                                \label{inst:0159}
\and Departamento de Astrof\'{i}sica, Centro de Astrobiolog\'{i}a (CSIC-INTA), ESA-ESAC. Camino Bajo del Castillo s/n. 28692 Villanueva de la Ca\~{n}ada, Madrid, Spain\relax                                \label{inst:0168}
\and Department of Physics and Astronomy, University of Leicester, University Road, Leicester LE1 7RH, United Kingdom\relax                                                                                  \label{inst:0170}
\and Deimos Space S.L.U. for ESA/ESOC, Robert-Bosch-Stra{\ss}e 5, 64293 Darmstadt, Germany\relax                                                                                                             \label{inst:0173}
\and University of Oviedo, Campus Universitario, 33203 Gij\'{o}n, Spain\relax                                                                                                                                \label{inst:0174}
\and University of C\'{a}diz, Avd. De la universidad, Jerez de la Frontera, C\'{a}diz, Spain\relax                                                                                                           \label{inst:0177}
\and Kapteyn Astronomical Institute, University of Groningen, Landleven 12, 9747 AD Groningen, The Netherlands\relax                                                                                         \label{inst:0191}
\and Consorci de Serveis Universitaris de Catalunya, C/ Gran Capit\`{a}, 2-4 3rd floor, 08034 Barcelona, Spain\relax                                                                                         \label{inst:0202}
\and University of Turin, Department of Computer Sciences, Corso Svizzera 185, 10149 Torino, Italy\relax                                                                                                     \label{inst:0204}
\and INAF - Osservatorio Astronomico di Roma, Via di Frascati 33, 00078 Monte Porzio Catone (Roma), Italy\relax                                                                                              \label{inst:0208}
\and CRAAG - Centre de Recherche en Astronomie, Astrophysique et G\'{e}ophysique, Route de l'Observatoire Bp 63 Bouzareah 16340 Algiers, Algeria\relax                                                       \label{inst:0225}
\and Universiteit Antwerpen, Onderzoeksgroep Toegepaste Wiskunde, Middelheimlaan 1, 2020 Antwerpen, Belgium\relax                                                                                            \label{inst:0228}
\and Department of Physics and Astronomy, University of Padova, Via Marzolo 8, I-35131 Padova, Italy\relax                                                                                                   \label{inst:0230}
\and INAF - Osservatorio Astronomico di Teramo, Via Mentore Maggini, 64100 Teramo, Italy\relax                                                                                                               \label{inst:0232}
\and INAF - Osservatorio Astronomico di Capodimonte, Via Moiariello 16, 80131, Napoli, Italy\relax                                                                                                           \label{inst:0234}
\and Instituto de Astronomia, Geof\`{i}sica e Ci\^{e}ncias Atmosf\'{e}ricas, Universidade de S\~{a}o Paulo, Rua do Mat\~{a}o, 1226, Cidade Universitaria, 05508-900 S\~{a}o Paulo, SP, Brazil\relax          \label{inst:0235}
\and Telespazio Vega Deutschland GmbH for ESA/ESOC, Robert-Bosch-Stra{\ss}e 5, 64293 Darmstadt, Germany\relax                                                                                                \label{inst:0241}
\and Department of Geosciences, Tel Aviv University, Tel Aviv 6997801, Israel\relax                                                                                                                          \label{inst:0249}
\and Astronomical Institute Anton Pannekoek, University of Amsterdam, PO Box 94249, 1090 GE, Amsterdam, The Netherlands\relax                                                                                \label{inst:0250}
\and Leibniz Institute for Astrophysics Potsdam (AIP), An der Sternwarte 16, 14482 Potsdam, Germany\relax                                                                                                    \label{inst:0253}
\and ATOS for CNES Centre Spatial de Toulouse, 18 avenue Edouard Belin, 31401 Toulouse Cedex 9, France\relax                                                                                                 \label{inst:0259}
\and School of Physics and Astronomy, Tel Aviv University, Tel Aviv 6997801, Israel\relax                                                                                                                    \label{inst:0262}
\and UNINOVA - CTS, Campus FCT-UNL, Monte da Caparica, 2829-516 Caparica, Portugal\relax                                                                                                                     \label{inst:0263}
\and Laboratoire G\'{e}oazur, Universit\'{e} Nice Sophia-Antipolis, UMR 7329, CNRS, Observatoire de la C\^{o}te d'Azur, 250 rue A. Einstein, F-06560 Valbonne, France\relax                                  \label{inst:0270}
\and RHEA for ESA/ESAC, Camino bajo del Castillo, s/n, Urbanizacion Villafranca del Castillo, Villanueva de la Ca\~{n}ada, E-28692 Madrid, Spain\relax                                                       \label{inst:0276}
\and Astronomical Institute, Academy of Sciences of the Czech Republic, Fri\v{c}ova 298, 25165 Ond\v{r}ejov, Czech Republic\relax                                                                            \label{inst:0278}
\and Barcelona Supercomputing Center - Centro Nacional de Supercomputaci\'{o}n, c/ Jordi Girona 29, Ed. Nexus II, 08034 Barcelona, Spain\relax                                                               \label{inst:0294}
\and Department of Mechanical Engineering, University of La Rioja, c/ San Jos\'{e} de Calasanz, 31, 26004 Logro\~{n}o, La Rioja, Spain\relax                                                                 \label{inst:0297}
\and ETSE Telecomunicaci\'{o}n, Universidade de Vigo, Campus Lagoas-Marcosende, 36310 Vigo, Galicia, Spain\relax                                                                                             \label{inst:0299}
\and SRON, Netherlands Institute for Space Research, Sorbonnelaan 2, 3584CA, Utrecht, The Netherlands\relax                                                                                                  \label{inst:0324}
\and Faculty of Mathematics and Physics, University of Ljubljana, Jadranska ulica 19, 1000 Ljubljana, Slovenia\relax                                                                                         \label{inst:0330}
\and Physics Department, University of Antwerp, Groenenborgerlaan 171, 2020 Antwerp, Belgium\relax                                                                                                           \label{inst:0332}
\and Harvard-Smithsonian Center for Astrophysics, 60 Garden Street, Cambridge MA 02138, USA\relax                                                                                                            \label{inst:0334}
\and Institut de Physique de Rennes, Universit{\'e} de Rennes 1, F-35042 Rennes, France\relax                                                                                                                \label{inst:0350}
\and Shanghai Astronomical Observatory, Chinese Academy of Sciences, 80 Nandan Rd, 200030 Shanghai, China\relax                                                                                              \label{inst:0359}
\and CSC Danmark A/S, Retortvej 8, 2500 Valby, Denmark\relax                                                                                                                                                 \label{inst:0362}
\and Las Cumbres Observatory Global Telescope Network, Inc., 6740 Cortona Drive, Suite 102, Goleta, CA  93117, USA\relax                                                                                     \label{inst:0363}
\and CGI for ESA/ESOC, Robert-Bosch-Stra{\ss}e 5, 64293 Darmstadt, Germany\relax                                                                                                                             \label{inst:0368}
\and Astrophysics Research Institute, Liverpool John Moores University, L3 5RF, United Kingdom\relax                                                                                                         \label{inst:0377}
\and LSE Space GmbH for ESA/ESOC, Robert-Bosch-Stra{\ss}e 5, 64293 Darmstadt, Germany\relax                                                                                                                  \label{inst:0379}
\and Konkoly Observatory, Research Centre for Astronomy and Earth Sciences, Hungarian Academy of Sciences, Konkoly Thege Mikl\'{o}s \'{u}t 15-17, 1121 Budapest, Hungary\relax                               \label{inst:0384}
\and Baja Observatory of University of Szeged, Szegedi \'{u}t III/70, 6500 Baja, Hungary\relax                                                                                                               \label{inst:0385}
\and Laboratoire AIM, IRFU/Service d'Astrophysique - CEA/DSM - CNRS - Universit\'{e} Paris Diderot, B\^{a}t 709, CEA-Saclay, F-91191 Gif-sur-Yvette Cedex, France\relax                                      \label{inst:0386}
\and GMV for ESA/ESOC, Robert-Bosch-Stra{\ss}e 5, 64293 Darmstadt, Germany\relax                                                                                                                             \label{inst:0394}
\and INAF - Osservatorio Astronomico di Trieste, Via G.B. Tiepolo 11, 34143, Trieste, Italy\relax                                                                                                            \label{inst:0400}
\and Laboratoire de l'Acc\'{e}l\'{e}rateur Lin\'{e}aire, Universit\'{e} Paris-Sud, CNRS/IN2P3, Universit\'{e} Paris-Saclay, 91898 Orsay Cedex, France\relax                                                  \label{inst:0402}
\and \'{E}cole polytechnique f\'{e}d\'{e}rale de Lausanne, SB MATHAA STAP, MA B1 473 (B\^{a}timent MA), Station 8, CH-1015 Lausanne, Switzerland\relax                                                       \label{inst:0409}
\and INAF/IASF-Bologna, Via P. Gobetti 101, 40129 Bologna, Italy\relax                                                                                                                                       \label{inst:0418}
\and Technical University of Madrid, Jos\'{e} Guti\'{e}rrez Abascal 2, 28006 Madrid, Spain\relax                                                                                                             \label{inst:0421}
\and EQUERT International for CNES Centre Spatial de Toulouse, 18 avenue Edouard Belin, 31401 Toulouse Cedex 9, France\relax                                                                                 \label{inst:0431}
\and AKKA for CNES Centre Spatial de Toulouse, 18 avenue Edouard Belin, 31401 Toulouse Cedex 9, France\relax                                                                                                 \label{inst:0439}
\and Villanova University, Dept. of Astrophysics and Planetary Science, 800 E Lancaster Ave, Villanova PA 19085, USA\relax                                                                                   \label{inst:0440}
\and Vitrociset Belgium for ESA/ESAC, Camino bajo del Castillo, s/n, Urbanizacion Villafranca del Castillo, Villanueva de la Ca\~{n}ada, E-28692 Madrid, Spain\relax                                         \label{inst:0445}
\and Fork Research, Rua do Cruzado Osberno, Lt. 1, 9 esq., Lisboa, Portugal\relax                                                                                                                            \label{inst:0469}
\and APAVE SUDEUROPE SAS for CNES Centre Spatial de Toulouse, 18 avenue Edouard Belin, 31401 Toulouse Cedex 9, France\relax                                                                                  \label{inst:0473}
\and Serco Services GmbH for ESA/ESOC, Robert-Bosch-Stra{\ss}e 5, 64293 Darmstadt, Germany\relax                                                                                                             \label{inst:0480}
\and Spanish Virtual Observatory\relax                                                                                                                                                                       \label{inst:0482}
\and Fundaci\'{o}n Galileo Galilei - INAF, Rambla Jos\'{e} Ana Fern\'{a}ndez P\'{e}rez 7, E-38712 Bre\~{n}a Baja, Santa Cruz de Tenerife, Spain\relax                                                        \label{inst:0493}
\and INSA for ESA/ESAC, Camino bajo del Castillo, s/n, Urbanizacion Villafranca del Castillo, Villanueva de la Ca\~{n}ada, E-28692 Madrid, Spain\relax                                                       \label{inst:0499}
\and Dpto. Arquitectura de Computadores y Autom\'{a}tica, Facultad de Inform\'{a}tica, Universidad Complutense de Madrid, C/ Prof. Jos\'{e} Garc\'{i}a Santesmases s/n, 28040 Madrid, Spain\relax            \label{inst:0500}
\and H H Wills Physics Laboratory, University of Bristol, Tyndall Avenue, Bristol BS8 1TL, United Kingdom\relax                                                                                              \label{inst:0503}
\and Stellar Astrophysics Centre, Aarhus University, Department of Physics and Astronomy, 120 Ny Munkegade, Building 1520, DK-8000 Aarhus C, Denmark\relax                                                   \label{inst:0506}
\and Applied Physics Department, University of Vigo, E-36310 Vigo, Spain\relax                                                                                                                               \label{inst:0509}
\and HE Space Operations BV for ESA/ESTEC, Keplerlaan 1, 2201AZ, Noordwijk, The Netherlands\relax                                                                                                            \label{inst:0521}
\and Warsaw University Observatory, Al. Ujazdowskie 4, 00-478 Warszawa, Poland\relax                                                                                                                         \label{inst:0532}
\and Instituto de Astrof\'{\i}sica de Canarias, E-38205 La Laguna, Tenerife, Spain\relax                                                                                                                     \label{inst:0541}
\and Universidad de La Laguna, Departamento de Astrof\'{\i}sica, E-38206 La Laguna, Tenerife, Spain\relax                                                                                                    \label{inst:0542}
\and RHEA for ESA/ESTEC, Keplerlaan 1, 2201AZ, Noordwijk, The Netherlands\relax                                                                                                                              \label{inst:0552}
\and Max Planck Institute for Solar System Research, Justus-von-Liebig-Weg 3, 37077 G\"{ o}ttingen, Germany\relax                                                                                            \label{inst:0556}
\and SISSA (Scuola Internazionale Superiore di Studi Avanzati), via Bonomea 265, 34136 Trieste, Italy\relax                                                                                                  \label{inst:0557}
\and Instituto Nacional de Pesquisas Espaciais/Minist\'{e}rio da Ciencia Tecnologia, Avenida dos Astronautas 1758, S\~{a}o Jos\'{e} Dos Campos, SP 12227-010, Brazil\relax                                   \label{inst:0566}
\and Argelander Institut f\"{ u}r Astronomie der Universit\"{ a}t Bonn, Auf dem H\"{ u}gel 71, 53121 Bonn, Germany\relax                                                                                     \label{inst:0570}
\and European Southern Observatory (ESO), Karl-Schwarzschild-Stra{\ss}e 2, 85748 Garching bei M\"{ u}nchen, Germany\relax                                                                                    \label{inst:0573}
\and Laboratory of Optics, Lasers and Systems, Faculty of Sciences, University of Lisbon, Campus do Lumiar, Estrada do Pa\c{c}o do Lumiar, 22, 1649-038 Lisboa, Portugal\relax                               \label{inst:0576}
\and Department of Physics and Astronomy, Notre Dame University, Louaize, PO Box 72, Zouk Mika\"{ e}l, Lebanon\relax                                                                                         \label{inst:0587}
\and University of Nova Gorica, Vipavska 13, 5000 Nova Gorica, Slovenia\relax                                                                                                                                \label{inst:0589}
\and Max Planck Institute for Extraterrestrial Physics, OPINAS, Gie{\ss}enbachstra{\ss}e, 85741 Garching, Germany\relax                                                                                      \label{inst:0593}
\and NASA/IPAC Infrared Science Archive, California Institute of Technology, Mail Code 100-22, 770 South Wilson Avenue, Pasadena, CA, 91125, USA\relax                                                       \label{inst:0600}
\and Center of Applied Space Technology and Microgravity (ZARM), c/o Universit\"{ a}t Bremen, Am Fallturm 1, 28359 Bremen, Germany\relax                                                                     \label{inst:0602}
\and RHEA System for ESA/ESOC, Robert Bosch Stra{\ss}e 5, 64293 Darmstadt, Germany\relax                                                                                                                     \label{inst:0603}
\and Tartu Observatory, 61602 T\~{o}ravere, Estonia\relax                                                                                                                                                    \label{inst:0609}
\and Sydney Institute for Astronomy, School of Physics A28, The University of Sydney, NSW 2006, Australia\relax                                                                                              \label{inst:0611}
\and Slovak Organisation for Space Activities, Zamocka 18, 85101 Bratislava, Slovak Republic\relax                                                                                                           \label{inst:0613}
\and National Astronomical Observatories, CAS, 100012 Beijing, China\relax                                                                                                                                   \label{inst:0617}
\and US Naval Observatory, Astrometry Department, 3450 Massachusetts Ave. NW, Washington DC 20392-5420 D.C., USA\relax                                                                                       \label{inst:0619}
\and European Southern Observatory (ESO), Alonso de C\'{o}rdova 3107, Vitacura, Casilla 19001, Santiago de Chile, Chile\relax                                                                                \label{inst:0622}
\and EJR-Quartz BV for ESA/ESTEC, Keplerlaan 1, 2201AZ, Noordwijk, The Netherlands\relax                                                                                                                     \label{inst:0635}
\and The Server Labs for ESA/ESAC, Camino bajo del Castillo, s/n, Urbanizacion Villafranca del Castillo, Villanueva de la Ca\~{n}ada, E-28692 Madrid, Spain\relax                                            \label{inst:0636}
\and Astronomical Observatory Institute, Faculty of Physics, A. Mickiewicz University, ul. S\l{}oneczna 36, 60-286 Pozna\'{n}, Poland\relax                                                                  \label{inst:0640}
\and CS Syst\`{e}mes d'Information for CNES Centre Spatial de Toulouse, 18 avenue Edouard Belin, 31401 Toulouse Cedex 9, France\relax                                                                        \label{inst:0651}
\and Praesepe BV for ESA/ESAC, Camino bajo del Castillo, s/n, Urbanizacion Villafranca del Castillo, Villanueva de la Ca\~{n}ada, E-28692 Madrid, Spain\relax                                                \label{inst:0654}
\and Sorbonne Universit\'{e}s UPMC et CNRS, UMR7095, Institut d'Astrophysique de Paris, F75014, Paris, France\relax                                                                                          \label{inst:0658}
\and GMV for ESA/ESAC, Camino bajo del Castillo, s/n, Urbanizacion Villafranca del Castillo, Villanueva de la Ca\~{n}ada, E-28692 Madrid, Spain\relax                                                        \label{inst:0664}
\and Institute of Theoretical Physics and Astronomy, Vilnius University, Sauletekio al. 3, Vilnius, LT-10222, Lithuania\relax                                                                                \label{inst:0673}
\and S[\&]T Corporation, PO Box 608, 2600 AP, Delft, The Netherlands\relax                                                                                                                                   \label{inst:0675}
\and Department of Space Studies, Southwest Research Institute (SwRI), 1050 Walnut Street, Suite 300, Boulder, Colorado 80302, USA\relax                                                                     \label{inst:0676}
\and Deutsches Zentrum f\"{ u}r Luft- und Raumfahrt, Institute of Space Systems, Am Fallturm 1, D-28359 Bremen, Germany\relax                                                                                \label{inst:0677}
\and University of Applied Sciences Munich, Karlstr. 6, 80333 Munich, Germany\relax                                                                                                                          \label{inst:0679}
\and Dipartimento di Fisica, Universit\`{a} di Roma Tor Vergata, via della Ricerca Scientifica 1, 00133 Rome, Italy\relax                                                                                    \label{inst:0681}
\and Department of Physics and Astronomy, University of the Western Cape, Robert Sobukwe Road, 7535 Bellville, Cape Town, South Africa\relax                                                                 \label{inst:0684}
\and INAF - Istituto di Radioastronomia, via Gobetti 101, 40129 Bologna, Italy\relax                                                                                                                         \label{inst:0685}
\and Department of Physics, Florida International University, 11200 SW 8th Street, Miami, FL 33199, USA\relax                                                                                                \label{inst:0688}
\and Hamburger Sternwarte, Gojenbergsweg 112, D-21029 Hamburg, Germany\relax                                                                                                                                 \label{inst:0694}
}

  \date{Received 2016-07-08; accepted 2016-08-18}

  \abstract{{\it Gaia} is a cornerstone mission in the science programme of the European Space Agency (ESA). The spacecraft construction was approved in 2006, following a study in which the original interferometric concept was changed to a direct-imaging approach. Both the spacecraft and the payload were built by European industry. The involvement of the scientific community focusses on data processing for which the international {\it Gaia} Data Processing and Analysis Consortium (DPAC) was selected in 2007. {\it Gaia} was launched on 19 December 2013 and arrived at its operating point, the second Lagrange point of the Sun-Earth-Moon system, a few weeks later. The commissioning of the spacecraft and payload was completed on 19 July 2014. The nominal five-year mission started with four weeks of special, ecliptic-pole scanning and subsequently transferred into full-sky scanning mode. We recall the scientific goals of {\it Gaia} and give a description of the as-built spacecraft that is currently (mid-2016) being operated to achieve these goals. We pay special attention to the payload module, the performance of which is closely related to the scientific performance of the mission. We provide a summary of the commissioning activities and findings, followed by a description of the routine operational mode. We summarise scientific performance estimates on the basis of in-orbit operations. Several intermediate {\it Gaia} data releases are planned and the data can be retrieved from the {\it Gaia} Archive, which is available through the {\it Gaia} home page at \url{http://www.cosmos.esa.int/gaia}.}


  \maketitle

  \titlerunning{The {\it Gaia} Mission}
  \authorrunning{{\it Gaia} Collaboration}

\section{Introduction}\label{sect:introduction}

Astrometry is the astronomical discipline concerned with the accurate measurement and study of the (changing) positions of celestial objects. Astrometry has a long history \citep{2012EPJH...37..745P} even before the invention of the telescope. Since then, advances in the instrumentation have steadily improved the achievable angular accuracy, leading to a number of important discoveries: stellar proper motion \citep{1717RSPT...30..736H}, stellar aberration \citep{1727RSPT...35..637B}, nutation \citep{1748RSPT...45....1B}, and trigonometric stellar parallax \citep{1838AN.....16...65B,1840MmRAS..11...61H,1840AN.....17..177V}. Obtaining accurate parallax measurements from the ground, however, remained extremely challenging owing to the difficulty to control systematic errors and overcome the disturbing effects of the Earth's atmosphere, and the need to correct the measured relative to absolute parallaxes. Until the mid-1990s, for instance, the number of stars for which ground-based parallaxes were available was limited to just over 8000 \citep[][but see \citealt{2016yCat.1333....0F}]{1995gcts.book.....V}.

This situation changed dramatically in 1997 with the {\it Hipparcos} satellite of the European Space Agency (ESA), which measured the absolute parallax with milli-arcsecond accuracy of as many as $117\,955$ objects \citep{1997ESASP1200.....E}. The {\it Hipparcos} data have influenced many areas of astronomy \citep[see the review by][]{2009aaat.book.....P}, in particular the structure and evolution of stars and the kinematics of stars and stellar groups. Even with its limited sample size and observed volume, {\it Hipparcos} also made significant advances in our knowledge of the structure and dynamics of our Galaxy, the Milky Way.

The ESA astrometric successor mission, {\it Gaia}, is expected to completely transform the field. The main aim of {\it Gaia} is to measure the three-dimensional spatial and the three-dimensional velocity distribution of stars and to determine their astrophysical properties, such as surface gravity and effective temperature, to map and understand the formation, structure, and past and future evolution of our Galaxy \citep[see the review by][]{2016arXiv160207702B}. The Milky Way contains a complex mix of stars (and planets), interstellar gas and dust, and dark matter. These components are widely distributed in age, reflecting their formation history, and in space, reflecting their birth places and subsequent motions. Objects in the Milky Way move in a variety of orbits that are determined by the gravitational force generated by the integrated mass of baryons and dark matter, and have complex distributions of chemical-element abundances, reflecting  star formation and gas-accretion history. Understanding all these aspects in one coherent picture is the main aim of {\it Gaia}. Such an understanding is clearly also relevant for studies of the high-redshift Universe because a well-studied template galaxy underpins the analysis of unresolved galaxies.

{\it Gaia} needs to sample a large, representative, part of the Galaxy, down to a magnitude limit of at least 20 in the {\it Gaia} $G$ band to meet its primary science goals and to reach various (kinematic) tracers in the thin and thick disks, bulge, and halo \citep[][Table~1]{2001A&A...369..339P}. For the 1000 million stars expected down to this limit, {\it Gaia} needs to determine their present-day, three-dimensional spatial structure and their three-dimensional space motions to determine their orbits and the underlying Galactic gravitational potential and mass distribution. The astrometry of {\it Gaia} delivers absolute parallaxes and transverse kinematics (see \citealt{2015PASP..127..994B} on how to derive distances from parallaxes). Complementary radial-velocity and photometric information complete the kinematic and astrophysical information for a subset of the target objects, including interstellar extinctions and stellar chemical abundances.

Following the {\it R{\o}mer} mission proposal from the early 1990s \citep[see][]{2008IAUS..248..300H}, the {\it Gaia} mission was proposed by Lennart Lindegren and Michael Perryman in 1993 \citep[for historical details, see][]{2014arXiv1408.4668H}, after which a concept and technology study was conducted. The resulting science case and mission and spacecraft concept are described in \cite{2001A&A...369..339P}. In the early phases, {\it Gaia} was spelled as {\it GAIA,} for Global Astrometric Interferometer for Astrophysics, but the spelling was later changed because the final design was non-interferometric and based on monolithic mirrors and direct imaging and the final operating principle was actually closer to a large {\it R{\o}mer} mission than the original {\it GAIA} proposal. After the selection of {\it Gaia} in 2000 as an ESA-only mission, followed by further preparatory studies, the implementation phase started in 2006 with the selection of the prime contractor, EADS Astrium (later renamed Airbus Defence and Space), which was responsible for the development and implementation of the spacecraft and payload. Meanwhile, the complex processing and analysis of the mission data was entrusted to the Data Processing and Analysis Consortium (DPAC), a pan-European, nationally funded collaboration of several hundred astronomers and software specialists. {\it Gaia} was launched in December 2013 and the five-year nominal science operations phase started in the summer of 2014, after a half-year period of commissioning and performance verification.

Unlike the {\it Hipparcos mission, the {\it Gaia} collaboration does not have data rights.} After processing, calibration, and validation inside DPAC, data are made available to the world without limitations; this also applies to the photometric and solar system object science alerts (Sect.~\ref{subsect:initial_data_treatment}). Several intermediate releases, with roughly a yearly cadence, have been defined and this paper accompanies the first of these, referred to as {\it Gaia Data Release~1} \citep[{\it Gaia DR1;}][]{DPACP-8}. The data, accompanied by several query, visualisation, exploration, and collaboration tools, are available from the {\it Gaia} Archive \citep{DPACP-19}, which is reachable from the {\it Gaia} home page at \url{http://www.cosmos.esa.int/gaia} and directly at \url{http://archives.esac.esa.int/gaia}.

This paper is organised as follows: Section~\ref{sect:scientific_goals} summarises the science goals of the mission. The spacecraft and payload designs and characteristics are described in Sect.~\ref{sect:spacecraft_and_payload}. The launch and commissioning phase are detailed in Sect.~\ref{sect:launch_and_commissioning}. Section~\ref{sect:mission_and_spacecraft_operations} describes the mission and mission operations. The science operations are summarised in Sect.~\ref{sect:science_operations}. Section~\ref{sect:dpac} outlines the structure and flow of data in DPAC. The science performance of the mission is discussed in Sect.~\ref{sect:scientific_performance}. A summary can be found in Sect.~\ref{sect:conclusions}. All sections are largely stand-alone descriptions of certain mission aspects and can be read individually. The use of acronyms in this paper has been minimised; a list can be found in Annex~\ref{sect:acronyms}.

\section{Scientific goals}\label{sect:scientific_goals}

The science case for the {\it Gaia mission} was compiled in the year 2000 \citep{2001A&A...369..339P}. The scientific goals of the design reference mission were relying heavily on astrometry, combined with its photometric and spectroscopic surveys. The astrometric part of the science case remains unique, and so do the photometric and spectroscopic data, despite various, large ground-based surveys having materialised in the last decade(s). The space environment and design of {\it Gaia} enable a combination of accuracy, sensitivity, dynamic range, and sky coverage, which is practically impossible to obtain with ground-based facilities targeting photometric or spectroscopic surveys of a similar scientific scope. The spectra collected by the radial-velocity spectrometer (Sect.~\ref{subsubsect:spectroscopic_instrument}) have sufficient signal to noise for bright stars to make the {\it Gaia} spectroscopic survey the biggest of its kind. The astrometric part of {\it Gaia} is unique simply because global, micro-arcsecond astrometry is possible only from space. Therefore, the science case outlined more than a decade ago remains largely valid and the {\it Gaia} data releases are still needed to address the scientific questions \citep[for a recent overview of the expected yield from Gaia, see][]{2014EAS....67....1W}. A non-exhaustive list of scientific topics is provided in this section with an outline of the most important {\it Gaia contributions.}

\subsection{Structure, dynamics, and evolution of the Galaxy}\label{subsect:structure_and_dynamics_of_the_Galaxy}

The fundamental scientific-performance requirements for {\it Gaia} stem, to a large extent, from the main scientific target of the mission: the Milky Way galaxy. {\it Gaia} is built to address the question of the formation and evolution of the Galaxy through the analysis of the distribution and kinematics of the luminous and dark mass in the Galaxy. By also providing measurements to deduce the physical properties of the constituent stars, it is possible to study the structure and dynamics of the Galaxy. Although the {\it Gaia} sample will only cover about 1\% of the stars in the Milky Way, it will consist of more than 1000 million stars covering a large volume (out to many kpc, depending on spectral type), allowing thorough statistical analysis work to be conducted. The dynamical range of the {\it Gaia} measurements facilitates reaching stars and clusters in the Galactic disk out to the Galactic centre as well as far out in the halo, while providing extremely high accuracies in the solar neighbourhood. In addition to using stars as probes of Galactic structure and the local, Galactic potential in which they move, stars can also be used to map the interstellar matter. By combining extinction deduced from stars, it is possible to construct the three-dimensional distribution of dust in our Galaxy. In this way, {\it Gaia} will address not only the stellar contents, but also the interstellar matter in the Milky Way.

\subsection{Star formation history of the Galaxy}\label{subsect:star_formation_history_of_the_Galaxy}

The current understanding of galaxy formation is based on a combination of theories and observations, both of (high-redshift) extragalactic objects and of individual stars in our Milky Way. The Milky Way galaxy provides the single possibility to study details of the processes, but the observational challenges are different in comparison with measuring other galaxies. From our perspective, the Galaxy covers the full sky, with some components far away in the halo requiring sensitivity, while stars in the crowded Galactic centre region require spatial resolving power. Both these topics can be addressed with the {\it Gaia} data. {\it Gaia} distances will allow the derivation of absolute luminosities for stars which, combined with metallicities, allow the derivation of accurate individual ages, in particular for old subgiants, which are evolving from the main-sequence turn-off to the bottom of the red giant branch. By combining the structure and dynamics of the Galaxy with the information of the physical properties of the individual stars and, in particular, ages, it is possible to deduce the star formation histories of the stellar populations in the Milky Way.

\subsection{Stellar physics and evolution}\label{subsect:stellar_astrophysics}

Distances are one of the most fundamental quantities needed to understand and interpret various astronomical observations of stars. Yet direct distance measurement using trigonometric parallax of any object outside the immediate solar neighbourhood or not emitting in radio wavelengths is challenging from the ground. The {\it Gaia} revolution will be in the parallaxes, with hundreds of millions being accurate enough to derive high-quality colour-magnitude diagrams and to make significant progress in stellar astrophysics. The strength of {\it Gaia} is also in the number of objects that are surveyed as many phases of stellar evolution are fast. With 1000 million parallaxes, {\it Gaia} will cover most phases of evolution across the stellar-mass range, including pre-main-sequence stars and (chemically) peculiar objects. In addition to parallaxes, the homogeneous, high-accuracy photometry will allow fine tuning of stellar models to match not only individual objects, but also star clusters and populations as a whole. The combination of {\it Gaia} astrometry and photometry will also contribute significantly to star formation studies.

\subsection{Stellar variability and distance scale}\label{subsect:stellar_variability_and_distance_scale}

On average, each star is measured astrometrically $\sim$70 times during the five-year nominal operations phase (Sect.~\ref{subsect:scanning_law}). At each epoch, photometric measurements are also made: ten in the {\it Gaia} $G$ broadband filter and one each with the red and blue photometer (Sect.~\ref{subsect:photometry}). For the variable sky, this provides a systematic survey with the sampling and cadence of the scanning law of {\it Gaia} (Sect.~\ref{subsect:scanning_law}). This full-sky survey will provide a census of variable stars with tens of millions of new variables, including rare objects. Sudden photometric changes in transient objects can be captured and the community can be alerted for follow-up observations. Pulsating stars, especially RR Lyrae and Cepheids, can easily be discovered from the {\it Gaia} data stream allowing, in combination with the parallaxes, calibration of the period-luminosity relations to better accuracies, thereby improving the quality of the cosmic-distance ladder and scale.

\subsection{Binaries and multiple stars}\label{subsect:binaries_and_multiple_stars}

{\it Gaia} is a powerful mission to improve our understanding of multiple stars. The instantaneous spatial resolution, in the scanning direction, is comparable to that of the {\it Hubble Space Telescope} and {\it Gaia} is surveying the whole sky. In addition to resolving many binaries, all instruments in {\it Gaia} can complement our understanding of multiple systems. The astrometric wobbles of unresolved binaries, seen superimposed on parallactic and proper motions, can be used to identify multiple systems. Periodic changes in photometry can be used to find (eclipsing) binaries and an improved census of double-lined systems based on spectroscopy will follow from the {\it Gaia} data. It is again the large number of objects that {\it Gaia} will provide that will help address the fundamental questions of mass distributions and orbital eccentricities among binaries.

\subsection{Exoplanets}\label{subsect:exoplanets}

From the whole spectrum of scientific topics that {\it Gaia} can address, the exoplanet research area has been the most dynamic in the past two decades. The field has expanded from hot, giant planets to smaller planets, to planets further away from their host star, and to multiple planetary systems. These advancements have been achieved both with space- and ground-based facilities. Nevertheless, the {\it Gaia} astrometric capabilities remain unique, probing a poorly explored area in the parameter space of exoplanetary systems and providing astrophysical parameters not obtainable by other means. A strong point of {\it Gaia} in the exoplanet research field is the provision of an unbiased, volume-limited sample of Jupiter-mass planets in multiyear orbits around their host stars. These are logical prime targets for future searches of terrestrial-mass exoplanets in the habitable zone in an orbit protected by a giant planet further out. In addition, the astrometric data of {\it Gaia} allow actual masses (rather than lower limits) to be measured. Finally, the data of {\it Gaia} will provide the detailed distributions of giant exoplanet properties (including the giant planet - brown dwarf transition regime) as a function of stellar-host properties with unprecedented resolution.

\subsection{Solar system}\label{subsect:solar_system}

Although {\it Gaia} is designed to detect and observe stars, it will provide a full census of all sources that appear point-like on the sky. The movement of solar system objects with respect to the stars smears their images and makes them less point-like. As long as this smearing is modest, {\it Gaia} will still detect the object. The most relevant solar system object group for {\it Gaia} are asteroids. Unlike planets, which are too big in size (and, in addition, sometimes too bright) to be detected by {\it Gaia}, asteroids remain typically point-like and have brightness in the dynamical range of {\it Gaia}. {\it Gaia} astrometry and photometry will provide a census of orbital parameters and taxonomy in a single, homogeneous photometric system. The full-sky coverage of {\it Gaia} will also provide this census far away from the ecliptic plane as well as for locations inside the orbit of the Earth. An alert can be made of newly discovered asteroids to trigger ground-based observations to avoid losing the object again. For near-Earth asteroids, {\it Gaia} is not going to be very complete as the high apparent motion of such objects often prevents {\it Gaia} detection, but in those cases where {\it Gaia} observations are made, the orbit determination can be very precise. {\it Gaia} will provide fundamental mass measurements of those asteroids that experience encounters with other solar system bodies during the {\it Gaia} operational lifetime.

\subsection{The Local Group}\label{subsect:the_local_group}

In the Local Group, the spatial resolution of {\it Gaia} is sufficient to resolve and observe the brightest individual stars. Tens of Local Group galaxies will be covered, including the Andromeda galaxy and the Magellanic Clouds. While for the faintest dwarf galaxies only a few dozen of the brightest stars are observed, this number increases to thousands and millions of stars in Andromeda and the Large Magellanic Cloud, respectively. In dwarf spheroidals such as Fornax, Sculptor, Carina, and Sextans, thousands of stars will be covered. A major scientific goal of {\it Gaia} in the Local Group concerns the mutual, dynamical interaction of the Magellanic Clouds and the interaction between the Clouds and the Galaxy. In addition to providing absolute proper motions for transverse-velocity determination, needed for orbits, it is possible to explore internal stellar motions within dwarf galaxies. These kinds of data may reveal the impact of dark matter, among other physical processes in the host galaxy, to the motions of its stars.

\subsection{Unresolved galaxies, quasars, and the reference frame}\label{subsect:Unresolved_galaxies,_quasars,_and_the_reference_frame}

{\it Gaia} will provide a homogeneous, magnitude-limited sample of unresolved galaxies. For resolved galaxies, the sampling function is complicated as the onboard detection depends on the contrast between any point-like, central element (bulge) and any extended structure, convolved with the scanning direction. For unresolved galaxies, the most valuable measurements are the photometric observations. Millions of galaxies across the whole sky will be measured systematically. As the same {\it Gaia} system is used for stellar work, one can anticipate that, in the longer term, the astrophysical interpretation of the photometry of extragalactic objects will be based on statistically sound fundaments obtained from Galactic studies. Quasars form a special category of extragalactic sources for {\it Gaia} as not only their intrinsic properties can be studied, but they can also be used in comparisons of optical and radio reference frames. Such a comparison will, among others, answer questions of the coincidence of quasar positions across different wavelengths.

\subsection{Fundamental physics}\label{subsect:fundamental_physics}

As explained in Sect.~\ref{subsect:Simulations,_supplementary_data_and_observations,_data publication}, relativistic corrections are part of the routine data processing for {\it Gaia}. Given the huge number of measurements, it is possible to exploit the redundancy in these corrections to conduct relativity tests or to use (residuals of) the {\it Gaia} data in more general fundamental-physics experiments. Specifically for light bending, it is possible to determine the $\gamma$ parameter in the parametrised post-Newtonian formulation very precisely. Another possible experiment is to explore light bending of star images close to the limb of Jupiter to measure the quadrupole moment of the gravitational field of the giant planet. A common element in all fundamental physics tests using {\it Gaia} data is the combination of large sets of measurements. This is meaningful only when all systematic effects are under control, down to micro-arcsecond levels. Therefore, {\it Gaia} results for relativistic tests can be expected only towards the end of the mission, when all calibration aspects have been handled successfully.

\section{Spacecraft and payload}\label{sect:spacecraft_and_payload}

\begin{figure}[t]
\includegraphics[width=\columnwidth]{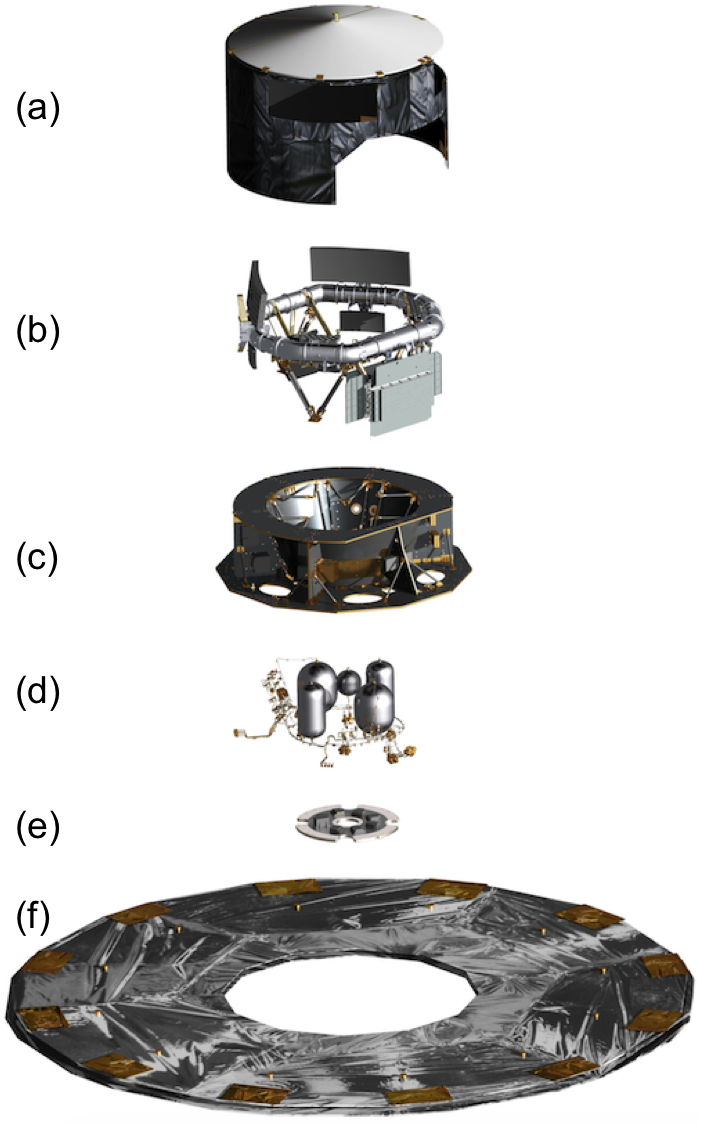}
\caption{Exploded, schematic view of {\it Gaia}.
  (a) Payload thermal tent (Sect.~\ref{subsect:payload_module});
  (b) payload module: optical bench, telescopes, instruments, and focal plane assembly (Sect.~\ref{subsect:payload_module});
  (c) service module (structure): also housing some electronic payload equipment, e.g. clock distribution unit, video processing units, and payload data-handling unit (Sect.~\ref{subsect:service_module});
  (d) propellant systems (Sect.~\ref{subsubsect:attitude_and_orbit_control});
  (e) phased-array antenna (Sect.~\ref{subsubsect:phased-array_antenna}); and
  (f) deployable sunshield assembly, including solar arrays (Sect.~\ref{subsect:service_module}).
  Credit: ESA, ATG Medialab.}
\label{fig:exploded_view}
\end{figure}

The {\it Gaia} satellite (Fig.~\ref{fig:exploded_view}) has been built under an ESA contract by Airbus Defence and Space (DS, formerly known as Astrium) in Toulouse (France). It consists of a payload module (PLM; Sect.~\ref{subsect:payload_module}), which was built under the responsibility of Airbus DS in Toulouse; a mechanical service module (M-SVM; Sect.~\ref{subsect:service_module}), which was built under the responsibility of Airbus DS in Friedrichshafen (Germany); and an electrical service module (E-SVM; Sect.~\ref{subsect:service_module}), which was built under the responsibility of Airbus DS in Stevenage (United Kingdom).

\subsection{Astrometric measurement principle and overall design considerations}\label{subsect:measurement_principle_and_overall_design_considerations}

\begin{figure*}[t]
\includegraphics[width=\textwidth]{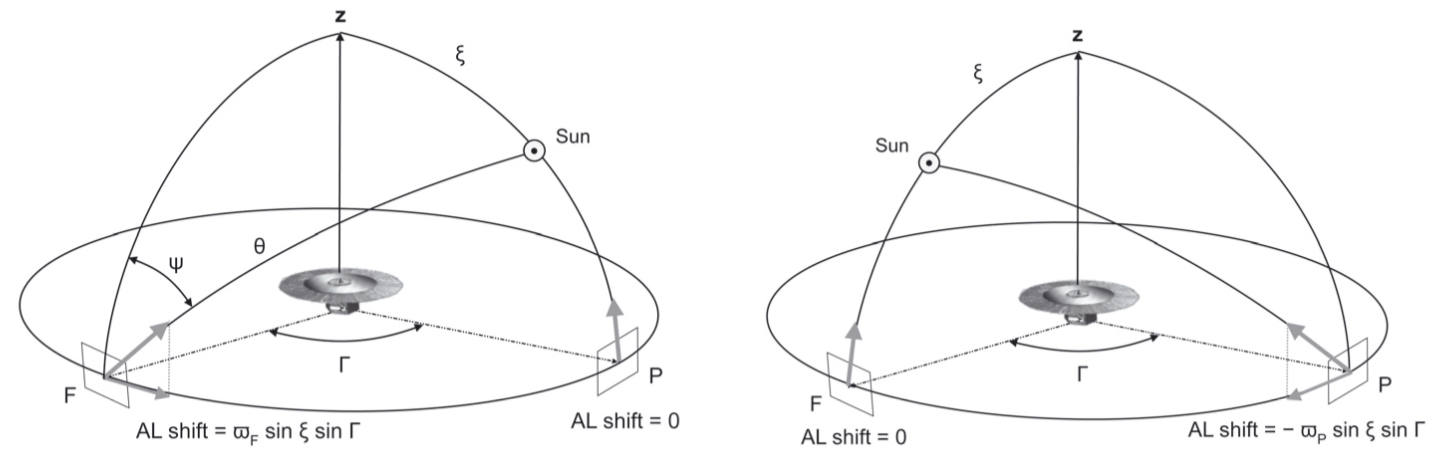}
\caption{Measurable, along-scan (AL) angle between the stars at P and F depends on their parallaxes $\varpi_{\rm P}$ and $\varpi_{\rm F}$ in different ways, depending on the position of the Sun. This allows us to determine their absolute parallaxes, rather than just the relative parallax $\varpi_{\rm P} - \varpi_{\rm F}$. Wide-angle measurements also guarantee a distortion-free and rigid system of coordinates and proper motions over the whole sky. Image from \cite{2011EAS....45..109L}.}
\label{fig:absolute_parallax}
\end{figure*}

The measurement principle of {\it Gaia} is derived from the global-astrometry concept successfully demonstrated by the ESA astrometric predecessor mission, {\it Hipparcos} \citep{1989hmps....1.....P}. This principle of scanning space astrometry \citep{2011EAS....45..109L} relies on a slowly spinning satellite that measures the crossing times of targets transiting the focal plane. These observation times represent the one-dimensional, along-scan (AL) stellar positions relative to the instrument axes. The astrometric catalogue is built up from a large number of such observation times, by an astrometric global iterative solution (AGIS) process \citep[e.g.][]{2012A&A...538A..78L,DPACP-14}, which also involves a simultaneous reconstruction of  the instrument pointing (attitude) as a function of time, and of the optical mapping of the focal plane detector elements (pixels) through the telescope(s) onto the celestial sphere (geometric calibration). The fact that the nuisance parameters to describe the attitude and geometric calibration are derived simultaneously with the astrometric source parameters from the regular observation data alone (without special, calibration data) means that {\it Gaia} is a self-calibrating mission.

Following in the footsteps of {\it Hipparcos}, {\it Gaia} is equipped with two fields of view, separated by a constant, large angle (the basic angle) on the sky along the scanning circle. The two viewing directions map the images onto a common focal plane such that the observation times can be converted into small-scale angular separations between stars inside each field of view and large-scale separations between objects in the two fields of view. Because the parallactic displacement (parallax factor) of a given source is proportional to $\sin\theta$, where $\theta$ is the angle between the star and the Sun, the parallax factors of stars inside a given field of view are nearly identical, suggesting only relative parallaxes could be measured. However, although scanning space astrometry makes purely differential measurements, absolute parallaxes can be obtained because the relative parallactic displacements can be measured between stars that are separated on the sky by a large angle (the basic angle) and, hence, have a substantially different parallax factor. To illustrate this further, consider an observer at one astronomical unit from the Sun. The apparent shift of a star owing to its parallax $\varpi$ then equals $\varpi\sin\theta$ and is directed along the great circle from the star towards the Sun. As shown in Fig.~\ref{fig:absolute_parallax} (left panel), the measurable, along-scan parallax shift of a star at position F (for following field of view) equals $\varpi_{\rm F}\sin\theta\sin\psi = \varpi_{\rm F}\sin\xi\sin\Gamma$, where $\xi$ is the angle between the Sun and the spin axis (the solar-aspect angle). At the same time, the measurable, along-scan parallax shift of a star at position P (for preceding field of view) equals zero. The along-scan measurement of F relative to P therefore depends on $\varpi_{\rm F}$ but not on $\varpi_{\rm P}$, while the reverse is true at a different time (right panel). So, scanning space astrometry delivers absolute parallaxes.

The sensitivity of {\it Gaia} to parallax, which means the measurable, along-scan effect, is proportional to $\sin\xi\sin\Gamma$. This has the following implications:
\begin{itemize}
\item Ideally, $\Gamma$ equals $90^\circ$. However, when scanning more or less along a great circle (as during a day or so), the accuracy with which the one-dimensional positions of stars along the great circle can be derived, as carried out in the one-day iterative solution (ODAS) as part of continuous payload health monitoring (Sect.~\ref{subsect:payload-health_monitoring}), is poor when $\Gamma = 360^\circ \times m/n$ for small integer values of $m$ and $n$ \citep{2011EAS....45..109L}; this can be understood in terms of the connectivity of stars along the circle \citep{1998A&A...340..309M}. Taking this into account, several acceptable ranges for the basic angle remain, for instance $99{\fdg}4 \pm 0{\fdg}1$ and $106{\fdg}5 \pm 0{\fdg}1$. Telescope accommodation aspects identified during industrial studies favoured $106{\fdg}5$ as the design value adopted for {\it Gaia}. During commissioning, using Tycho-2 stars, the actual in-flight value was measured to be $1\farcs{3}$ larger than the design value. For the global-astrometry concept to work, it is important to either have an extremely stable basic angle (i.e. thermally stable payload) on timescales of a few revolutions and/or to continuously measure its variations with high precision. Therefore, {\it Gaia} is equipped with a basic angle monitor (Sect.~\ref{subsubsect:basic-angle_monitor}).
\item Ideally, $\xi$ equals $90^\circ$. However, this would mean that sunlight would enter the telescope apertures. To ensure optimum thermal stability of the payload, in view of minimising basic angle variations, it is clear that $\xi$ should be chosen to be constant. For {\it Gaia}, $\xi = 45^\circ$ represents the optimal point between astrometric-performance requirements, which call for a large angle, and implementation constraints, such as the required size of the sunshield to keep the payload in permanent shadow and solar-array-efficiency and sizing arguments, which call for a small angle.
\end{itemize}
Finally, the selected spin rate of {\it Gaia}, nominally $60\arcsec$~s$^{-1}$ (actual, in-flight value: $59\farcs{9605}$~s$^{-1}$), is a complex compromise involving arguments on mission duration and these arguments: revisit frequency, attitude-induced point spread function blurring during detector integration, signal-to-noise ratio considerations, focal plane layout and detector characteristics, and telemetry rate.

\subsection{Service module}\label{subsect:service_module}

The mechanical service module comprises all mechanical, structural, and thermal elements supporting the instrument and the spacecraft electronics. The service module physically accommodates several electronic boxes including the video processing units (Sect.~\ref{subsubsect:video_processing_unit_and_algorithms}), payload data-handling unit (Sect.~\ref{subsubsect:payload_data_handling_unit}), and clock distribution unit (Sect.~\ref{subsubsect:clock_distribution_unit}), which functionally belong to the payload module but are housed elsewhere in view of the maintenance of the thermal stability of the payload. The service module also includes the chemical and micro-propulsion systems, deployable-sunshield assembly, payload thermal tent, solar-array panels, and electrical harness. The electrical services also support functions to the payload and spacecraft for attitude control, electrical power control and distribution, central data management, and communications with the Earth through low gain antennae and a high-gain phased-array antenna for science data transmission. In view of their relevance to the science performance of {\it Gaia}, the attitude and orbit control and phased-array antenna subsystems are described in more detail below.

\subsubsection{Attitude and orbit control}\label{subsubsect:attitude_and_orbit_control}

The extreme centroiding needs of the payload make stringent demands on satellite attitude control over the integration time of the payload detectors (of order a few seconds). This requires in particular that rate errors and relative-pointing errors be kept at the milli-arcsecond per second and milli-arcsecond level, respectively. These requirements prohibit the use of moving parts, such as conventional reaction wheels, on the spacecraft, apart from moving parts within thrusters. The attitude- and orbit-control subsystem (AOCS) is therefore based on a custom design \citep[e.g.][]{Chapman2011,2012ExA....34..669R} including various sensors and actuators. The sensors include two autonomous star trackers (used in cold redundancy), three fine Sun sensors used in hot redundancy (i.e. with triple majority voting), three fibre-optic gyroscopes (internally redundant), and low-noise rate data provided by the payload through measurements of star transit speeds through the focal plane. {\it Gaia} contains two flavours of actuators: two sets of eight bi-propellant (NTO oxidiser and MMH fuel) newton-level thrusters (used in cold redundancy) forming the chemical-propulsion subsystem (CPS) for spacecraft manoeuvres and back-up modes, including periodic orbit maintenance (Sect.~\ref{subsubsect:orbit_prediction,_monitoring,_and_control}); and two sets of six proportional-cold-gas, micro-newton-level thrusters forming the micro-propulsion subsystem (MPS) for fine attitude control required for nominal science operations. In nominal operations (AOCS normal mode), only the star-tracker and payload-rate data are used in a closed-loop, three-axes control with the MPS thrusters, which are operated with a commanded thrust bias; the other sensors are only used for failure detection, isolation, and recovery. Automatic, bi-directional mode transitions between several coarse and fine pointing modes have been implemented to allow efficient operation and autonomous settling during transient events, such as micro-meteoroid impacts (Sect.~\ref{subsect:orbit_and_environment}).

\subsubsection{Phased-array antenna}\label{subsubsect:phased-array_antenna}

Extreme centroiding requirements of the payload prohibit the use of a conventional, mechanically steered dish antenna for science data downlink because moving parts in {\it Gaia} would cause unacceptable degradation of the image quality through micro-vibrations. {\it Gaia} therefore uses a high-gain phased-array antenna (PAA), allowing the signal to be directed towards Earth as the spacecraft rotates (and as it moves through its orbit around the L$_2$ Lagrange point; Sect.~\ref{subsect:orbit_and_environment}) by means of electronic beam steering (phase shifting). The antenna is mounted on the Sun- and Earth-pointing face of the service module, which is perpendicular to the rotation axis. The radiating surface resembles a 14-sided, truncated pyramid. Each of the 14 facets has two subarrays and each comprises six radiating elements. Each subarray splits the incoming signal to provide the amplitude weighting that determines the radiation pattern of the subarray. The overall antenna radiation pattern is obtained by combining the radiation patterns from the 14 subarrays. The equivalent isotropic radiated power (EIRP) of the antenna exceeds 32~dBW over most of the $30^{\circ}$ elevation range (Sect.~\ref{subsect:orbit_and_environment}), allowing a downlink information data rate of 8.7 megabits per second (Sect.~\ref{subsubsect:ground_stations}) in the X band. The phased-array antenna is also used with orbit reconstruction measurements made from ground (Sect.~\ref{subsubsect:orbit_prediction,_monitoring,_and_control}).

\subsection{Payload module}\label{subsect:payload_module}

\begin{figure*}[t]
  \includegraphics[width=\columnwidth]{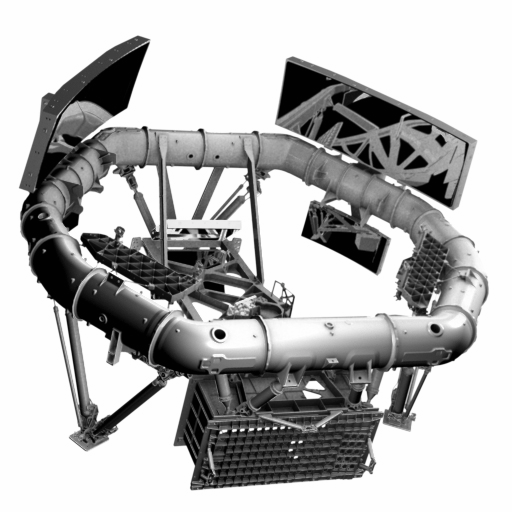}
  \includegraphics[width=\columnwidth]{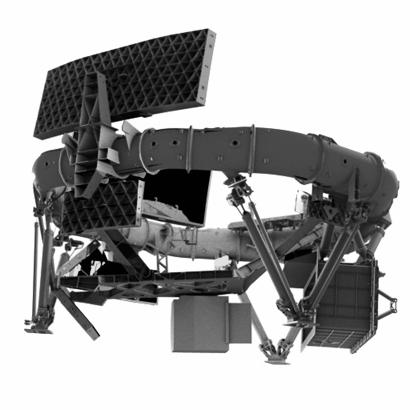}
  \caption{Schematic payload overview without protective tent. Most electronic boxes, e.g. clock distribution unit, video processing units, or payload data-handling unit, are physically located in the service module and hence not visible here. Credit: ESA.}
\label{fig:payload_overview}
\end{figure*}

The payload module (Fig.~\ref{fig:payload_overview}) is built around an optical bench that provides structural support for the two telescopes (Sect.~\ref{subsubsect:telescope}) and the single integrated focal plane assembly (Sect.~\ref{subsubsect:focal-plane_assembly}) that comprises, besides wave-front-sensing and basic angle metrology (Sects.~\ref{subsubsect:wave-front_sensor} and \ref{subsubsect:basic-angle_monitor}), three science functions: astrometry (Sect.~\ref{subsubsect:astrometric_instrument}), photometry (Sect.~\ref{subsubsect:photometric_instrument}), and spectroscopy (Sect.~\ref{subsubsect:spectroscopic_instrument}). The payload module is mounted on top of the service module via two (parallel) sets of three, V-shaped bipods. The first set of launch bipods is designed to withstand mechanical launch loads and these have been released in orbit to a parking position to free the second set of glass-fibre-reinforced polymer in-orbit bipods; the latter have low conductance and thermally decouple the payload from the service module. The payload is covered by a thermal tent based on a carbon-fibre-reinforced-polymer and aluminium sandwich structure with openings for the two telescope apertures and for the focal plane, warm-electronics radiator. The tent provides thermal insulation from the external environment and protects the focal plane and mirrors from micro-meteoroid impacts. The payload module furthermore contains the spacecraft master clock (Sect.~\ref{subsubsect:clock_distribution_unit}) and all necessary electronics for managing the instrument operation and processing and storing the science data (Sects.~\ref{subsubsect:video_processing_unit_and_algorithms} and \ref{subsubsect:payload_data_handling_unit}); these units, however, are physically located in the service module.

\subsubsection{Telescope}\label{subsubsect:telescope}

{\it Gaia} is equipped with two identical, three-mirror anastigmatic (TMA) telescopes, with apertures of 1.45~m $\times$ 0.50~m pointing in directions separated by the basic angle ($\Gamma = 106\fdg5$). These telescopes and their associated viewing directions (lines of sight) are often referred to as 1 and 2 or preceding and following, respectively, where the latter description refers to objects that are scanned first by the preceding and then by the following telescope. In order to allow both telescopes to illuminate a shared focal plane, the beams are merged into a common path at the exit pupil and then folded twice to accommodate the 35 m focal length. The total optical path hence encounters six reflectors: the first three (M1--M3 and M1'--M3') form the TMAs, the fourth is a flat beam combiner (M4 and M4'), and the final two are flat folding mirrors for the common path (M5--M6). All mirrors have a protected silver coating ensuring high reflectivity and a broad bandpass, starting around 330~nm. Asymmetric optical aberrations in the optics cause tiny yet significant chromatic shifts of the diffraction images and thus of the measured star positions. These systematic displacements are calibrated out as part of the on-ground data processing \citep{DPACP-14} using colour information provided by the photometry collected for each object (Sect.~\ref{subsubsect:photometric_instrument}).

The telescopes are mounted on a quasi-octagonal optical bench of  $\sim$3~m in diameter. The optical bench (composed of 17 segments, brazed together) and all telescope mirrors are made of sintered silicon carbide. This material combines high specific strength and thermal conductivity, providing optimum passive thermo-elastic stability (but see Sect.~\ref{subsect:commissioning_and_performance_verification}).

The (required) optical quality of {\it Gaia} is high, with a total wave-front error budget of 50~nm. To reach this number in orbit, after having experienced launch vibrations and gravity release, alignment and focussing mechanisms have been incorporated at the secondary (M2) mirrors. These devices, called M2 mirror mechanisms (M2MMs), contain a set of actuators that are capable of orienting the M2 mirrors with five degrees of freedom, which is sufficient for a rotationally symmetric surface. The in-orbit telescope focussing is detailed in \citet[][see also Sect.~\ref{subsect:payload_calibrations_and_special_operations}]{2014SPIE.9143E..0XM} and has been inferred from a combination of the science data themselves (size and shape of the point spread function) combined with data from the two wave-front sensors (WFSs; Sect.~\ref{subsubsect:wave-front_sensor}).

\subsubsection{Focal plane assembly}\label{subsubsect:focal-plane_assembly}

\begin{figure}[t]
\includegraphics[width=\columnwidth]{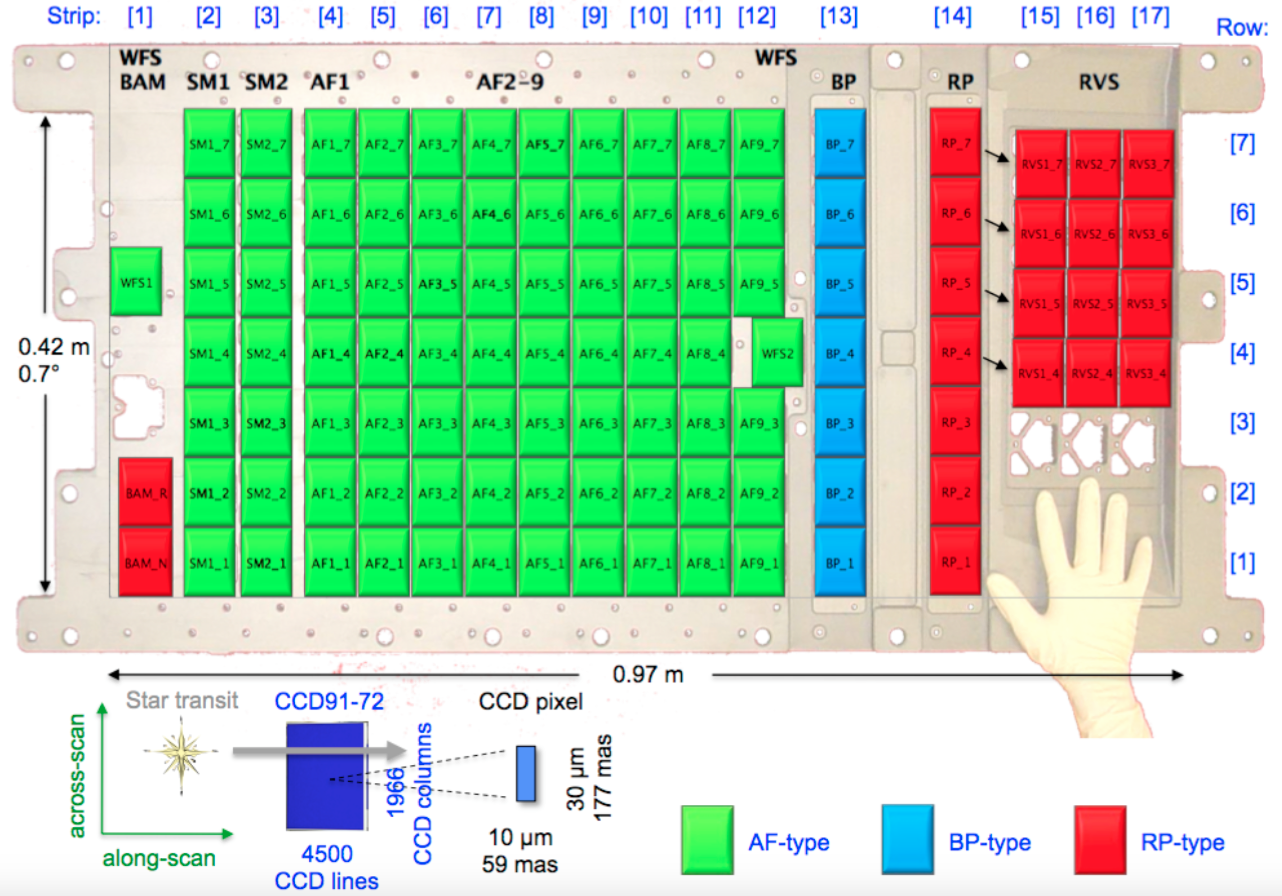}
\caption{Schematic image of the focal plane assembly, superimposed on a real picture of the CCD support structure (with a human hand to indicate the scale), with {\it Gaia}-specific terminology indicated (e.g. CCD strip and row, TDI line and pixel column). The RVS spectrometer CCDs are displaced vertically (in the across-scan direction) to correct for a lateral optical displacement of the light beam caused by the RVS optics such that the RVS CCD rows are aligned with the astrometric and photometric CCD rows on the sky; the resulting semi-simultaneity of the astrometric, photometric, and spectroscopic transit data is advantageous for stellar variability, science alerts, spectroscopic binaries, etc. Image from \cite{2010SPIE.7731E..1CD,2012SPIE.8442E..1PK}, courtesy Airbus DS and Boostec Industries.}
\label{fig:focal_plane}
\end{figure}

The focal plane assembly of {\it Gaia} \citep[for a detailed description, see][]{2012SPIE.8442E..1PK,dpacp-20} is common to both telescopes and has five main functions: (i) metrology (wave-front sensing [WFS] and basic angle monitoring [BAM]; Sects.~\ref{subsubsect:wave-front_sensor} and \ref{subsubsect:basic-angle_monitor}), (ii) object detection in the sky mapper (SM; Sect.~\ref{subsubsect:astrometric_instrument}), (iii) astrometry in the astrometric field (AF; Sect.~\ref{subsubsect:astrometric_instrument}), (iv) low-resolution spectro-photometry using the blue and red photometers (BP and RP; Sect.~\ref{subsubsect:photometric_instrument}), and (v) spectroscopy using the radial-velocity spectrometer (RVS; Sect.~\ref{subsubsect:spectroscopic_instrument}). The focal plane is depicted in Figure~\ref{fig:focal_plane} and carries 106 charge-coupled device (CCD) detectors, arranged in a mosaic of 7 across-scan rows and 17 along-scan strips, with a total of 938 million pixels. These detectors come in three different types, which are all derived from CCD91-72 from e2v technologies Ltd: the default, broadband CCD; the blue(-enhanced) CCD; and the red(-enhanced) CCD. Each of these types has the same architecture but differ in their anti-reflection coating and applied surface-passivation process, their thickness, and the resistivity of their silicon wafer. The broadband and blue CCDs are both 16~$\mu$m thick and are manufactured from standard-resistivity silicon (100~$\Omega$~cm); they differ only in their anti-reflection coating, which is optimised for short wavelengths for the blue CCD (centred on 360~nm) and optimised to cover a broad bandpass for the broadband CCD (centred on 650~nm). The red CCD, in contrast, is based on high-resistivity silicon (1000~$\Omega$~cm), is 40~$\mu$m thick, and has an anti-reflection coating optimised for long wavelengths (centred on 750~nm). The broadband CCD is used in SM, AF, and the WFS. The blue CCD is used in BP. The red CCD is used in BAM, RP, and the RVS.

\begin{figure}[t]
\includegraphics[width=\columnwidth]{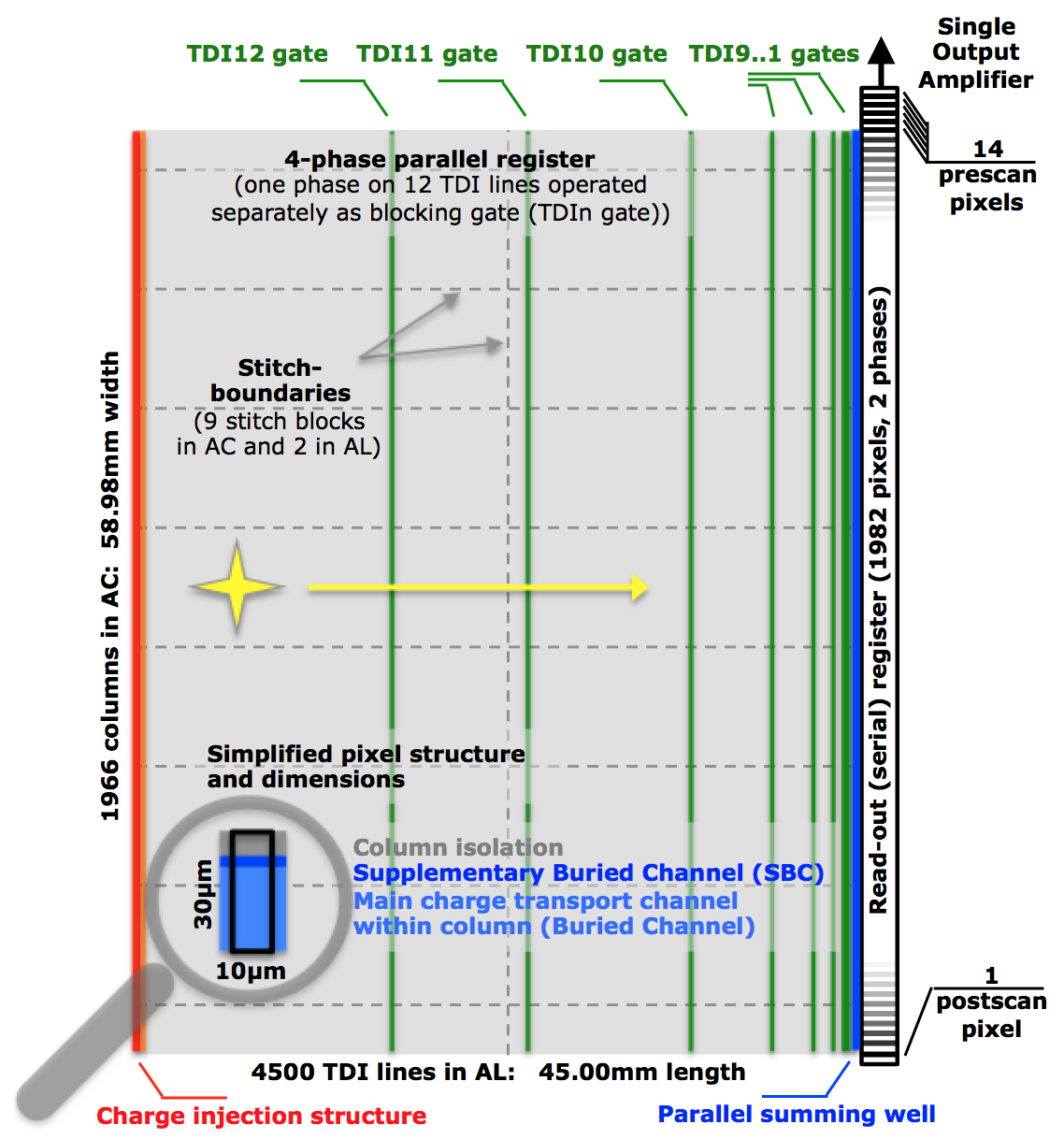}
\caption{Schematic view of a {\it Gaia} CCD detector. Stars move from left to right in the along-scan direction (yellow arrow). Charges in the readout register are clocked from bottom to top. The first line of the CCD (left) contains the charge-injection structure (red). The last line of the CCD before the readout register (right) contains the summing well and transfer gate (blue). Dashed, grey lines indicate stitch-block boundaries. Solid, green vertical lines indicate TDI gates (the three longest lines are labelled at the top of the CCD). The inset shows some details of an individual pixel. See Sect.~\ref{subsubsect:focal-plane_assembly} for details.}
\label{fig:ccd}
\end{figure}

The detectors (Fig.~\ref{fig:ccd}; \citealt{dpacp-20}) are back-illuminated, full-frame devices with an image area of 4500 lines along-scan and 1966 columns across-scan; each pixel is 10~$\mu$m $\times$ 30~$\mu$m in size (corresponding to 58.9~mas $\times$ 176.8~mas on the sky), balancing along-scan resolution and pixel full-well capacity (around $190\,000$~e$^{-}$). All CCDs are operated in time-delayed integration (TDI) mode to allow collecting charges as the object images move over the CCD and transit the focal plane as a result of the spacecraft spin. The fundamental line shift period of 982.8~$\mu$s is derived from the spacecraft atomic master clock (Sect.~\ref{subsubsect:clock_distribution_unit}); the focus of the telescopes is adjusted to ensure that the speed of the optical images over the CCD surface matches the fixed speed at which the charges are clocked inside the CCD. The 10~$\mu$m pixel in the along-scan direction is divided into four clock phases to minimise the blurring effect of the discrete clocking operation on the along-scan image quality. The integration time per CCD is 4.42~s, corresponding to the 4500 TDI lines along-scan; actually, only 4494 of these lines are light sensitive. The CCD image area is extended along-scan by a light-shielded summing well with adjacent transfer gate to the two-phase serial (readout) register, permitting TDI clocking (and along-scan binning) in parallel with register readout. The serial register ends with a non-illuminated post-scan pixel and begins with several non-illuminated pre-scan pixels that are connected to a single, low-noise output-amplifier structure, enabling across-scan binning on the high-charge-handling capacity ($\sim$$240\,000$~e$^{-}$) output node. Total noise levels of the full detection chain vary from 3 to 5 electrons RMS per read sample (except for SM and AF1, which have values of 11 and 8~electrons RMS, respectively), depending on the CCD operating mode.

The CCDs are composed of 18 stitch blocks, originating from the mask employed in the photo-lithographic production process with eight across-scan and one along-scan boundaries (Fig.~\ref{fig:ccd}). Each block is composed of 250 columns (and 2250 lines) except for the termination blocks, which have 108 columns. Whereas pixels inside a given stitch block are typically well-aligned, small misalignments between adjacent stitch blocks necessitate discontinuities in the small-scale geometric calibration of the CCDs \citep{DPACP-14}. The mask-positioning accuracy for the individual stitch blocks also produces discontinuities in several response vectors, such as charge-injection non-uniformity and column-response non-uniformity. At distinct positions along the 4500 TDI lines, a set of 12 special electrodes (TDI gates) are connected to their own clock driver. In normal operation, these electrodes are clocked synchronously with the other electrodes. These TDI-gate electrodes can, however, be temporarily (or permanently) held low such that charge transfer over these lines in the image area is inhibited and TDI integration time is effectively reduced to the remaining number of lines between the gate and the readout register. While the full 4500-lines integration is normally used for faint objects, TDI gates are activated for bright objects to limit image-area saturation. Available integration times are 4500, 2900, 2048, 1024, 512, 256, 128, 64, 32, 16, 8, 4, and 2 TDI lines. The choice of which gate to activate is user-defined, based on configurable look-up tables depending on the brightness of the object, the CCD, the field of view, and the across-scan pixel coordinate. Because the object brightness that is measured on board in the sky mapper (Sect.~\ref{subsubsect:payload_data_handling_unit}) has an error of a few tenths of a magnitude, a given (photometrically-constant) star, in particular when close in brightness to a gate-transition magnitude, is not always observed with the same gate on each transit. This mixing of gates is beneficial for the astrometric and photometric calibrations of the gated instruments.

The {\it Gaia} CCDs are n-channel devices, i.e. built on p-type silicon wafers with n-type channel doping. Displacement damage in the silicon lattice, caused by non-ionising irradiation, creates defect centres (traps) in the channel that act as electron traps during charge transfer, leading to charge-transfer inefficiency (CTI). Under the influence of radiation, n-channel devices are susceptible to develop a variety of trap species with release-time constants varying from micro-seconds to tens of seconds. Traps, in combination with TDI operation, affect the detailed shape of the point spread function of all instruments in subtle yet significant ways through continuous trapping and de-trapping \citep{2012MNRAS.422.2786H,2012MNRAS.419.2995P}, removing charge from the leading edge and releasing it in the trailing edge of the images and spectra. The resulting systematic biases of the image centroids and the spectra will be calibrated in the on-ground data processing, for instance using a forward-modelling approach based on a charge-distortion model \citep[CDM;][]{2013MNRAS.430.3078S}. The CCDs are passively cooled to 163~K to reduce dark current and minimise (radiation-induced) along- and across-scan CTI. To further mitigate CTI, two features have been implemented in the detector design: first, a charge-injection structure to periodically inject a line of electronic charge into the last CCD line (furthest from the readout register), which is then transferred by the TDI clocks through the device image area along with star images, thereby (temporarily) filling traps; and, second, a supplementary buried channel \citep[SBC;][]{2013MNRAS.430.3155S} in each CCD column to reduce the effect of CTI for small charge packets by confining the transfer channel in the across-scan direction, thereby exposing the signal to fewer trapping centres.

The CCDs are mounted on a support structure integrated into a cold-radiator box, which provides a radiative surface to the internal payload cavity (which is around 120~K), CCD shielding against radiation, and mounting support for the photometer prisms (Sect.~\ref{subsubsect:photometric_instrument}) and straylight vanes and baffles. Each CCD has its own proximity-electronics module (PEM), located behind the CCD (support structure) on the warm side of the focal plane assembly. Power from the warm electronics is dissipated directly to cold space through an opening in the thermal tent that encloses the payload module. Low-conductance bipods and thermal shields provide thermal isolation between the warm and cold parts of the focal plane assembly. The PEMs provide digital correlated double sampling and contain an input stage, a low-noise pre-amplifier with two programmable gain stages (low gain for full dynamic range or high gain for limited dynamic range and minimum noise), a bandwidth selector, and a 16-bit analogue-to-digital converter (ADC). The PEMs allow for adjustment of the CCD operating points, which might become necessary at some point as a result of flat-band voltage shifts induced by ionising radiation (monitoring of which is described in Sect.~\ref{subsect:payload_calibrations_and_special_operations}). All CCD-PEM couples of a given row of CCDs are connected through a power- and command-distribution interconnection module to a video processing unit (VPU; Sect.~\ref{subsubsect:video_processing_unit_and_algorithms}), which is in charge of generating the CCD commanding and acquiring the science data.

Operating the 100+ CCDs, comprising nearly a billion pixels, in TDI mode with a line period of $\sim$1~ms would generate a data rate that is orders of magnitude too high to be transmitted to ground. Three onboard reduction processes are hence applied:
\begin{enumerate}
\item Not all pixel data are read from the CCDs but only small areas, windows, around objects of interest; remaining pixel data are flushed at high speed in the serial register. This has an associated advantage of decreased read noise for the desired pixels;
\item The two-dimensional images (windows) are, except for bright stars, binned in the across-scan direction, nevertheless preserving the scientific information content (timing / along-scan centroid, total intensity / magnitude, and spectral information);
\item The resulting along-scan intensity profiles, such as line-spread functions or spectra, are compressed on board without loss of information; the typical gain in data volume is a factor 2.0--2.5.
\end{enumerate}
Windows are assigned by the VPU on-the-fly following autonomous object detection in the sky mapper (Sect.~\ref{subsubsect:astrometric_instrument}) and therefore the readout configuration of flushed and read (binned or unbinned) pixels is constantly changing with the sky passing by. This, together with the high-frequency pixel shift in the readout register and the interleaving of the TDI image-area clocking, causes a systematic fluctuation of the electronic bias level along the same TDI line during readout (known as the [CCD-]PEM [bias] non-uniformity), which is calibrated on ground \citep{DPACP-7}.

\subsubsection{Wave-front sensor}\label{subsubsect:wave-front_sensor}

The focal plane of {\it Gaia} is equipped with two wave-front sensors \citep[WFSs;][]{2009SPIE.7439E..14V}. These allow monitoring of the optical performance of the telescopes and deriving information to drive the M2 mirror mechanisms to (re-)align and (re-)focus the telescopes (Sect.~\ref{subsubsect:telescope}). The WFSs are of Shack-Hartmann type and sample the output pupil of each telescope with an array of $3 \times 11$ microlenses. These microlenses focus the light of bright stars transiting the focal plane on a CCD. Comparison of the stellar spot pattern with the pattern of a built-in calibration source (used during initial tests after launch) and with the pattern of stars acquired after achieving best focus (used afterwards) allows reconstruction of the wave front in the form of a series of two-dimensional Legendre polynomials \citep[Zernike polynomials are less appropriate for a rectangular pupil;][]{2012SPIE.8442E..1QM}. The location of the microlenses within the telescope pupils is inferred from the flux collected by the surrounding, partially-illuminated lenslets. The M2 mirror-mechanism actuations are derived using a telescope-alignment tool based on modelled sensitivities for each degree of freedom. The number of actuators to use and the weight given to each Legendre coefficient are adjustable. The corrections applied so far after each decontamination campaign (Sect.~\ref{subsect:payload_calibrations_and_special_operations}) have consisted of pure focus displacements.

\subsubsection{Basic angle monitor}\label{subsubsect:basic-angle_monitor}

As explained in Sect.~\ref{subsect:measurement_principle_and_overall_design_considerations}, the measurement principle of {\it Gaia} relies on transforming transit-time differences between stars observed in both telescopes into angular measurements. This requires the basic angle $\Gamma$ between the two fields of view either to be stable or to be monitored continuously at $\mu$as level and observed variations corrected as part of the data processing. Whereas low-frequency variations that are longer than, for instance two spin periods, i.e. 12~h (Sect.~\ref{subsect:scanning_law}), are absorbed in the geometric instrument calibration \citep{DPACP-14}, short-term variations, on timescales of minutes to hours, are non-trivial to calibrate and can introduce systematic errors in the astrometric results. In particular, a Sun-synchronous, periodic basic angle variation is known to be (nearly) fully degenerate with the parallax zero point \citep[e.g.][]{1992A&A...258...18L}. For this reason, the payload of {\it Gaia} was designed to be stable on these timescales to within a few $\mu$as (but see Sect.~\ref{subsect:commissioning_and_performance_verification}) and arguably carries the most precise interferometric metrology system ever flown, the basic angle monitor \citep[BAM; e.g.][]{2009SPIE.7439E..15M,2013SPIE.8863E..0GG,2014SPIE.9143E..0XM}. The BAM is composed of two optical benches fed by a common laser source that introduces two parallel, collimated beams per telescope. The BAM creates one Young-type fringe pattern per telescope in the same detector in the focal plane. The relative along-scan displacement between the two fringe patterns allows monitoring of the changes in the line of sight of each telescope and, thus, the basic angle. The (short-term) precision achieved in the differential measurement is 0.5~$\mu$as each 10--15~min, which corresponds to picometer displacements of the primary mirrors. A spare laser unit is kept in cold redundancy in case the primary source were to fail. The BAM exposures are continuously acquired with a period of 23~s (18.7~s stare-mode integration plus 4.4~s TDI-mode readout). A forward-modelling approach, which is based on a mathematical model representing the BAM image that is fitted using a least-squares algorithm, is  applied in the daily preprocessing pipeline \citep{DPACP-7} to monitor basic angle variations; basic angle variations are also monitored independently on a daily basis using cross-correlation techniques.

\subsubsection{Astrometric instrument}\label{subsubsect:astrometric_instrument}

The astrometric instrument comprises the two telescopes (Sect.~\ref{subsubsect:telescope}), a dedicated area of $7+7$ CCDs in the focal plane devoted to the sky mappers of the preceding and following telescope, and a dedicated area of 62 CCDs in the focal plane where the two fields of view are combined onto the astrometric field (AF). The wavelength coverage of the astrometric instrument, defining the unfiltered, white-light photometric $G$ band (for {\it Gaia}), is 330--1050~nm \citep{DPACP-9,DPACP-12}. These photometric data have a high signal-to-noise ratio and are particularly suitable for variability studies \citep{DPACP-15}.

Unlike its predecessor mission {\it Hipparcos}, which selected its targets for observation based on a predefined input catalogue loaded on board \citep{1993BICDS..43....5T}, {\it Gaia} performs an unbiased, flux-limited survey of the sky. This difference is primarily motivated by the fact that an all-sky input catalogue at the spatial resolution of {\it Gaia} that is complete down to 20$^{\rm th}$ mag, does not exist. Hence, autonomous, onboard object detection has been implemented through the Sky Mapper (Sect.~\ref{subsubsect:video_processing_unit_and_algorithms}), with the advantage that transient sources such as supernovae and near-Earth asteroids are observed too. Every object crossing the focal plane is first detected either by SM strip 1 (SM1) or SM strip 2 (SM2). These CCDs exclusively record, respectively, the objects from the preceding or from the following telescope. This is achieved through a physical mask that is placed in each telescope intermediate image, at the M4/M4' beam-combiner level (Sect.~\ref{subsubsect:telescope}).

The SM CCDs are read out in full-frame TDI mode, which means without windowing. Read samples, however, have a reduced spatial resolution with an on-chip binning of 2~pixels along-scan $\times$ 2~pixels across-scan per sample. Windows are assigned to detected objects and transmitted to ground; they measure $40 \times 6$~samples of $2 \times 2$~pixels each for stars brighter than $G = 13$~mag and $20 \times 3$~samples of $4 \times 4$~pixels each for fainter objects. The SM CCD has the longest TDI gate, with 2900 TDI lines (2.85~s) effective integration time, permanently active to reduce image degradation caused by optical distortions (which are significant at the edge of the field of view), and to reduce the CCD effective area susceptible to false detections generated by cosmic rays and solar protons.

The astrometric data acquired in the 62 CCDs in the AF field are binned on-chip in the across-scan direction over 12 pixels, except in the first AF strip (AF1) and for stars brighter than 13~mag. For these stars, unbinned, single-pixel-resolution windows are often used in combination with temporary TDI-gate activation, during the period of time that corresponds to the bright-star window length, to shorten the CCD integration time and avoid pixel-level saturation. In AF1, across-scan information is maintained at the CCD readout, but later binned by the onboard software before transmission to ground; this permits the measuring of the actual velocities of objects through the focal plane to feed the attitude and control subsystem,  to allow along- and across-scan window propagation through the focal plane, and to identify suspected moving objects, which receive a special, additional window either right on top or right below the nominal window in the photometric instrument (Sect.~\ref{subsubsect:photometric_instrument}). The AF1 data are also used on board for confirming the presence of detected objects. The along-scan window size in AF is 18 pixels for stars that are brighter than 16~mag and 12 pixels for fainter objects. The astrometric instrument can handle object densities up to $1\,050\,000$~objects~deg$^{-2}$ (Sect.~\ref{subsect:survey_coverage_and_completeness}). In denser areas, only the brightest stars are observed.

\subsubsection{Photometric instrument}\label{subsubsect:photometric_instrument}

The photometric instrument measures the spectral energy distribution (SED) of all detected objects at the same angular resolution and at the same epoch as the astrometric observations. This serves two goals:
\begin{enumerate}
\item The instrument provides astrophysical information for all objects \citep{2013A&A...559A..74B}, in particular astrophysical classification (for instance object type such as star, quasar, etc.) and astrophysical characterisation (for instance interstellar reddenings, surface gravities, metallicities, and effective temperatures for stars, photometric redshifts for quasars, etc.).
\item The instrument enables chromatic corrections of the astrometric centroid data induced by optical aberrations of the telescope (Sect.~\ref{subsubsect:telescope}).
\end{enumerate}

Like the spectroscopic instrument (Sect.~\ref{subsubsect:spectroscopic_instrument}), the photometric instrument is highly integrated with the astrometric instrument, using the same telescopes, the same focal plane (albeit using a dedicated section of it), and the same sky-mapper (and AF1) function for object detection (and confirmation). The photometry function is achieved through two fused-silica prisms dispersing light entering the fields of view. One disperser, called BP for blue photometer, operates in the wavelength range 330–-680~nm; the other disperser, called RP for red photometer, covers the wavelength range 640–-1050~nm. Sometimes, BP and RP are collectively referred to as XP. Optical coatings deposited on the prisms, together with the telescope transmission and detector quantum efficiency, define the bandpasses. The prisms are located in the common path of the two telescopes, and mounted on the CCD cold radiator, directly in front of the focal plane. Both photometers are equipped with a dedicated strip of seven CCDs each, which cover the full astrometric field of view in the across-scan direction (see Sect.~\ref{subsubsect:focal-plane_assembly} for details on the photometric CCDs). This implies that the photometers see the same (number of) transits as the astrometric instrument.

The prisms disperse object images along the scan direction and spread them over $\sim$45~pixels (for 15-mag objects): the along-scan window size is chosen as 60~pixels to allow for background subtraction (and window-propagation and window-placement quantisation errors). The spectral dispersion, which matches the earlier photometric-filter design described in \cite{2006MNRAS.367..290J}, results from the natural dispersion curve of fused silica and varies in BP from 3 to 27~nm~pixel$^{-1}$ over the wavelength range 330–-680 nm and in RP from 7 to 15 nm pixel$^{-1}$ over the wavelength range 640–-1050~nm. The 76\% energy extent of the along-scan line-spread function varies along the BP spectrum from 1.3~pixels at 330~nm to 1.9~pixels at 680~nm and along the RP spectrum from 3.5~pixels at 640~nm to 4.1~pixels at 1050~nm.

For the majority of objects, BP and RP spectra are binned on-chip in the across-scan direction over 12 pixels to form one-dimensional, along-scan spectra.  Unbinned, single-pixel-resolution windows (of size $60 \times 12$~pixels$^{2}$) are only used for stars brighter than $G = 11.5$~mag; this is often in combination with temporary TDI-gate activation, during the period of time corresponding to the bright-star window length, to shorten the CCD integration time and avoid pixel-level saturation. The object-handling capability of the photometric instrument is limited to $750\,000$~objects~deg$^{-2}$ (Sect.~\ref{subsect:survey_coverage_and_completeness});  only the brightest objects receive a window in areas exceeding this density. The data quality, however, is already affected at lower densities by contamination from the point spread function wings of nearby sources falling outside the window (degrading flux and background estimation) and by blending with sources falling inside the window \citep[leading to window truncation and necessitating a deblending procedure;][]{2012SPIE.8442E..42B}.

\subsubsection{Spectroscopic instrument}\label{subsubsect:spectroscopic_instrument}

The spectroscopic instrument, known as the radial-velocity spectrometer (RVS), obtains spectra of the bright end of the {\it Gaia} sample to provide:
\begin{enumerate}
\item radial velocities through Doppler-shift measurements using cross-correlation for stars brighter than $G_{\rm RVS} \approx 16$~mag \citep[Sect.~\ref{subsect:survey_coverage_and_completeness};][]{2014A&A...562A..97D}, which are required for kinematical and dynamical studies of the Galactic populations and for deriving good astrometry of nearby, fast-moving sources which show perspective acceleration \citep[e.g.][]{2012A&A...546A..61D};
\item coarse stellar parametrisation for stars brighter than $G_{\rm RVS} \approx 14.5$~mag \citep[e.g.][]{2016A&A...585A..93R};
\item astrophysical information, such as interstellar reddening, atmospheric parameters, and rotational velocities, for stars brighter than $G_{\rm RVS} \approx 12.5$~mag \citep[e.g.][]{2016A&A...585A..93R};
\item individual element abundances for some elements (e.g. Fe, Ca, Mg, Ti, and Si) for stars brighter than $G_{\rm RVS} \approx 11~$mag \citep[e.g.][]{2016A&A...585A..93R},
\end{enumerate}
where $G_{\rm RVS}$ denotes the integrated, instrumental magnitude in the spectroscopic bandpass (defined below).

The spectroscopic instrument \citep{2011EAS....45..181C}, like the photometric instrument (Sect.~\ref{subsubsect:photometric_instrument}), is highly integrated with the astrometric instrument, using the same telescopes, the same focal plane (using a dedicated section of it), and the same sky-mapper (and AF1) function for object detection (and confirmation). The actual (faint-end) selection of an object for RVS, however, is based on an onboard estimate of $G_{\rm RVS}$ that is generally derived from the RP spectrum collected just before the object enters RVS. The RVS is an integral-field spectrograph and the spectral dispersion of objects in the fields of view is materialised through an optical module with unit magnification, which is mounted in the common path of the two telescopes between the last telescope mirror (M6) and the focal plane. This module contains a blazed-transmission grating plate (used in transmission in order $+1$), four fused-silica prismatic lenses (two with flat surfaces and two with spherical surfaces), and a multilayer-interference bandpass-filter plate to limit the wavelength range to 845--872~nm. This range was selected to cover the \ion{Ca}{II} triplet, which is suitable for radial-velocity determination over a wide range of metallicities, signal-to-noise ratios, temperatures, and luminosity classes in particular for abundant FGK stars, and which is also a well-known metallicity indicator and stellar parametriser \citep[e.g.][]{1989Ap&SS.157...15T,2011A&A...535A.106K}. For early-type stars, the RVS wavelength range covers the hydrogen Paschen series from which radial velocities can be derived. In addition, the wavelength range covers a diffuse interstellar band (DIB), located at 862~nm, which traces out interstellar reddening \citep[e.g.][]{2002Ap&SS.280..119K,2008A&A...488..969M}.

The dispersed light from the RVS illuminates a dedicated area of the focal plane containing 12 CCDs arranged in three strips of four CCD rows (see Sect.~\ref{subsubsect:focal-plane_assembly} for details on the spectroscopic CCDs). This implies that an object observed by RVS has 43\% ($1 - 4/7$) fewer RVS focal plane transits than astrometric and photometric focal plane transits. The grating plate disperses object images along the scan direction and spreads them over $\sim$1100~pixels ($R = \lambda / \Delta\lambda \approx 11\,700$, dispersion $0.0245$~nm~pixel$^{-1}$); the along-scan window size is 1296~pixels to allow for background subtraction (and window-propagation and window-placement quantisation errors).

For the majority of objects, RVS spectra are binned on-chip in the across-scan direction over 10~pixels to form one-dimensional, along-scan spectra. The onboard software (Sect.~\ref{subsubsect:video_processing_unit_and_algorithms}) contains a provision to adapt this size to the instantaneous, straylight-dominated background level (Sect.~\ref{subsect:commissioning_and_performance_verification}), in view of optimising the signal-to-noise ratio of the spectra, but this functionality is not being used. Single-pixel-resolution windows (of size $1296 \times 10$~pixels$^{2}$) are only used for stars brighter than $G_{\rm RVS} = 7$~mag. The object-handling capability of RVS is limited to $35\,000$~objects~deg$^{-2}$ (Sect.~\ref{subsect:survey_coverage_and_completeness}); in areas exceeding this density, only the brightest objects receive a window. As for the photometers, however, the data quality will be severely compromised in dense areas by contamination from and blending with nearby sources.

\subsubsection{Video processing unit and algorithms}\label{subsubsect:video_processing_unit_and_algorithms}

Each CCD row in the focal plane (Sect.~\ref{subsubsect:focal-plane_assembly}) is connected to its own video processing unit (VPU), essentially a computer in charge of commanding the CCDs and collecting the science data and transmitting it to the onboard storage (Sect.~\ref{subsubsect:payload_data_handling_unit}).
The VPUs run the video processing algorithms \citep[VPAs;][]{2007ESASP.638E..39P}, which are a collection of software routines configurable through a set of parameters that can be changed by telecommand. The seven VPUs are fully independent although each one runs the same set of VPAs albeit (possibly) with different parameter sets. Parameter updates are possible but require a transition from VPU operational mode to VPU service mode, which means a loss of science data of a few dozen seconds. The VPUs and VPAs have a large number of functions such as CCD command generation, including deriving the TDI-line signals from the spacecraft master clock (Sect.~\ref{subsubsect:clock_distribution_unit}) for the synchronisation of the CCD sequencing. The CCD TDI (line) period is defined as 19,656 master-clock cycles and hence lasts 982.8~$\mu$s. The VPAs are also responsible for the detection, selection, and confirmation of objects. The detection algorithm uses full-frame SM data to discriminate stars from spurious objects, such as cosmic rays and solar protons, autonomously  using PSF-based criteria; the parameter settings adopted for operations guarantee a high level of completeness down to the faint limit at $G = 20.7$~mag (Sect.~\ref{subsect:survey_coverage_and_completeness}) at the expense of spurious detections in the (diffraction) wings of bright stars essentially passing unfiltered \citep[][; in May 2016, a new set of parameters was uploaded that accepts fewer false detections at the expense of a reduced detection efficiency of objects beyond 20~mag]{2015A&A...576A..74D}. After detection in SM, (the brightest) accepted objects are allocated a window from the pool of available windows. A final confirmation of each detection is enabled by the CCD detectors in the first AF strip (AF1); this step eliminates false detections in SM caused by cosmic rays or solar protons. Whether a detected object is actually selected or not for observation, i.e. receives a window, depends on a number of factors. Several limitations exist, for example in dense areas or when multiple bright stars, each requiring single-pixel-resolution windows, are present in the same TDI line(s); in particular this is caused by the fact that the total number of samples in the serial register that {\it Gaia} can observe simultaneously per CCD is 20 in AF, 71 in BP and RP, and 72 in RVS (Sect.~\ref{subsubsect:focal-plane_assembly}). In case of a shortage of windows, object selection (or resource allocation, where resource refers to serial samples) is based on object priority; the latter is a user-defined attribute which, in practice, is only a function of magnitude, where bright stars have higher priority.
The VPAs assign windows based on the onboard measured position and brightness of the object propagate windows through the focal plane, along-scan in line with the spin rate and across-scan to follow the small, across-scan motion of objects induced by the scanning law (Sect.~\ref{subsect:scanning_law}). The window management, meaning the collection of CCD sample data, the truncation of samples in case windows of nearby sources (partially) overlap, and packetisation and lossless compression of the science data is also driven by the VPAs. In addition, the VPAs feed the (closed) attitude control loop with rate measurements based on the measured transit velocities of 13--18-mag objects between SM and AF1 (Sect.~\ref{subsubsect:attitude_and_orbit_control}). They also govern the activation of TDI gates for the along-scan duration of bright-star windows in AF, BP, and RP, and the periodic activation of charge-injection lines in AF, BP, and RP (Sect.~\ref{subsubsect:focal-plane_assembly}). The VPAs collect health and housekeeping data, such as pre-scan data for CCD-bias monitoring, detection-confirmation-selection statistics, object logs to enable CCD-readout reconstruction for PEM non-uniformity calibration (Sect.~\ref{subsubsect:focal-plane_assembly}), etc., collect BAM and WFS data (Sects.~\ref{subsubsect:basic-angle_monitor} and \ref{subsubsect:wave-front_sensor}), and collect service-interface-function (SIF) data. The SIF function provides direct access to the synchronous dynamic random-access memory of the VPU, allowing monitoring, debugging, or extracting data that is not available in the nominal telemetry (for instance post-scan pixels or full-frame SM data). Finally, the VPAs govern special features such as user-commanded virtual objects (inserted into the stream of real, detected objects, useful for CCD-health monitoring, background monitoring, etc.), calibration faint stars (a small, user-configurable fraction of faint stars receive full-pixel-resolution windows for calibration purposes), and suspected moving objects (objects in a certain, user-defined across-scan speed range receive an extra window in BP and RP to increase the probability of measuring moving objects).

The VPUs generate three different kinds of data packets: auxiliary science data (ASD) packets, star packets (SPs), and SIF packets. The SPs contain the (generally raw) sample data of all scientific CCDs and form the core of the {\it Gaia} science data; they have nine flavours (SP1–-SP9), but only SP1 (SM/AF/BP/RP data), SP2 (RVS data), SP3 (suspect-moving-object windows in BP/RP), SP4 (BAM data), and SP5 (WFS data) are produced during nominal science operations. The ASD packets are essential for interpreting and processing the SPs and come in seven flavours (ASD1–-ASD7); they provide pre-scan data, logs of TDI-gate activations, charge injections, object logs to ease on-ground data processing, etc.

\subsubsection{Payload data-handling unit}\label{subsubsect:payload_data_handling_unit}

Science data generated by the VPUs (Sect.~\ref{subsubsect:video_processing_unit_and_algorithms}) is not directly transmitted to ground but first stored in the payload data-handling unit (PDHU). The PDHU is a solid-state mass memory with a storage capacity of $61\,440$ sectors, each of size 2~megabytes, providing $\sim$120~gigabytes effective in total (after subtraction of Reed-Solomon error-correction bits). The VPAs (Sect.~\ref{subsubsect:video_processing_unit_and_algorithms})
contain a FILE\_ID function that generates  an 8-bit file identification
(FILE\_ID in the range $[0, 255]$), which is stored in the packet header,
for each SP/ASD/SIF science packet.
The FILE\_ID is assigned in the VPAs, through user-configurable settings, based on VPA-derived attributes such as VPU number (1--7), packet type (SP, ASD, SIF), field of view (preceding or following), object type (virtual object, calibration faint star, suspected moving object, normal star), magnitude, and/or window-truncation flags.
This maximises early availability on ground of high-priority data, protects such data from being deleted on board, balances onboard data losses between astrometry plus photometry and spectroscopy, and minimises the latency of astrometric data for bright(er) objects, which is essential for the science alert pipelines \citep{2016P&SS..123...87T,2016arXiv160102827W}.

In practice, around 80 different FILE\_IDs are in use, allowing us to discriminate between critical data (ASD packets), high-priority data (SIF packets, virtual objects, calibration faint stars, and bright stars, i.e. $G < 16$~mag in astrometry and photometry and $G_{\rm RVS} < 10.5$~mag in spectroscopy), medium-priority data (narrow magnitude bins covering the full magnitude range to cover requirements for First-Look payload health monitoring; Sect.~\ref{subsect:payload-health_monitoring}), and low-priority data (faint stars). The FILE\_ID of a packet determines where it is stored in the PDHU: each FILE\_ID uniquely corresponds to a dedicated PDHU file with the same identifier. At the PDHU level, user-configurable prioritisation of science data is achieved through the following:
\begin{itemize}
\item Downlink-priority table: important data are downloaded first.
\item Deletion-priority table: in case of PDHU overflow (for instance during Galactic plane scans;\ Sect.~\ref{subsubsect:ground_stations}), less important data are deleted first to provide free space for more important data. The default deletion priority is the inverse of the downlink priority.
\item Data loss target table: deletion of data from a file, in case of PDHU saturation, is authorised only if the accumulated data loss of that file does not exceed the
target; this data loss is estimated as the ratio between the number of sectors deleted in the file and the total number of sectors allocated to the file since the start of the mission (or since the last PDHU reset). Typically, data loss targets are 0\% for high- and medium-priority data and gradually increase to 100\% for the faintest objects ($G > 20.5$~mag and $G_{\rm RVS} > 15.8$~mag).
\end{itemize}
A PDHU file can either be dynamic or cyclic in nature:
\begin{itemize}
\item A cyclic file has a fixed, user-defined size, meaning that the oldest data are overwritten after the file fills up and wraps around. Cyclic files hence have the property that, after data are transmitted to ground, the data are temporarily left accessible (until overwritten after wrap-around) meaning that data replay is possible, if needed. Cyclic files hence normally store critical data, which means ASD packets. The criticality of these data stems from the fact that one missing packet can inhibit ground processing of thousands of observations.
\item A dynamic file grows and shrinks in size as data are added and removed, respectively. Dynamic files have the property that, after data are transmitted to ground, the sector is immediately freed up for new data, meaning that data replay is typically not possible. Dynamic files hence normally store non-critical data, which means SP and SIF packets. For dynamic files, the maximum number of sectors to be transmitted during each access of the file has to be defined by the user.
\end{itemize}
Because {\it Gaia} is not in permanent ground-station contact (Sect.~\ref{subsubsect:ground_stations}), the PDHU occupancy level typically varies over a day, (partially) filling up outside ground-station contact periods and subsequently emptying during ground-station contacts. During contact, the memory manager cyclically goes through the downlink-priority table, which first contains the cyclic files with critical data, followed by high-, medium-, and low-priority dynamic files with non-critical data. High-, medium-, and low-priority files have assigned a finite (maximum) number of sectors to download before moving to the next file in the table, with the number reflecting their relative importance.
Using a small number of sectors forces rapid multiplexing of all files because, after reaching the end of the priority table and returning to the start, cycling through the cyclic files is rapidly completed because only the data acquired since the files were last visited have to be transmitted. In short, the adopted approach means that all (new) critical data comes down at the start of each ground-station contact period, after which the downlink rapidly multiplexes between non-critical data, taking into account their relative priorities, while keeping up to date with the critical data as it is generated on board; the typical cycle time of the full table is $\sim$30~min.

\subsubsection{Clock distribution unit}\label{subsubsect:clock_distribution_unit}

As explained in \citet[][see also Sect.~\ref{subsect:measurement_principle_and_overall_design_considerations}]{DPACP-14}, the fundamental astrometric measurements of {\it Gaia} are the observation times at which the image centroids pass the fiducial observation lines of the CCD detectors. Therefore, the architecture of the onboard timing chain has been carefully designed. Central in the time management and time distribution subsystem is the clock distribution unit. This unit maintains a 20 MHz satellite master clock, which is directly derived from an internal, 10 MHz rubidium atomic frequency standard (RAFS) based on the atomic reference given by the spectral absorption line of the $^{87}$Rb isotope. This clock is stable to within a few ns over a six-hour spacecraft revolution and has very small temperature, magnetic field, and input voltage sensitivities. The free running onboard time (OBT) counter, which is the time tag of all science data, is generated from the 20 MHz master clock and is coded on 64 bits; the resolution of this counter is hence 50~ns. The spacecraft-elapsed time counter in the central data management unit, which is used for time tagging housekeeping data and for spacecraft operations, is continuously synchronised to OBT using a pulse-per-second mechanism.

\section{Launch and commissioning}\label{sect:launch_and_commissioning}

\subsection{Launch and early-orbit phase}\label{subsect:launch_and_early-orbit_phase}

{\it Gaia} was launched from the European space port in French Guiana by a Soyuz-STB launch vehicle with Fregat upper stage on 19 December 2013 at 09:12:19.6 UTC. Initially, the coupled Fregat-{\it Gaia} upper composite was placed on a 180 km altitude parking orbit, after which a single Fregat boost injected {\it Gaia} into its transfer orbit towards the second Lagrange (L$_2$) point of the Sun-Earth-Moon system. Soon after {\it Gaia} separated from the Fregat, it autonomously pointed itself towards the Sun and initiated the deployment of the sunshield assembly and the release of the launch bipods between the service and payload modules. One day later, a turn-and-burn orbit manoeuvre with size $\Delta V = 23.5$~m~s$^{-1}$ was conducted to remove the stochastic launcher dispersion, after which the switch-on of service-module units and the ten-day payload decontamination (heating) phase was started. The launch and early orbit phase with extra ground-station coverage and 24-hour manned shifts of operators at the Mission Control Centre (Sect.~\ref{subsect:mission_operations}) lasted four days. After completion of the decontamination phase, on 3 January 2014 the scientific payload was switched on and overall system tuning began (Sect.~\ref{subsect:commissioning_and_performance_verification}). The transfer to L$_2$ took 26 days from launch and the insertion burn into the Lissajous orbit (Sect.~\ref{subsect:orbit_and_environment}) was split in two parts, separated by one week (7 and 14 January 2014), with a total $\Delta V$ of 166.3~m~s$^{-1}$.

\subsection{Commissioning and performance verification}\label{subsect:commissioning_and_performance_verification}

The commissioning and performance-verification phase was coordinated by ESA and the industrial prime contractor, Airbus DS, and was supported by scientists in the data processing and analysis consortium, in particular, the payload experts (Sect.~\ref{sect:science_operations}). This phase started with a period during which {\it Gaia} was initialised and its performance was iteratively improved \citep{2016AcAau.127..394M} and ended with a period during which {\it Gaia} was operated for a few weeks in ecliptic-pole scanning mode (Sect.~\ref{subsect:scanning_law}) to allow the science ground-segment verification of the scientific performance of {\it Gaia} \citep{2014SPIE.9150E..0AE}. Early activities in this process included (initial) focussing, spin-rate adjustments, and tuning the various settings of the onboard attitude-control loop, which matches the rotation of {\it Gaia} with the fixed TDI rate. The overall conclusion of the commissioning phase was that nearly all subsystems behaved nominally and some even better than expected. Examples of properly functioning subsystems are the focal plane assembly (noise, linearity, bias, cross-talk, etc.), the onboard data handling (including compression and prioritisation of science data), the onboard detection and windowing of sources, the phased-array antenna and link budget, the pointing and spin-rate performance achieved by the combined attitude control and micro-propulsion subsystems, and the Rubidium atomic master clock. This is the case despite the latter showing occasional, presumably stress-relief induced steps in the Rubidium lamp light level, at the level of $10^{-3}$~V, and occasional frequency jumps, at the level of $\Delta f/f \sim 5\cdot10^{-12}$, sometimes correlated with light level changes.

However, three particular points came to light. First, the latch valve of chemical thruster 3B was found stuck in closed position, later found, most likely as a result of a tiny leak of propellant inside the valve cap, i.e. the mechanical housing containing the valve circuitry. The leak itself is thought to be caused by a minuscule crack in a flexure sleeve, allowing NTO or MMH chemical propellant to leak into the actuator electronic assembly, causing circuit failure to the valve actuator coils and the microswitch. As an immediate mitigation, thruster 3B was removed from the onboard failure detection, isolation, and recovery logic such that thruster 3A would always have been used in case of safe mode. To recover robustness against the loss of redundancy of chemical thruster 3A, which had become a single-point failure in this configuration, a new AOCS survival mode was developed and implemented in the central software; this mode uses the torque authority provided by a slight misalignment of thrusters (originally designed only for attitude-control manoeuvres) to maintain three-axis spacecraft control in case of thruster 3A failure.
  
Second, the biases of the mass-flow sensors of the micro-propulsion thrusters, which means the offsets achieved with zero cold-gas mass flow, were found to be drifting. Such a drift in itself can be calibrated, although initially at the expense of science time because such a calibration initially required switching to chemical-propulsion control. Although the fear was that the drift would exceed the dynamic range of the offset measurement circuit, which is $\pm$400~mV. With time progressing and constant operation of the B branch, however, the offset drifts of the various thrusters have stabilised to values within the range that can be calibrated. The most probable root cause of the drift is incomplete pre-launch annealing of a (or some) resistor(s) in the mass-flow measurement chain.

Third, the spacecraft rotation rate was frequently found, of order once per minute, to be changing rapidly by typically up to a few milliarcsec per second and then quickly back to the rate before the excursion. From the characteristics of the rate-change signature, it is clear that these events are caused by sudden, minute structural changes (mass displacements) within the spacecraft causing a quasi-instantaneous discontinuity in the spacecraft attitude; these rate spikes are hence named micro-clanks in contrast to micro-meteoroid hits, which cause a sudden input of angular momentum and hence permanently change the spin rate of the spacecraft. Micro-clanks are observed both in the along-scan and across-scan direction, and they are often repeated (quasi-)periodically with the spin period of the spacecraft. In the along-scan direction, the vast majority of micro-clanks affect both fields of view equally and simultaneously, with no discernible effect in the BAM data, suggesting their origin is outside of the optical instrument. For a small fraction of events, however, the occurrence times coincide with jumps in the BAM fringe-position data (see below), suggesting that these events originate within the mechanical structure of the optics. Micro-clanks have also been detected in {\it Hipparcos} data \citep{2007ASSL..350.....V} and,  for that mission, have been attributed to small mechanical adjustments in the hinges of the solar-panel wings created by the varying amount of sunlight falling upon the wings over the rotation period of the {\it Hipparcos} spacecraft. For {\it Gaia}, something similar is likely happening but then primarily involving the bottom of the spacecraft (i.e. sunshield, launcher-interface ring, and/or phased-array antenna). Micro-clanks are easy to identify from data derived from CCD transit times and will be calibrated out in the preprocessing of the attitude data \citep[see][]{DPACP-14}.

In addition to the above, three issues affecting payload performance were uncovered during commissioning: contamination (Sect.~\ref{subsubsect:contamination}), straylight (Sect.~\ref{subsubsect:straylight}), and periodic basic angle variations (Sect.~\ref{subsubsect:periodic_basic-angle_variations}).

\subsubsection{Contamination}\label{subsubsect:contamination}

Soon after launch, it was discovered that the optics have time-variable, degrading transmission because of continued contamination by water ice. The transmission loss is wavelength dependent and, hence, different in the different instruments and also this loss varies with detector position in the focal plane within a given instrument. To restore the telescope throughput, three payload decontaminations were performed during commissioning (7 February, 13 March, and 30 June 2014) in which the focal plane and/or (selected) telescope mirrors were actively heated to sublimate the contaminant and let it escape through the apertures to space. With each decontamination, the rate of contamination was reduced and the stable period without noticeable contamination build-up lasted longer (see also Sect.~\ref{subsect:payload_calibrations_and_special_operations} and Fig.~\ref{fig:throughput}). This indicates that the source of the contamination, suspected to be slowly releasing trapped air (water vapour) within multilayer insulation blankets and/or carbon fibre-reinforced polymer structural parts, is drying up.

\subsubsection{Straylight}\label{subsubsect:straylight}

Soon after payload was switched on, it was discovered that straylight levels are modulated with the spacecraft rotation and some two orders of magnitude higher than expected. The origin of the straylight has been traced back foremost to scattered sunlight and secondly to the integrated brightness (and extremely bright stars) of the Milky Way, the light of which can reach the focal plane through a few unbaffled straylight paths. Although the telescope apertures are mostly shielded from direct illumination of the Sun by the double layer, deployable sunshield assembly, sunlight can scatter into the apertures through outward-protruding (bundles of) Nomex fibres; these fibres are present at the edges of the foldable sunshield blankets, which do not have a kapton tape finishing, and are present between each pair of fixed sunshield frames. Although the increased background levels themselves can be dealt with in the data processing, the associated noise negatively impacts the performance of faint objects, in particular in the spectroscopic instrument, which operates at very low signal levels. Therefore, a partial mitigation was implemented in the RVS windowing strategy allowing one to adaptively adjust the across-scan size of the windows to the instantaneous straylight level to increase the signal-to-noise ratio of the spectra. In addition, the RVS object selection was modified to adapt the faint limit to the instantaneous background level, which does not directly mitigate the straylight, but optimises the overall science quality of the RVS data acquired on board and downlinked to ground.

\subsubsection{Periodic basic angle variations}\label{subsubsect:periodic_basic-angle_variations}

Periodic fluctuations of the basic angle were measured from the switch on of the basic angle monitor (BAM), which is designed to monitor basic angle variations at the $\mu$as level with a sampling of $\sim$23~s and typical random noise of 12--15~$\mu{\rm as}~ {\rm Hz}^{-1/2}$ (Sect.~\ref{subsubsect:basic-angle_monitor}).

These fluctuations are two orders of magnitude larger than expected based on pre-launch calculations and show a strong asymmetry between both telescopes; the line of sight of the preceding telescope fluctuates within a $\pm$1000~$\mu$as range, while the other line-of-sight range is only $\pm$200~$\mu$as. These fluctuations show a modulation with the 6-hour spacecraft rotation as well as a smaller, 24-hour modulation. Fluctuations are also seen on longer timescales, which is consistent with the change of solar irradiance caused by the periodic change of the {\it Gaia}-Sun distance and sunshield aging, and correlated with Galactic plane crossings of the preceding telescope (30~$\mu$as amplitude). The periodic variations seen in the BAM signal are strongly coupled to the heliotropic spin phase $\Omega$ of the satellite (i.e. with respect to the Sun; Sect.~\ref{subsect:scanning_law}). In-orbit tests and experience have furthermore shown that variations disappear with a solar aspect angle of 0$^\circ$ or when the spin is stopped; in the latter case, the variations reappear within minutes after restarting the spin. A dedicated working group, involving ESA and Airbus DS, has come to the following conclusions. First, the BAM data are reliable; they measure and reflect true basic angle variations at least at the level of accuracy that is relevant for the first intermediate data release, which is a few dozen $\mu$as. Second, a purely mechanical root cause can be ruled out. Third, there is strong evidence of a thermo-elastic origin. In particular, the 24-hour variation originates from the central data management unit, transponders, and payload data-handling unit in the service module, which have varying power dissipations and temperatures as a result of the daily downlink operations (Sect.~\ref{subsubsect:ground_stations}), while the reaction of the preceding field of view to the Galactic plane crossings shows a perfect, delayed correlation with thermal variations of the VPUs (with a coupling coefficient of $\sim$500~$\mu$as~K$^{-1}$). A detailed sensitivity analysis was carried out in flight, applying thermal pulses to many service- and payload-module components during one decontamination campaign. When combined with the typical, cyclic temperature changes experienced during a revolution, a number of candidates were identified as possible originators of the six-hour periodic basic angle variations. These candidates are mostly located in the Sun-illuminated part of the spacecraft. However, the (probably thermo-elastic) coupling mechanism needed to efficiently and quickly translate those perturbations to the payload module is still unclear.

In addition to the periodic variations, the BAM fringe position data also show jumps. Their size distribution follows a power law with an exponent of $-0.8$. There is on average one jump per day (in either field of view) exceeding $30~\mu$as, although jumps are much more frequent after perturbations such as decontamination and (re-)focussing. There are about equal numbers of jumps affecting either both fields of view at the same time, only the preceding field, or only the following field. Large jumps show up in the astrometric residuals, allowing for the possibility that most of them are real, for instance originating within the mechanical structure of the payload.

\section{Mission and spacecraft operations}\label{sect:mission_and_spacecraft_operations}

\subsection{Orbit and environment}\label{subsect:orbit_and_environment}

{\it Gaia} operates at the second Lagrange (L$_2$) point of the Sun-Earth-Moon system. This saddle point is located $\sim$1.5 million km from Earth, in the anti-Sun direction, and co-rotates with the Earth in its one-year orbit around the Sun. {\it Gaia} moves around L$_2$ in a Lissajous-type orbit with amplitudes of $120\,000$~km $\times$ $340\,000$~km and $180\,000$~km (in and perpendicular to the ecliptic plane, respectively) and an orbital period of $\sim$180 days; these numbers guarantee that the Earth is maximally $15^\circ$ away from the boresight of the phased-array antenna, which is used for science data transmission (Sect.~\ref{subsubsect:phased-array_antenna}). An L$_2$ Lissajous orbit has several advantages over an Earth-bound orbit, for instance stable thermal conditions, a benign radiation environment, and a high observing efficiency in which the Sun, Earth, and Moon are outside the fields of view of the instrument. By design, the orbit of {\it Gaia} is not impacted by eclipses of the Sun by the Earth during its five-year nominal lifetime, although small (percent level) partial eclipses by the Moon typically occur once per year. In July 2019, a $14$~m~s$^{-1}$ eclipse-avoidance manoeuvre is scheduled to ensure continued Earth-eclipse-free operations through to 2025.

After more than two years in orbit, the L$_2$ environment has not (yet) provided surprises:
\begin{enumerate}
\item The temperature of the spacecraft is generally as expected, with long-term trends visible owing to the seasonal variation in the spacecraft-Sun distance.
\item The rate of micro-meteoroid fluxes is generally as expected (with the attitude-control system reacting to momentum disturbances exceeding $\sim$$10^{-5}$~N~m~s), essentially following the Gr\"un interplanetary flux model \citep{1985Icar...62..244G}.
\item The radiation damage to the CCD detectors is measured through first-pixel response in the along-scan profiles of lines of charge injection and through onboard counters registering the number of rejected detections in the sky mapper; this radiation damage is generally as expected, essentially showing a slow yet persistent degradation, caused by the omnipresent flux of high-energy Galactic cosmic rays (a handful of particles~cm$^{-2}$~s$^{-1}$), combined with a handful of discontinuities corresponding to solar activity and associated proton events \citep{2014SPIE.9154E..06K,2016arXiv160801476C}. The data also show a clear sign of the well-known anti-correlation between cosmic-ray fluxes and solar activity \citep[e.g.][]{2001ICRC...10.3971D}. Owing to the (so far) benign nature of the current solar cycle 24, however, the accumulated radiation dose has been significantly lower than expected, with only $\sim$4\% of the predicted end-of-mission (five-year) dose having materialised so far during the first two years of operations \citep[for a detailed description, see][]{dpacp-20}.
Radiation damage is hence not expected to strongly affect the long-term performance of {\it Gaia}.
\end{enumerate}

\subsection{Scanning law}\label{subsect:scanning_law}

\begin{figure}[t]
\includegraphics[width=\columnwidth]{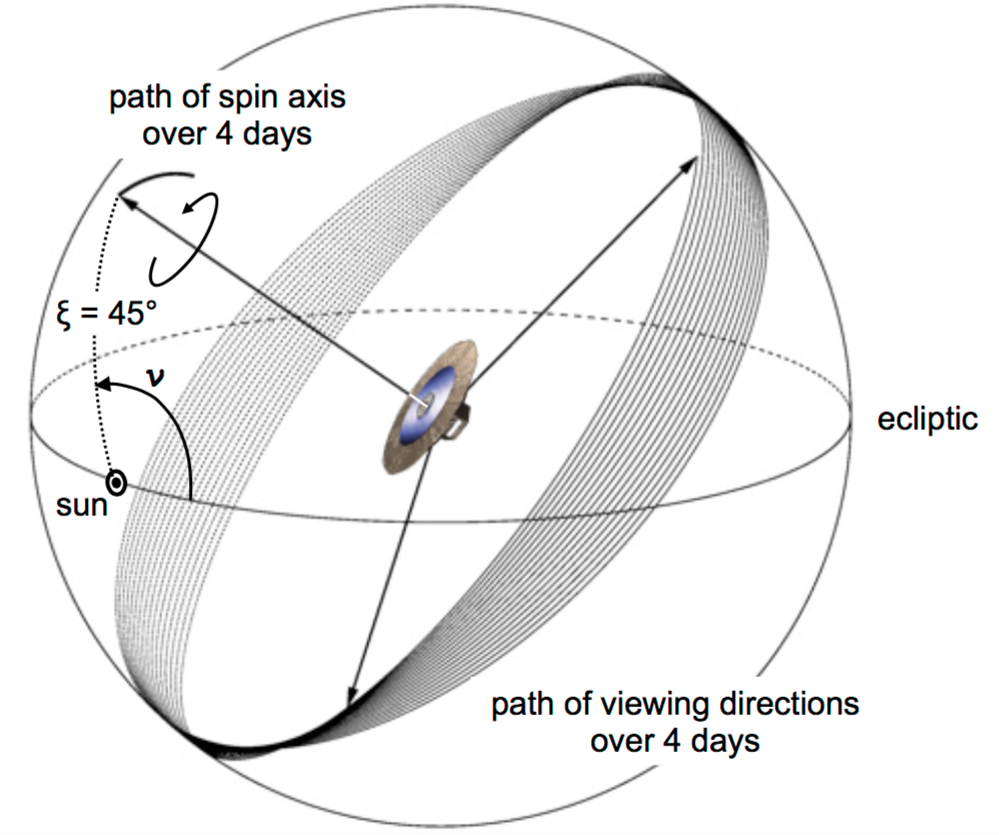}
\caption{Illustration of the scanning law of {\it Gaia}, showing the path of the spin axis ($z$), and the corresponding path of the preceding viewing direction, during four days. For clarity, the path of the following viewing direction is not shown. Image courtesy Lennart Lindegren.}
\label{fig:nsl1}
\end{figure}

\begin{figure}[t]
\includegraphics[width=\columnwidth]{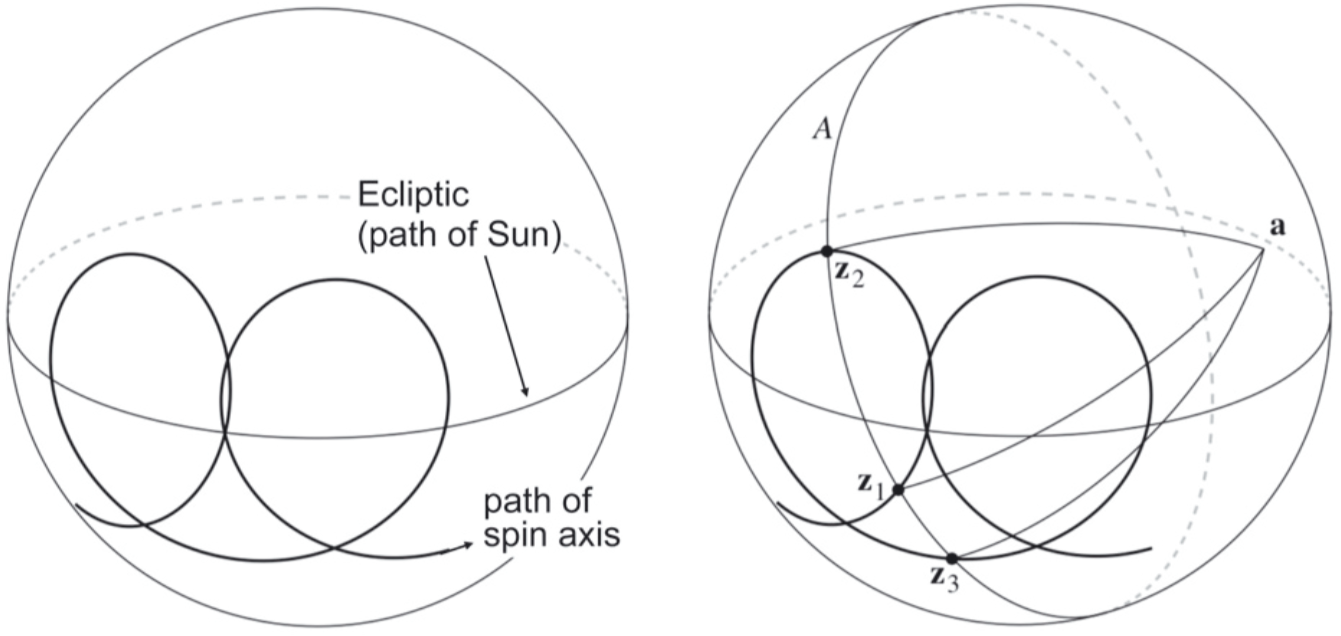}
\caption{Spin axis $z$ makes loops around the Sun, which must overlap. A star at point $a$ may be scanned whenever $z$ is $90^\circ$ from $a$, i.e. on the great circle A at $z_1$, $z_2$, $z_3$, etc. The scans intersect at a large angle, which allows the determination of two-dimensional object coordinates from one-dimensional measurements. Image courtesy Lennart Lindegren.}
\label{fig:nsl2}
\end{figure}

The scanning law, which describes the intended spacecraft pointing as a function of time, is one of the keys to the astrometric performance of {\it Gaia}. Following principles already worked out in the 1970s and employed for {\it Hipparcos} in the 1990s \citep[e.g.][]{1981A&A...101..228H}, {\it Gaia} scans the sky using uniform revolving scanning. Such a scanning law optimises the astrometric accuracy in the sense of maximising the uniformity of the sky coverage. Key ingredients of the nominal scanning law are as follows \citep[Fig.~\ref{fig:nsl1};][]{2011EAS....45..109L}:
\begin{itemize}
\item There is a fixed spin rate $\omega_z = 60\arcsec$~s$^{-1}$ around the spacecraft spin axis ($z$) to ensure that the optical stellar images move over the detector surface with the same speed as the electrons are transferred inside the CCD during the TDI operation (Sect.~\ref{subsect:measurement_principle_and_overall_design_considerations}).
\item The solar-aspect angle, $\xi = 45^\circ$, between the Sun and the instrument $z$ axis, is fixed to ensure maximum parallax sensitivity (because the measurable, along-scan parallax displacement of an object is proportional to $\sin\xi$ ; Sect.~\ref{subsect:measurement_principle_and_overall_design_considerations}) and maximum thermal stability of the payload and basic angle in particular. In practice, the scanning law is defined with respect to a fictitious, nominal Sun: this allows unambiguous specification of the scanning law with respect to the ICRS, independent of the orbital motion of {\it Gaia} around L$_2$. The difference between the actual and nominal Sun is never larger than a few arcminutes.
\item A slow precession of the spin axis around the Sun results in a series of loops around the solar direction (Fig.~\ref{fig:nsl2}). A side effect of the precession is an across-scan speed of stellar images during their focal plane transits. The speed of the precession is as small as possible to limit the across-scan smearing of images when they transit the focal plane yet (just) large enough to ensure that subsequent loops overlap, in which case there are at least six distinct epochs of observations per year for any object in the sky. For {\it Gaia}, this is achieved with 5.8~revolutions per year ($4^\circ$~day$^{-1}$ precession relative to the stars), which means that the precession period is $365.25 / 5.8 = 63$ days and that the across-scan speed of images transiting the focal plane varies sinusoidally with time, with a nominal period of six~hours and an amplitude of 173~mas~s$^{-1}$.
\end{itemize}
Given the apparent path of the Sun on the celestial sphere and the fixed value of $\xi$, the scanning law, i.e. the orientation of the instrument in the heliotropic frame, which has the nominal ecliptic as its fundamental plane and the Sun as origin, is described by two heliotropic angles. First, the revolving phase $\nu(t)$ (also known as precession phase), which is the angle between the ecliptic plane and the plane containing the Sun and the instrument $z$ axis. Second,  the spin phase $\Omega(t)$, which is the angle between the plane containing the Sun and the instrument $z$ axis and the instrument $zx$ plane; the fundamental $xy$ plane is the plane through the two viewing directions (see \cite{DPACP-14} for a definition of the scanning reference system). The governing equation for $\nu(t)$ equals
\begin{equation}
\dot{\nu} \sin\xi = \dot{\lambda_\odot} \sqrt{S^2 - \cos^2 \nu} + \dot{\lambda_\odot} \cos\xi \sin\nu,
\end{equation}
where $\lambda_\odot$ denotes the ecliptic longitude of the (nominal) Sun and $S = |\vec{\omega} \times \vec{z}|~ \dot{\lambda_\odot}^{-1} = 4.220745$ is a dimensionless constant (corresponding to 5.8 revolutions per year). The spin phase $\Omega(t)$ then follows from
\begin{equation}
\dot{\Omega} = \omega_z - \dot{\lambda_\odot}\sin\xi \sin\nu - \dot{\nu}\cos\xi.
\end{equation}
The above two equations have only two free parameters: the initial spin phase and the initial precession angle, at the start of science operations. Both angles have been initialised to observe the most favourable passages of bright stars very close to the limb of Jupiter, aiming to measure the light deflection owing to the quadrupole component of the gravitational field of this planet \citep{2010IAUS..261..331D}. In practice, the scanning law (i.e. the intended orientation of the scanning reference system with respect to the ICRS as a function of time) is generated by Runge-Kutta integration of the above equations, converted from heliotropic angles $\nu$ and $\Omega$ to celestial coordinates, and finally approximated by piecewise polynomials using the Chebyshev representation.

During the first weeks of nominal science operations (between 25 July and 21 August 2014), {\it Gaia} was operating with a special scanning law, known as the ecliptic poles scanning law (EPSL). In this mode, $\nu(t)$ stayed constant at $180^\circ$, which means that the spin axis followed the Sun on the ecliptic (the alternative EPSL solution, with $\nu(t) = 0^\circ$, was not used). This means that the fields of view scan through both the south and the north ecliptic poles in every six-hour scan, and that the direction of scanning changes by the same rate as the Sun (and the spin axis) moves along the ecliptic, which is $\sim$$1^\circ$ per day. Advantages of the EPSL are that a limited set of objects near the ecliptic poles are observed very frequently and with a (four times) smaller across-scan image motion than nominally. The EPSL observations have hence been used to bootstrap astrometric and photometric calibrations and for performance verification during the commissioning phase (Sect.~\ref{subsect:commissioning_and_performance_verification}).

The scanning law returns maximum sky-coverage uniformity after the nominal, five-year mission. Nonetheless, the number of times an object is observed ($N_{\rm obs}$) depends on its position in the sky, in particular on its ecliptic latitude $\beta$ (see Tab.~\ref{tab:sky_dependence} in Sect.~\ref{subsect:astrometry}). For high-priority bright stars, data losses are small (Sect.~\ref{subsubsect:ground_stations}) and the sky average, end-of-mission number of focal plane transits, for both fields of view combined, is around 80 in AF, BP, and RP (and a factor $4 / 7 = 0.57$ smaller in RVS; Sect.~\ref{subsubsect:spectroscopic_instrument}). At the faint end ($G \gtrsim 20$~mag or $G_{\rm RVS} \gtrsim 14$~mag), however, data losses can grow to $\sim$20\% such that the number of transits reduces to $\sim$70 (and $\sim$40 in RVS).

\subsection{Mission operations}\label{subsect:mission_operations}

The mission operations of {\it Gaia} are conducted by the Mission Operations Centre (MOC) located at the European Space Operations Centre (ESOC) in Darmstadt, Germany. The core spacecraft operations of {\it Gaia}, which are similar to other ESA missions, include:
\begin{itemize}
\item mission planning, also based on input from the SOC (Sect.~\ref{sect:science_operations});
\item regular upload of the planning products to the mission timeline of {\it Gaia};
\item acquisition and distribution of science telemetry;
\item acquisition, monitoring and analysis, and distribution of health, performance (voltage, current, temperature, etc.), and resource (power, propellant, link budget, etc.) housekeeping data of all spacecraft units via $\sim$$40\,000$ telemetry parameters;
\item performing and monitoring operational time synchronisation (time correlation; Sect.~\ref{subsubsect:time_correlation});
\item anomaly investigation, mitigation, and recovery;
\item orbit prediction, reconstruction, monitoring, and control (Sect.\ref{subsubsect:orbit_prediction,_monitoring,_and_control});
\item spacecraft calibrations (e.g. star-tracker alignment, micro-propulsion offset calibration, etc.); and 
\item onboard software maintenance.
\end{itemize}
The {\it Gaia} avionics module (AVM), with flight-representative electronics hardware, is located at MOC to support spacecraft operations and trouble shooting.

\subsubsection{Ground stations}\label{subsubsect:ground_stations}

In order to have a high-quality link budget and high science data rate, the three 35-meter deep-space dishes in the ESA tracking station network (ESTRACK) are used. These stations are located at Malarg\"ue (Argentina), Cebreros (Spain), and New Norcia (Australia) and hence provide nearly 24 h coverage. The daily telecommunications period is adjusted to the expected data volume to be downlinked each day, which is predicted based on a sky model combined with the operational scanning law. The typical downlink time of $\sim$12.5~hours is normally covered by two of the three antennae. In times of enhanced data rates, typically when the scanning law makes {\it Gaia} scan along, or at small angles to, the Galactic plane (loosely referred to as a Galactic plane scan), required downlink times increase, up to, and exceeding, the maximum possible 24 hours per day, which means that three antennae are used sequentially. The science data is telemetered to ground in the X band through the high-gain phased-array antenna (Sect.~\ref{subsubsect:phased-array_antenna}) using Gaussian minimum-shift keying (GMSK) modulation. Error correction in the downlink telemetry stream is achieved through the use of concatenated convolutional punctured coding. In practice, a 7/8 convolutional encoding rate is used as baseline so that the downlink information data rate (including packetisation and error-correction overheads) is 8.7~megabits per second. The typical amount of (compressed) science data downlinked to ground is some 40~gigabytes per day. Actual onboard data losses caused by shortage of ground-station contact periods in times of Galactic plane scans are modest: for astrometry and photometry (star packet 1, SP1; Sect.~\ref{subsubsect:video_processing_unit_and_algorithms}), the data loss is zero for bright objects ($G < 16$~mag), a few percent between 16 and 20~mag, around 10\% between 20 and 20.5~mag, and $\sim$25\% for fainter objects. For spectroscopy (star packet 2, SP2), the data loss is zero for stars brighter than $G_{\rm RVS} = 10.5$~mag, a few percent between 10.5 and 14~mag, and grows to 5--10\% around 16~mag.

\subsubsection{Orbit prediction, reconstruction, monitoring, and control}\label{subsubsect:orbit_prediction,_monitoring,_and_control}

 The spacecraft must periodically undergo orbit maintenance (station-keeping) manoeuvres to ensure that it does not escape from L$_2$. These events occur a few times per year and necessitate roughly a one-hour pause of the science operations because they involve the chemical propulsion system. Velocity changes ($\Delta V$) are at the level of a (few) dozen cm~s$^{-1}$ and are aimed at removing the escape component from the spacecraft orbit, so that the actual orbit only drifts over a few 1000~km over five years compared with the reference orbit. The amount of chemical propellant available for orbit maintenance manoeuvres allows the spacecraft to be kept at L$_2$ for several decades. In contrast, the cold-gas consumption of the micro-propulsion subsystem (Sect.~\ref{subsubsect:attitude_and_orbit_control}) will limit the lifetime of Gaia to $10\pm1$~years, under the assumption of no hardware failures.

The requirements for the reconstructed orbit of {\it Gaia} are stringent: its near-Earth asteroid science case requires the positional error to be smaller than 150~m whereas aberration corrections for astrometric data collected for bright stars require the velocity error to be smaller than 2.5~mm~s$^{-1}$. To ensure that these requirements are met at all times, four types of data enter the orbit reconstruction process at MOC:
\begin{enumerate}
\item Two-way ranging data, typically acquired both at the beginning and at the end of a communications period with the spacecraft (pass);\item Doppler data, acquired continuously during spacecraft contact with the ground station;
\item Delta differential one-way-range ($\Delta$-DOR) data, occasionally obtained while tracking the spacecraft position on the sky with respect to a background quasar with known coordinates in the ICRF using two ESA ground stations simultaneously;
\item Optical, plane-of-sky measurements of the spacecraft location with respect to the background stars {\it Gaia} itself is measuring. These data are routinely being collected through the Ground-Based Optical Tracking (GBOT) programme \citep{2014SPIE.9149E..0PA} and are critical when the spacecraft is at low declinations ($|\delta| \la 15^\circ$). The GBOT uses a network of small-to-medium telescopes (the ESO VLT Survey Telescope, which has contributed 400~hours of time to date thanks to an agreement between ESA and ESO, the Liverpool Telescope, and the Las Cumbres telescopes) and aims to deliver a daily measurement with 20~mas precision. The GBOT data can only be used in full in combination with {\it Gaia} DR1 because including these data in the orbit reconstruction process requires the {\it Gaia} catalogue positions of the background stars against which the position of {\it Gaia} was measured.
\end{enumerate}
After processing at MOC, the reconstructed orbit is periodically delivered to the Science Operations Centre for inclusion in the science data processing cycles.

\subsubsection{Time synchronisation}\label{subsubsect:time_correlation}

Time synchronisation denotes the establishment of a relation between the onboard time (OBT) reading of the free-running, atomic master clock on board {\it Gaia} (Sect.~\ref{subsubsect:clock_distribution_unit}) and a reference timescale such as Universal Coordinate Time (UTC) or Barycentric Coordinate Time (TCB). Two time-synchronisation chains are active for {\it Gaia}:
\begin{enumerate}
\item MOC maintains a low-accuracy, on-the-fly service that is based on the predicted spacecraft orbit and uses simplified algorithms. This time synchronisation is used for spacecraft operations, for instance for the interpretation of housekeeping data or the definition of onboard command-execution times, and also for initial (first-look) science data processing at the Science Operations Centre. The formal (required) accuracy of this product is 1~ms, but the typical accuracy in practice is 50--100~$\mu$s.
\item For the science cases of {\it Gaia} to be met, the absolute time accuracy requirement at mission level is $2~\mu$s, which means that the 1-ms MOC service is insufficient. Therefore, a second time-synchronisation and onboard clock calibration chain is operated by DPAC (Sect.~\ref{sect:dpac}). This scheme is based on a time-source, packet-based, one-way clock synchronisation scheme \citep{2015jsrs.conf...55K} and achieves the 2 $\mu$s requirement with significant margin during nominal operations. This scheme is fully relativistic, includes onboard delays, propagation delays, and ground-station delays but unavoidably produces results with a delay of several weeks, for instance because the reconstructed spacecraft orbit is one of the inputs.
\end{enumerate}

\begin{figure}[t]
\includegraphics[width=\columnwidth]{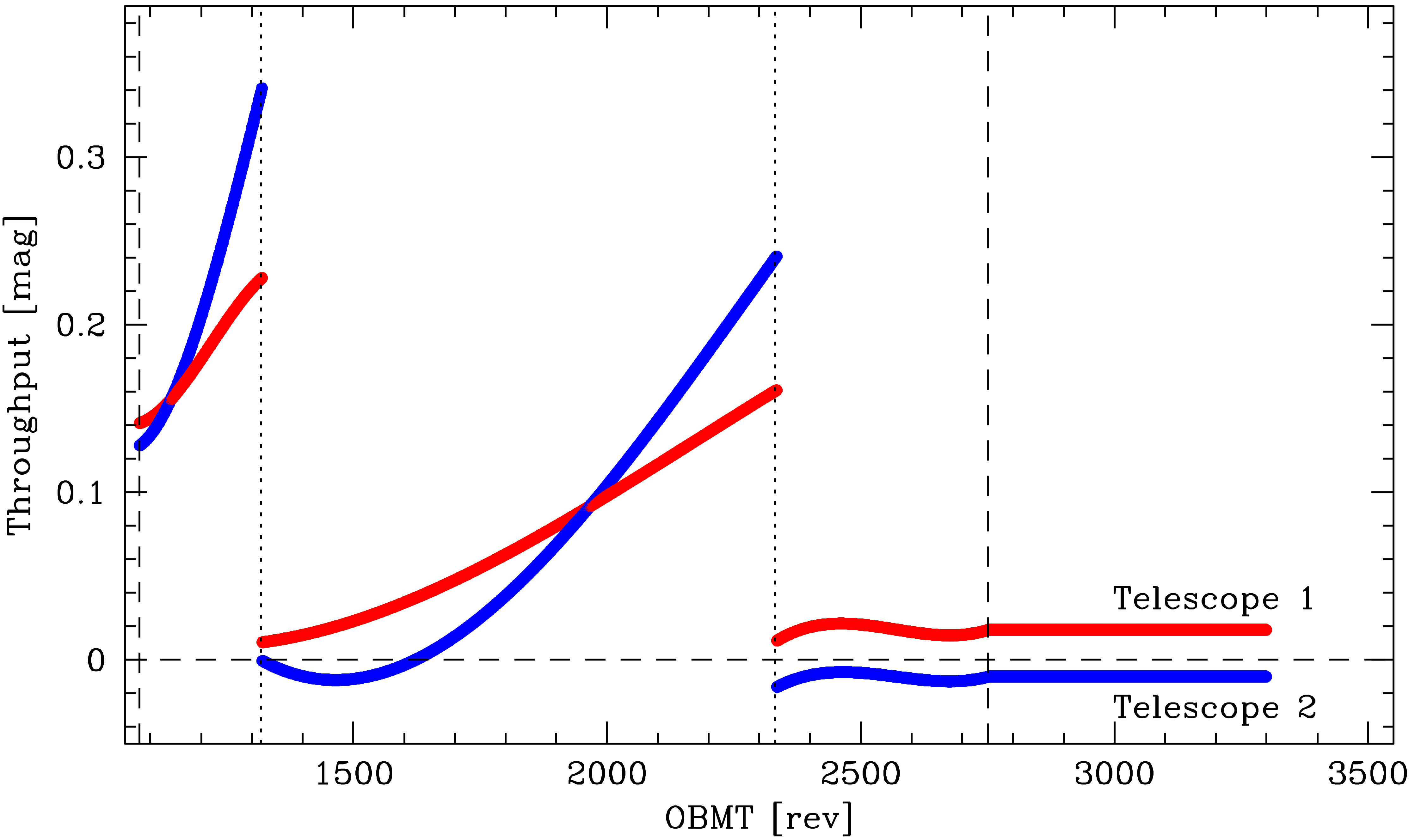}
\caption{Throughput evolution of the astrometric fields of view (averaged over all AF CCDs) as a function of time as revealed by the $G$-band photometric time-link calibration derived in \cite{DPACP-10}. The curves are Chebyshev polynomials of the actual calibrations. Red lines refer to the preceding field of view; blue lines to the following field of view. The discontinuities and dotted vertical lines refer to the decontaminations (Sect.~\ref{subsect:payload_calibrations_and_special_operations}). OBMT stands for onboard mission timeline in units of six-hour revolutions since launch (Sect.~\ref{subsubsect:time_correlation}). {\it Gaia} Data Release~1 is based on data covering the interval OBMT = 1078.3795--2751.3518, delimited by the dashed black lines.}
\label{fig:throughput}
\end{figure}

To avoid ambiguity in case of clock resets, OBT counts are first transformed in the initial processing of the data \citep{DPACP-7} into the OBMT (onboard mission timeline) by adding a constant offset. The OBMT is conventionally expressed in units of six-hour (21,600~s) spacecraft revolutions since launch. For instance for the purpose of interpreting the timeline of figures (e.g. Fig.~\ref{fig:throughput}), OBMT can be converted into TCB at {\it Gaia} (expressed in Julian years) by the approximate relation,
\begin{equation}
{\rm TCB} \approx {\rm J}2015.0 + ({\rm OBMT} - 1717.6256~{\rm rev}) / (1461~{\rm rev}).
\end{equation}
This relation is valid only for the interval OBMT = 1078.3795--2751.3518, which covers {\it Gaia} Data Release~1.

\section{Science operations}\label{sect:science_operations}

The science operations of {\it Gaia} are conducted by the Science Operations Centre (SOC) located at the European Space Astronomy Centre (ESAC) in Villafranca del Castillo, Spain. The {\it Gaia} SOC is an integral part of the {\it Gaia} DPAC (Sect.~\ref{sect:dpac}) and therefore has, in addition to the classical roles outlined in this section, a number of other responsibilities towards DPAC. These other responsibilities are 
\begin{itemize}
\item forming the main hub among the six data processing centres (DPCs; Sect.~\ref{subsect:organisation}) of DPAC;
\item providing system architecture and support functions for DPAC and hosting the main database (MDB) of the mission;
\item operating the daily pipeline (Sect.~\ref{subsect:daily_data_processing}) and disseminating all data products to other DPCs (hub-and-spokes topology);
\item operating and co-developing the astrometric global iterative solution (AGIS; Sect.~\ref{subsect:cyclic_data_processing});
\item operating and co-developing the mission archive (Sect.~\ref{subsect:Simulations,_supplementary_data_and_observations,_data publication}).
\end{itemize}

\subsection{Interface to the Mission Operations Centre}\label{subsect:interface_to_the_mission_operations_centre}

The SOC is the prime interface to the MOC for all payload- and science-related matters. Normal work includes
\begin{itemize}
\item generating the scanning law (Sect.~\ref{subsect:scanning_law}), including the associated calibration of the representation of the azimuth of the Sun in the scanning reference system in the VPU software;
\item generating the science schedule, i.e. the predicted onboard data rate according to the operational scanning law and a sky model, to allow for adaptive ground-station scheduling (Sect.~\ref{subsubsect:ground_stations});
\item generating the avoidance file containing time periods when interruptions to science collection would prove particularly detrimental to the final mission products;
\item generating payload operation requests \citep[PORs;][]{2014SPIE.9149E..0QM}, i.e. VPU-parameter updates (e.g. TDI-gating scheme or CCD-defect updates);
\item tracking the status and history of payload configuration parameters in the configuration database (CDB) through the mission timeline and telecommand history;
\item hosting the science-telemetry archive;
\item generating event anomaly reports (EARs) to inform downstream processing systems of bad time intervals, outages in the science data, or any (onboard) events that may have an impact on the data processing and/or calibration;
\item monitoring (and recalibrating as needed) the star-packet-compression performance;
\item monitoring (and recalibrating as needed) the BAM laser beam waist location inside the windows;
\item reformatting the optical observations of {\it Gaia} received from GBOT for MOC processing in the orbit reconstruction (Sect.~\ref{subsubsect:orbit_prediction,_monitoring,_and_control});
\item disseminating to DPAC (Sect.~\ref{sect:dpac}) meteorological ground-station data received from MOC and required for delay corrections in the high-accuracy time synchronisation (Sect.~\ref{subsubsect:time_correlation}).
\end{itemize}

\subsection{Daily processing}\label{subsect:initial_data_treatment}

As soon as science telemetry arrives from MOC at SOC, its processing in the daily pipeline is initiated \citep{2014SPIE.9149E..2ES}. After the MOC interface task (MIT), which provides a first-level data reconstruction into source-packet groups, and the decompression and calibration services (DCS),  which decompresses, identifies, and stores the various data packets in a relational database, the initial data treatment \citep[IDT;][]{DPACP-7} is run. The IDT processing includes reconstructing all details for each observation window, for example location, shape, and TDI gating, and calculating image parameters (flux and along-scan location, i.e. observation time) and preliminary colours. Also, IDT constructs a coarse version of the spacecraft attitude.  This on-ground attitude version
1 (OGA-1) has sufficient accuracy that the observed sources are either identified in the initial {\it Gaia} source list \citep[][IGSL]{2014A&A...570A..87S} or added as new sources during the cross-matching. The IDT data, which has positional accuracies below $0\farcs{1}$, forms the basis of both the asteroid and photometric science alert pipelines \citep{2016P&SS..123...87T,2016arXiv160102827W}. The IDT data is disseminated to the data processing centres in DPAC (Sect.~\ref{sect:dpac}) for further downstream processing through the main database \citep[MDB;][]{2014SPIE.9149E..2ES}.

\begin{figure}[t]
\includegraphics[width=\columnwidth]{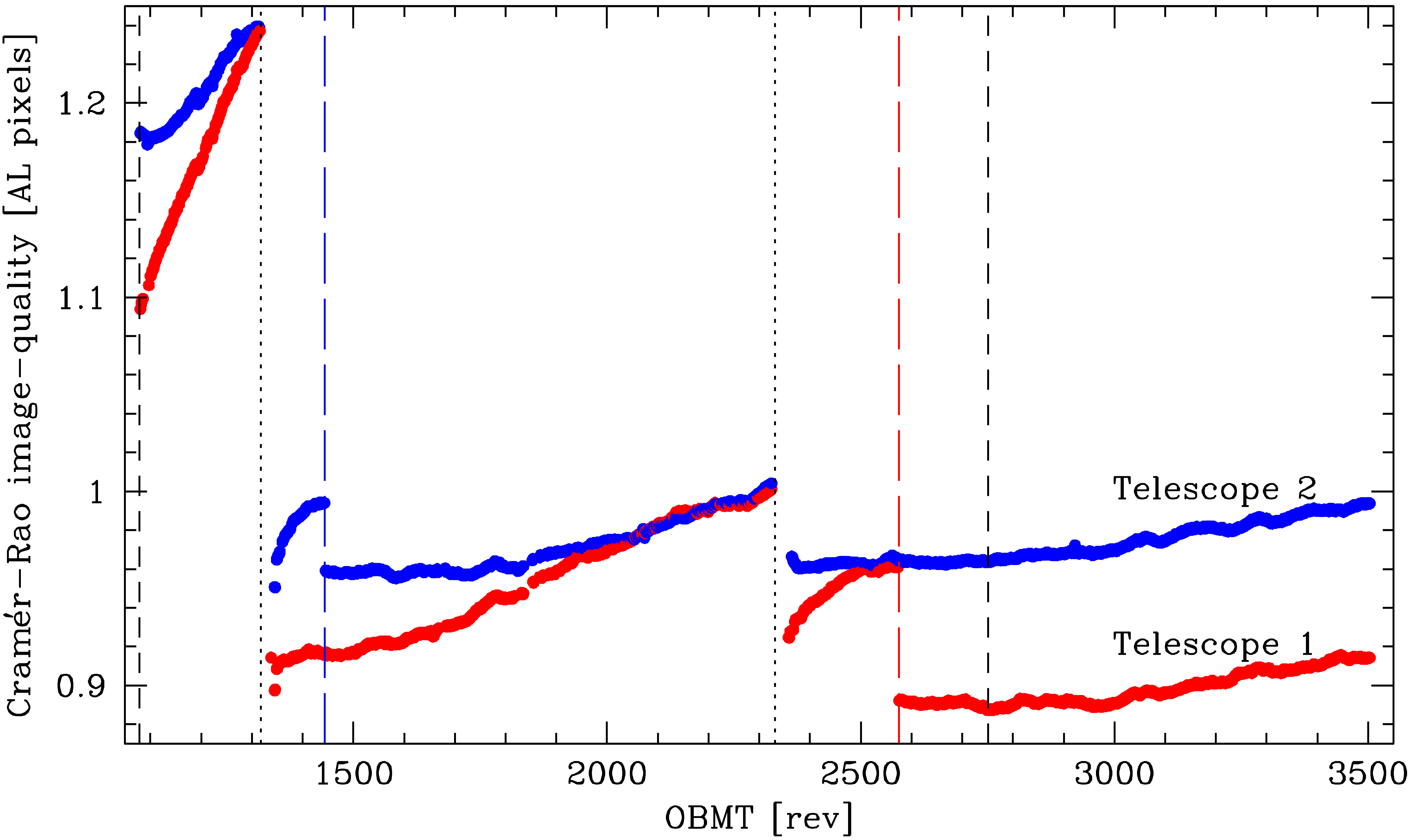}
\caption{Evolution of the astrometric image quality (averaged over all AF CCDs) as a function of time as quantified by the Cram\'er-Rao centroiding diagnostic. A lower value indicates a better image quality and centroiding performance. Red lines refer to the preceding field of view; blue lines to the following field of view. The discontinuities refer to the decontaminations (dotted vertical lines) and refocussings (dashed blue and red lines; Sect.~\ref{subsect:payload_calibrations_and_special_operations}). OBMT stands for onboard mission timeline in units of six-hour revolutions since launch (Sect.~\ref{subsubsect:time_correlation}). {\it Gaia} Data Release~1 is based on data covering the interval OBMT = 1078.3795--2751.3518, delimited by dashed black lines.}
\label{fig:focus}
\end{figure}

\subsection{Payload health monitoring}\label{subsect:payload-health_monitoring}

{\it Gaia} has a payload health and scientific data-quality monitoring system, called First Look \citep[FL;][]{DPACP-7}, that is sophisticated and close to real time. This system is deployed in the daily pipeline. First Look operates on IDT and spacecraft housekeeping data and produces thousands of summary diagnostic quantities as well as daily instrument calibrations and the second on-ground attitude reconstruction (OGA-2). These diagnostics are reported daily and inspected by qualified First-Look scientists to identify possible actions to improve the performance of the spacecraft and/or of the on-ground data processing. A group of DPAC-wide payload experts provide expertise in interpreting First-Look diagnostics, monitor the health of the satellite based on all daily processing systems (Sect.~\ref{subsect:daily_data_processing}), and provide advice to the project scientist and mission manager on identified improvements.

\subsection{Payload calibrations and special operations}\label{subsect:payload_calibrations_and_special_operations}

Although {\it Gaia} is essentially a self-calibrating mission, which means that the required instrument calibrations are derived from the science data themselves (Sect.~\ref{subsect:measurement_principle_and_overall_design_considerations}), dedicated payload calibrations are periodically performed \citep{dpacp-20} in close collaboration with the DPAC payload experts group. These calibrations include the following.
\begin{enumerate}
\item Post-scan-pixel acquisitions for serial-CTI monitoring \citep{2014SPIE.9154E..06K}: This calibration uses SIF data (Sect.~\ref{subsubsect:video_processing_unit_and_algorithms}) with the VPUs in operational mode but with the SM object detection disabled; one virtual object (Sect.~\ref{subsubsect:video_processing_unit_and_algorithms}) is defined to trigger the image-area readout.
\item Virtual object acquisitions for CCD-PEM non-uniformity monitoring (Sect.~\ref{subsubsect:focal-plane_assembly}): This calibration uses virtual objects with the VPUs in operational mode but with the SM object detection disabled. In addition, SIF is activated for short times to verify the TDI-gate settings;
\item SIF acquisitions for flat-band, voltage-shift monitoring induced by ionising radiation damage \citep{2014SPIE.9154E..06K}: This calibration exclusively uses SIF data with the VPUs in service mode.
\item Charge-injection calibration in the sky mappers for monitoring the radiation state of these detectors: This calibration exclusively uses SIF data with the VPUs in operational mode.
\end{enumerate}
In addition to these nominal calibrations, two types of special operations have been performed occasionally on {\it Gaia}, namely refocussing and decontamination of the optics.

During commissioning, three decontamination campaigns were conducted to sublimate contaminating water ice from the optics (Sect.~\ref{subsect:commissioning_and_performance_verification}). During nominal operations, throughput evolution continued (Fig.~\ref{fig:throughput}) and two more decontaminations were performed: the first on 23 September 2014 (OBMT $\sim$ 1317) and the second on 3 June 2015 (OBMT $\sim$ 2330). At the time of writing (mid-2016), the total build-up of throughput loss has been modest ($\sim$0.1~mag in AF) and the current rate of transmission loss is $\sim$0.1~mmag per six-hour revolution.

During the commissioning phase, both telescopes were aligned and focussed, first in April 2014 and then again, after reaching a more stable thermal equilibrium, in July 2014 \citep{2014SPIE.9143E..0XM}. Continuous image-quality monitoring was performed via the evolution of the full width at half maximum of the point spread function and was quantified through the Cram\'er-Rao centroid diagnostic \citep[e.g.][]{2013PASP..125..580M} applied to bright-star data; this monitoring has revealed slow image-quality degradations of both telescopes, which is strongly correlated with the build-up of contamination (Fig.~\ref{fig:focus}). This can be understood because uneven, patchy mirror apodisation introduced by the contamination broadens the point spread function. Although the image quality improved with both decontaminations carried out, there has been a need for two active refocussings to improve the image quality further, once on 24 October 2014 (OBMT $\sim$ 1444) for the following field of view and once on 3 August 2015 (OBMT $\sim$ 2575) for the preceding field of view. At the time of writing (mid 2016), both telescopes are optimally focussed within a few percent with slow degrading trends, most likely caused by the slow build-up of contamination.

\begin{figure}[t]
\includegraphics[width=\columnwidth]{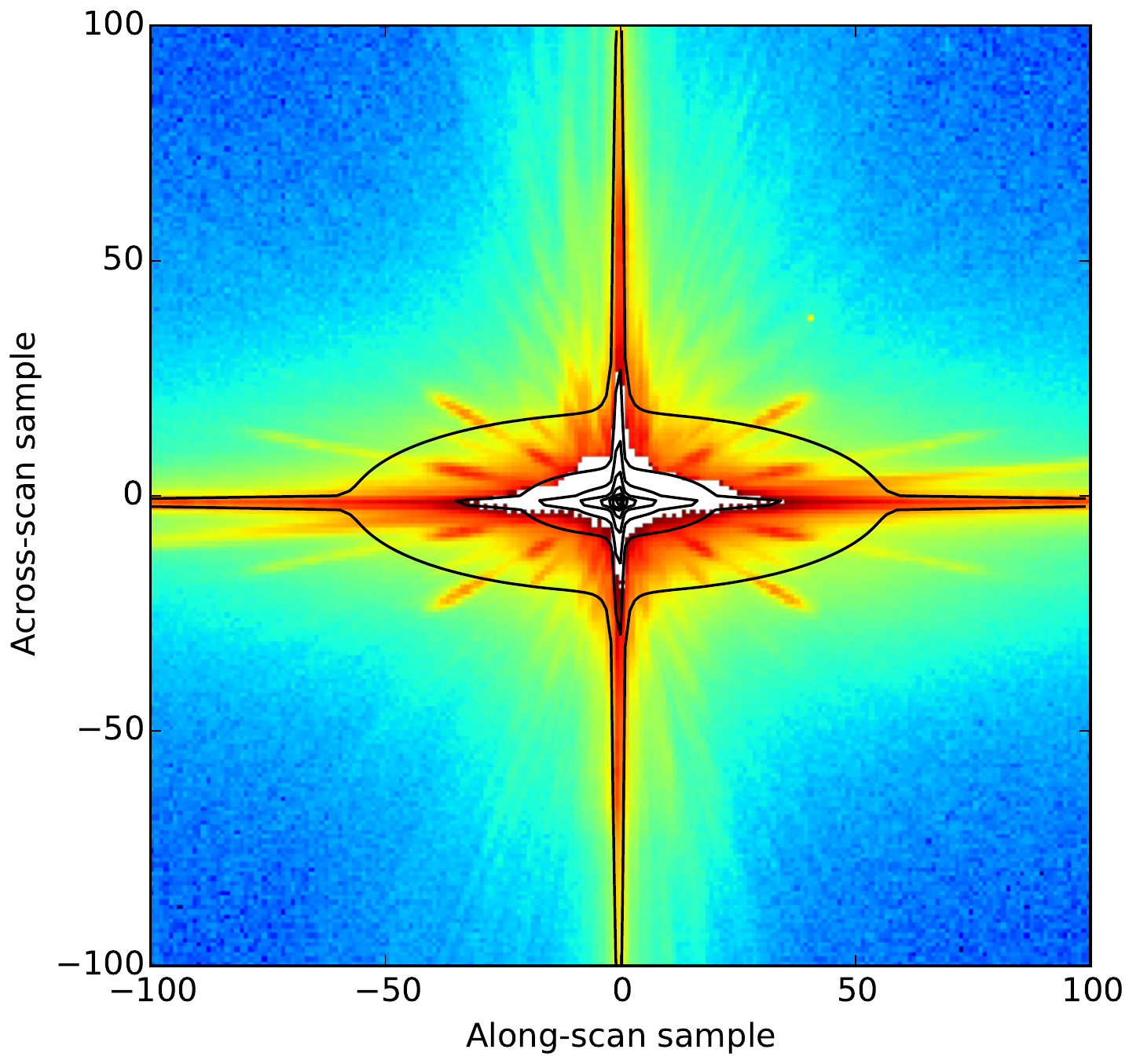}
\caption{Example sky-mapper SIF image of a very bright star (HIP~48036, R~Leo, a Mira-type variable star). The image shows the central $200 \times 200$ samples (each composed of $2 \times 2$ pixels) only. The colour scale is logarithmic and denotes counts, with white indicating saturation. The large, asymmetric saturation blob and the saw-edge pattern to the saturation are not optical properties of the point spread function but the result of known saturation behaviour of the readout node. The contours, evenly spaced in log-intensity space, indicate a model point spread function, aligned by eye, for a solar-type star in the astrometric field. Differences between the data and the model are due to the different spectral type, different part of the focal plane (sky mapper versus astrometric field), different across-scan motion, and the absence of read-out effects, saturation, charge-transfer inefficiency, and a high-frequency, wave-front-error feature of the primary mirror (quilting effect, resulting in diagonal spikes) in the model.}
\label{fig:bright_star}
\end{figure}
  
\subsection{Bright-star handling}\label{subsect:bright-star_handling}

The onboard detection is effective at the bright end down to magnitude $G \sim 3$~mag: the detection efficiency is $\sim$94\% at $G = 3$~mag and drops rapidly for brighter stars to below 10\% for $G = 2$~mag. The 230 brightest stars in the sky ($G < 3$~mag, loosely referred to as very bright stars) receive a special treatment to ensure complete
sky coverage at the bright end \citep{2014SPIE.9143E..0YM,2016arXiv160508347S}. Using the {\it Gaia} observing schedule tool (GOST; \url{https://gaia.esac.esa.int/gost/index.jsp}), their transit times and across-scan transit positions are predicted, based on propagated {\it Hipparcos} astrometry and the operational scanning law, and SIF data are acquired for these stars in the sky mapper (SM) and subsequently downloaded. These data comprise raw, two-dimensional images ($2540 \times 983$ samples, each composed of $2 \times 2$ pixels, integrated over 2.85~s), which are heavily saturated in the stellar core (Fig.~\ref{fig:bright_star}). The reduction and analysis of these data are special, off-line activities, which are not yet operational. The ultimate scientific quality of these data will primarily depend on the achievable quality of the calibration of the sky-mapper detectors and point spread functions and, in particular, at large distances from the stellar core far beyond the extension of regular SM windows, which are needed to avoid saturation. Because centroiding of these images in the uncalibrated detector frame can be carried out to within $50~\mu$as, it is expected that a single-measurement precision of $100~\mu$as will ultimately be achievable, which corresponds to end-of-life astrometry with standard errors of a few dozen $\mu$as. A virtual object-based scheme to acquire full focal plane transit data for the brightest 175 stars in the sky is currently in preparation.

\subsection{Dense area handling}\label{subsect:dense-area_handling}

\begin{figure}[t]
\includegraphics[width=\columnwidth]{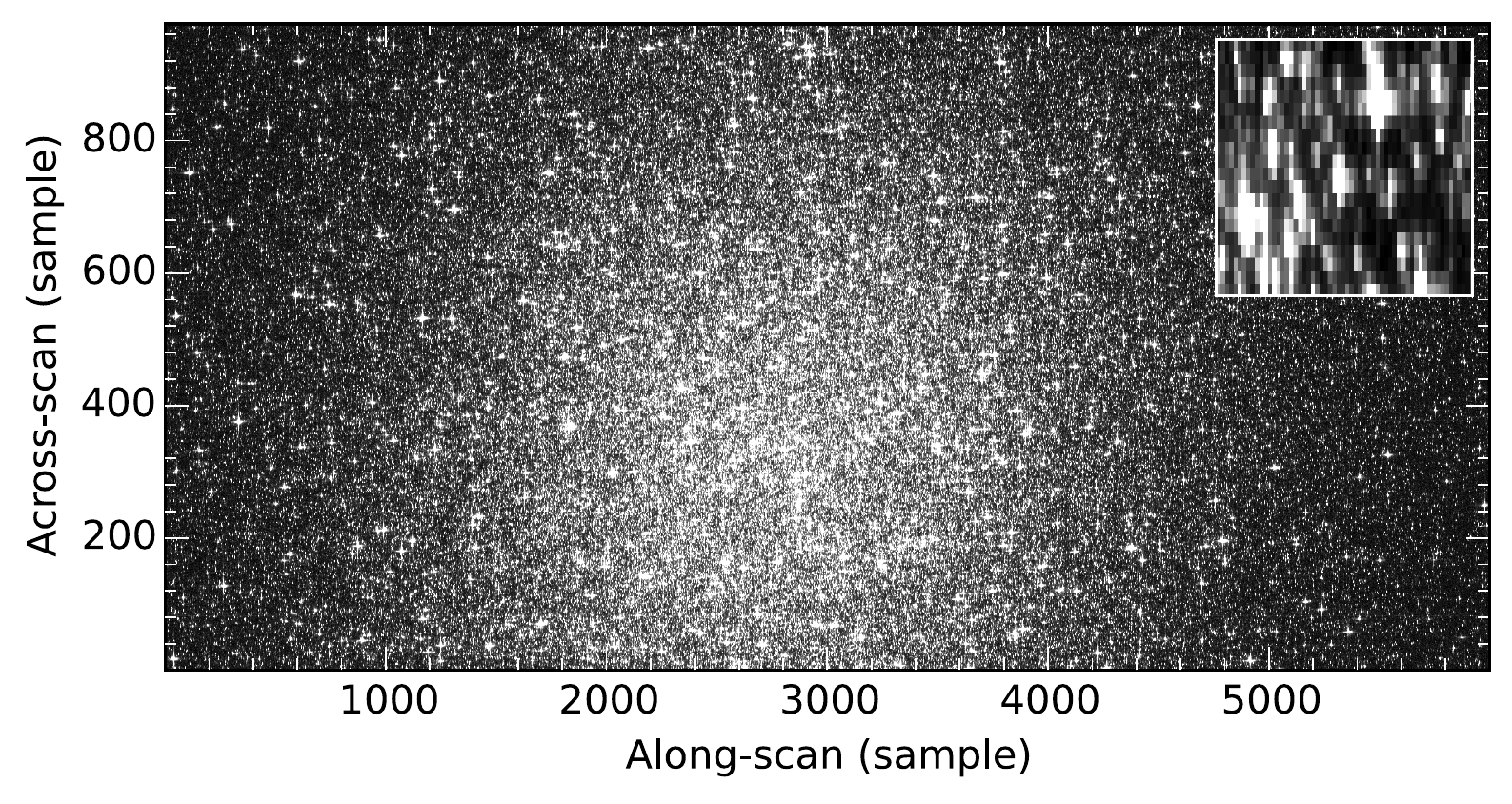}
\caption{Example sky-mapper SIF image of a dense area (part of the globular cluster $\omega$ Cen). The image shows $6000 \times 1000$ samples (each composed of $2 \times 2$ pixels) only, corresponding to $11\farcm{8} \times 5\farcm{8}$ on the sky. The inset shows a $60 \times 20$~sample ($7\arcsec \times 7\arcsec$) zoom of the central part of the cluster.}
\label{fig:dense_area}
\end{figure}

The astrometric crowding limit of {\it Gaia} is around $1\,050\,000$~objects~deg$^{-2}$; BP/RP photometry is limited to $750\,000$~objects~deg$^{-2}$ and RVS spectroscopy to $35\,000$~objects~deg$^{-2}$ (see Sect.~\ref{subsect:survey_coverage_and_completeness} for more details). At such (and) high(er) densities, window overlap and truncation is common and, because the onboard detection prioritises bright stars, faint-star completeness is impacted. In order to deblend the data observed in crowded regions, in particular for the BP/RP photometry and RVS spectroscopy, it is mandatory to have knowledge about the positions and fluxes of all (contaminating) sources in the field, down to a limit that is a few magnitudes fainter than the survey limit. The source environment analysis package \citep[Sect.~\ref{subsect:cyclic_data_processing};][]{2011ExA....31..157H} provides this information, combining the one-dimensional, along-scan {\it Gaia} data for each source obtained under different orientation angles into a two-dimensional image of its immediate surroundings. In order to support the deblending in extremely dense areas with high scientific importance, in particular Baade's Window into the Galactic bulge and the $\omega$~Centauri globular cluster (NGC5139), special sky-mapper SIF data are acquired for these areas (Fig.~\ref{fig:dense_area}). These data facilitate reaching a fainter completeness limit than the onboard detection limit, for instance 2~mag deeper in the core of $\omega$~Cen. The reduction and analysis of these (undersampled) data are special, off-line activities that are not yet operational. The ultimate aim also is to derive astrometry and photometry from the SIF data for faint stars not entering the nominal data reduction of {\it Gaia}  because of crowding. Also, an extension of the dense area SIF scheme to cover the Sgr~I bulge window, the Large and Small Magellanic Clouds, and four other globular clusters (M22 = NGC6656, M4 = NGC6121, NGC104 = 47~Tuc, and NGC4372) was activated in mid-2016.

\section{Data processing and analysis}\label{sect:dpac}

In order to address the science cases described in
Sect.~\ref{sect:scientific_goals}, the {\it Gaia} CCD-level
measurements need to be processed, i.e. calibrated and transformed
into astrophysically meaningful quantities. This is being carried out under
the remit of the {\it Gaia} Data Processing and Analysis Consortium
(DPAC). The DPAC evolved out of the {\it Gaia} community working
groups that were formed after the selection of {\it Gaia} in 2000, and
formally started its activities in 2006.  The DPAC subsequently
responded to the ESA announcement of opportunity for the {\it Gaia}
data processing, and was officially entrusted with this responsibility
in 2007. The DPAC currently consists of some 450 astronomers, software
engineers, and project management specialists, based in approximately
25, mostly European countries. The remainder of this section provides
a summary description of the {\it Gaia} data processing, which serves
as the context for more detailed expositions of how specific parts of
a {\it Gaia} data release are derived from the raw observations.

The {\it Gaia} data processing, which is summarised below, is a very complex
task, serving a wide variety of scientific goals. The data processing can be split
into two broad categories: daily and cyclic. The
daily tasks produce the preprocessed data that are needed by the
cyclic systems, provide payload health monitoring, and feed the
alerts systems. The daily systems process the {\it Gaia} telemetry
in near-real time as it comes down from the spacecraft. In contrast,
the cyclic processing iterates between calibration and the
determination of source parameters (to be interpreted here in a broad
sense, ranging from astrometry to astrophysical characteristics), by
repeatedly processing all the {\it Gaia} data until the system
converges. This is a consequence of the self-calibrating nature of
{\it Gaia}
(Sect.~\ref{subsect:measurement_principle_and_overall_design_considerations}).

In addition, the DPAC is responsible for the data simulations that
were used to support the mission preparations, the provision of
ground-based data needed for the calibration of the {\it Gaia} data, and the
validation, documentation, and publication (through the {\it Gaia} Archive,
developed, hosted, and operated by ESA) of the data processing
results.

\subsection{Daily data processing}\label{subsect:daily_data_processing}

The following daily tasks run within the DPAC:
\begin{description}
  \item[\em Initial Data Treatment and First Look.] Following the
    reception and reconstruction of the telemetry stream at the SOC
    (see Sect.~\ref{subsect:initial_data_treatment}), the IDT and
    First-Look systems together take care of the preprocessing of the
    incoming raw {\it Gaia} observations (so that they can be treated in the
    subsequent cyclic processing) and the daily payload health
    monitoring. These systems run at the DPAC data processing centre
    hosted at the {\it Gaia} SOC and are briefly
    described in Sects.~\ref{subsect:initial_data_treatment} and
    \ref{subsect:payload-health_monitoring} above, while the detailed
    description is provided by \cite{DPACP-7}.
  \item[\em Astrometric Verification Unit.] Two systems that    independently treat the raw data served by IDT are run as part of
    the astrometric verification unit, which aims to independently
    verify all the critical components contributing to the {\it
    Gaia} astrometric error budget. The basic angle monitor unit
    \citep{2014SPIE.9150E..1ZR} provides an independent monitoring of
    the BAM data and calibrated measurements of the basic angle
    variations. The astrometric instrument model unit
    \citep{2014RMxAC..45...39B} is a scaled-down counterpart of IDT
    and First Look that is restricted to a subset of the astrometric elements
    of the daily processing \citep[see][]{DPACP-7}, and it provides
    calibrations of the point spread function that are independent from the IDU system described below.
  \item[\em RVS daily.] Although IDT does some very basic processing of
    RVS data \citep[see][]{DPACP-7}, and First Look produces
    diagnostic plots that permit a first verification of the health of
    the RVS instrument, a special pipeline handles the full daily
    preprocessing of RVS data, including the establishment of initial
    calibrations and health monitoring of the RVS instrument. For
    further information on the RVS processing pipeline, see
    \cite{2011EAS....45..189K}.
  \item[\em Science alerts.] Because {\it Gaia} repeatedly scans the entire
    sky and quasi-simultaneously measures the position (to a spatial
    resolution of 50--100 mas), apparent brightness, and spectral
    energy distribution of sources, it forms a unique transient survey
    machine, for example capable of discovering many thousands of
    supernovae over the course of its lifetime
    \citep[e.g.][]{2013RSPTA.37120239H}. Therefore, DPAC runs a system
    that uses the IDT outputs, including source positions (already
    accurate to better than 100~mas) and fluxes, to build up a history
    of the observed sky to enable the discovery of transient
    phenomena, for which spectro-photometry is immediately available
    at each epoch.  The candidate transients are then filtered and the
    most interesting candidates are published as alerts, including the
    relevant {\it Gaia} data to enable rapid follow-up with
    ground-based telescopes. The science alerts system has been tested
    during an extended validation campaign \citep[leading, for
    example, to the discovery of an eclipsing AM
    CVn system;][]{2015MNRAS.452.1060C}, and is now routinely
    producing alerts that can be accessed through
    \url{http://gsaweb.ast.cam.ac.uk/alerts}.
  \item[\em Solar system alerts] This system processes the daily IDT
    outputs to search for new solar system objects (mostly
    main-belt asteroids and near-Earth objects) that can be
    recognised by their fast motion across the sky \citep[a typical
      main-belt asteroid moves at $\sim$$10$~mas~s$^{-1}$ with respect
    to the stars;][]{2016P&SS..123...87T}. At the instantaneous
    measurement precision of {\it Gaia}, these objects can be seen to
    move on the sky between successive scans by {\it Gaia} across the
    same region, and in a fraction of the cases the motion can be
    detected during a focal plane transit. The observations of
    candidate moving objects are matched to each other and preliminary
    orbits are determined. If it is established that an unknown
    solar system object is found, the orbit is used to predict where
    it should appear on the sky over the weeks following its
    discovery. This information is published as an alert (and will be
    made available through \url{https://gaiafunsso.imcce.fr/})  to enable ground-based follow-up observations. These
    observations are essential to establish an accurate orbit of the
    newly discovered object by observing it over a longer time
    baseline because, depending on the {\it Gaia} actual orbit and
    scanning law, it may never be observed by {\it Gaia} again. More
    details can be found in \cite{2016P&SS..123...87T}.
\end{description}

\subsection{Cyclic data processing}\label{subsect:cyclic_data_processing}

The self-calibrating nature of {\it Gaia}
(Sect.~\ref{subsect:measurement_principle_and_overall_design_considerations})
is reflected in the iterative processing of the data, which aims to
derive both the source parameters and the calibration (or nuisance)
parameters that together best explain the raw observations. This
iterative or cyclic processing is bootstrapped by the initial data
treatment in the daily systems
(Sect.~\ref{subsect:daily_data_processing}) and subsequently proceeds
by iterating the following steps:
\begin{enumerate}
  \item Update the basic calibrations, such as the
    point spread function (PSF) model (including detector
    charge-transfer-inefficiency effects), wavelength calibrations,
    the CCD-PEM non-uniformity calibration, the straylight and background
    model, etc.
  \item Reprocess all the raw observations using the latest
    calibrations. This step includes the improvement of the matching
    of {\it Gaia} observations to sources (which includes the creation of
    new sources where necessary), that is the cross-match.
  \item Use the results from the preceding preprocessing step to
    derive improved astrometry, photometry, and spectroscopy for each
    source.
\end{enumerate}
Steps 1 and 2 above both take into account the most up-to-date source
parameters, attitude model, geometric calibrations, etc., thus closing
the iterative loop. A concrete example to illustrate the processing
steps above is the iteration between PSF modelling and astrometry. A
given PSF model combined with predicted source positions in the focal
plane (which involves the source astrometry, the spacecraft attitude,
and the geometric calibration) can be used to predict the observed
sample values of the source image. The comparison to the actual sample
values can be used to improve the model. Conversely, the improved PSF
model can be used to derive an improved
estimate of the image location and the image flux from the raw sample values. In both steps, the
source colour should also be accounted for  to properly
calibrate the chromatic shifts of source images (see
Sect.~\ref{subsubsect:telescope}).

The following cyclic processing tasks execute the three steps above:
\begin{description}
  \item[\em Intermediate Data Update (IDU)] 
    This task \citep{2015hsa8.conf..792C} updates core calibrations,
    such as the PSF model, and improves the source to observation
    cross-matching. Subsequently it repeats all higher level functions
    of IDT (in particular the estimation of astrometric image
    parameters) on the (unchanging) raw astrometric and photometric
    observations, using better geometric calibrations, attitude and
    astrometric source parameters from AGIS, and source colours from
    the photometric pipeline. The results of the updated cross-match
    form the basis of the source list used by all the DPAC processing
    systems, including the daily systems described above.
  \item[\em Astrometric Global Iterative Solution (AGIS)] The
    astrometry for each source is derived within this system, which is
    described in \cite{2012A&A...538A..78L}, with the specifics for
    the processing for Gaia Data Release~1 ({\it Gaia} DR1) described in
    \cite{DPACP-14}. AGIS also produces the attitude model for {\it
      Gaia}, the geometric calibration of the SM and AF parts of the
    focal plane, and the calibration of a number of global parameters,
    such as the value and time evolution of the basic angle \citep[for
      the first data release, the basic angle variations are
      obtained from the BAM measurements;][]{DPACP-14}.
  \item[\em Global Sphere Reconstruction (AVU-GSR)] The astrometric
    verification unit provides an independent method, the global
    sphere reconstruction, for solving for the astrometry from the raw
    image locations. The GSR system will produce astrometry for the
    so-called primary sources in the AGIS solution
    \citep[cf.][]{2012A&A...538A..78L}. These results will
    then be compared to the AGIS astrometry as a strong form of
    quality control on the main DPAC astrometric outputs that are
    derived from AGIS. More details on the GSR can be found in
    \cite{2012SPIE.8451E..3CV}.
  \item[\em Photometric pipeline] This system processes the
    photometric data from the SM, AF, BP, and RP parts of the focal
    plane. The source fluxes in the $G$ band, derived from the AF
    images by the IDU (or for the most recently received data, the
    IDT) process, are turned into calibrated epoch photometry in the
    $G$ band. The integrated fluxes measured from the BP- and RP-prism
    spectra are also turned into calibrated epoch photometry
    ($G_\mathrm{BP}$ and $G_\mathrm{RP}$;
    Sect.~\ref{subsect:photometry}), while the spectra themselves are
    wavelength and flux calibrated. All photometric data are
    calibrated against standard stars for which high-quality,
    ground-based spectro-photometry is available
    \citep{2012MNRAS.426.1767P}. The calibration
    process delivers the actual photometric passbands and the
    physical flux and wavelength scales. The photometric processing
    for {\it Gaia} DR1 is described in
    \cite{DPACP-9,DPACP-10,DPACP-11,DPACP-12}.
  \item[\em RVS pipeline] The processing of the data from the RVS
    instrument is carried out within the RVS pipeline. The pipeline takes
    care of the basic spectroscopic calibrations, such as the
    wavelength scale, geometric calibration for the RVS focal plane,
    and the treatment of straylight and
    charge-transfer-inefficiency effects. The calibrations
      are subsequently used to stack the noise-dominated transit
      spectra of faint objects and derive mission-average radial
      velocities through cross-correlation techniques
      \citep[Sect.~\ref{subsect:spectroscopy};][]{2014A&A...562A..97D}.
      For the brightest subset, epoch spectra and epoch radial
      velocities will be available. The iterations between
    calibrations and spectra and source parameters is performed entirely
    within the RVS pipeline \citep{2011EAS....45..189K}.
\end{description}

The above processing steps provide the basic data needed for further
analysis and improvement of the astrometric, photometric, and
spectroscopic results from {\it Gaia}. For example, AGIS treats all sources
as point sources and will thus produce suboptimal astrometric
solutions for astrometric binaries. Likewise, the alignment of the
{\it Gaia} reference frame to the ICRF relies on a high-quality selection of
QSOs from the {\it Gaia} data itself, which will include many previously
unknown QSOs from poorly surveyed areas of the sky. AGIS itself does
not have the means to decide whether or not a source is a QSO. Such a
selection requires further analysis of the {\it Gaia} astrometry and
photometry combined. The DPAC therefore also carries out a number of
data analysis tasks that, on the one hand, provide higher level
scientific data products (such as source astrophysical parameters or
variable star characterisation) and, on the other hand, serve to
refine the astrometric, photometric, and spectroscopic processing
(e.g. by properly treating binaries or providing a clean selection of
QSOs for reference-frame alignment). The following cyclic processing
tasks provide this higher level data analysis:
\begin{description}
  \item[\em Non-single-star (NSS) treatment] This pipeline
    \citep{2011AIPC.1346..122P} provides a sophisticated treatment of
    all binary or multiple sources, including exoplanets. The
    astrometric data for any source that is found not to conform to
    the single-star source model will be treated with source models of
    increasing complexity, varying from the addition of a
    perspective-acceleration term to the derivation of orbital
    parameters for astrometric binaries. This pipeline, in addition,
    deals with resolved double or multiple stars and the astrophysical
    characterisation of eclipsing binaries.
  \item[\em Solar system object (SSO) treatment] This pipeline
    \citep{2012P&SS...73....5T} treats the {\it Gaia} data for solar system
    objects, primarily from the main asteroid belt. Orbital elements
    of known and newly discovered asteroids are determined, and their
    spectro-photometric properties are derived from the AF and BP/RP
    photometry. The results will also include mass measurements for
    the $\sim$100 largest asteroids, direct size measurements for some
    1000 asteroids, parametrised shapes, spin periods, and
    surface-scattering properties for some $10\,000$ asteroids
    \citep{2015P&SS..118..221C}, and taxonomic classifications from
    the BP/RP photometry.
  \item[\em Source environment analysis (SEA)] This task
    \citep{2011ExA....31..157H} performs a combined analysis of all
    the SM and AF images collected over the mission lifetime for a
    given source. A stacking of the images allows a slightly deeper
    survey of the surroundings of each source and in particular
    enables the identification of neighbouring sources that are not
    visible in the individual images. The non-trivial aspect of the
    task is the combination of one-dimensional images obtained at
    different scan angles across a source. The results will be used to
    refine the astrometric and photometric processing (by taking into
    account potentially disturbing sources near a target source) and
    will also feed into the non-single-star treatment and the
    treatment of extended objects described below.
  \item[\em Extended object (EO) analysis] This task treats all
    sources that are considered to be extended, such as galaxies or
    the cores thereof, and attempts to morphologically classify the
    sources and to quantify their morphology. This task also analyses
    all sources that are classified as QSOs to look, for
    example, for features that could prevent a QSO from being used as a
    source for the reference-frame alignment (e.g. host galaxy or
    lensing effects). More details can be found in
    \cite{2013A&A...556A.102K}.
  \item[\em Variable star analysis] The {\it Gaia} survey will
    naturally produce a photometric time series for each source
    spanning the lifetime of the mission and containing some 70--80
    observations on average (Sect.~\ref{subsect:astrometry}). The
    variable star processing takes all the epoch photometry and, for
    sources showing photometric variability, provides a classification
    of the variable type and a quantitative characterisation of
    the light curve. An overview of the variable star processing can
    be found in \cite{2014EAS....67...75E}. The treatment of variable
    stars for {\it Gaia} DR1 is described in \cite{DPACP-15} and
    \cite{DPACP-13}.
  \item[\em Astrophysical parameter inference] This system analyses
    the combination of astrometry, photometry, and spectroscopy from
    {\it Gaia} to derive discrete source classifications and infer the
    astrophysical properties of the sources. The source classification
    distinguishes single stars, white dwarfs, binaries, quasars, and
    galaxies, with the quasar selection feeding back into the astrometric
    processing as described above. The astrophysical parameters
    derived from the BP/RP data for stars are the effective
    temperature, surface gravity, metallicity, and extinction. From
    the RVS spectra of the brightest stars, $\alpha$-element
    enhancements and individual elemental abundances can additionally
    be derived. Descriptions of the status of this system
    can be found in \cite{2013A&A...559A..74B} and
    \cite{2016A&A...585A..93R}.
\end{description}

All the results from the analysis tasks above will also be published
as part of the {\it Gaia} data releases, thus complementing the basic
astrometry, photometry, and spectroscopy to provide a rich data set,
ready for exploitation by the astronomical community. Moreover, these
analysis tasks feed back into the preceding processing steps and thus
provide an essential and strong internal quality control of the DPAC
results.

The analysis tasks above will not necessarily lead to results that are
mutually consistent or consistent with the preceding processing
steps. This necessitates an additional, complex task within DPAC,
called catalogue integration, which is in charge of integrating, at
the end of each iterative cycle, all the results from the various
processing chains into a consistent list of sources and their
observational and astrophysical parameters. This list then forms the
basis for a next cycle of iterative processing and also for the public
{\it Gaia} data releases.

In order to achieve the iterative improvement of the {\it Gaia} processing
results, within a given processing cycle all the data collected from
the start of the mission are processed again to derive the
basic astrometric, photometric, and spectroscopic data (thus involving
the upstream systems IDU, AGIS, AVU-GSR, photometric, and RVS
processing). To keep this process manageable, the data collected by
{\it Gaia} are split into data segments, where each segment covers
a certain time range. During a given data processing cycle $n$,
typically all the data segments treated during cycle $n-1$ plus the
latest complete segment are processed again by the upstream systems.
During the same cycle $n$, the downstream systems (NSS, SSO, SEA,
EO, variable star analysis, and astrophysical-parameter inference)
will process the astrometric, photometric, and spectroscopic data
derived from the data segments used during cycle $n-1$.

The {\it Gaia} intermediate data release schedule is thus driven by the
lengths of the processing cycles, while the quality of the published
data at each release is related to the amount of data segments treated
(sky, scan-direction, and time coverage; signal-to-noise) and the
amount of iterative cycles completed (calibration quality).

\subsection{Simulations, supplementary data and observations, data publication}
\label{subsect:Simulations,_supplementary_data_and_observations,_data publication}

Apart from the processing tasks mentioned in the previous sections,
the DPAC responsibilities also include supporting tasks that are
necessary for the successful preparation and execution of the {\it Gaia}
data processing and the publication of its results:
\begin{description}
  \item[\em Simulations.] Data simulations formed an essential element
    of the DPAC preparations for the operational lifetime of both {\it Gaia}
    and the data processing. The simulations spanned the three levels
    from the pixels in the focal plane
    \citep[GIBIS;][]{2011ascl.soft07002B}, through simulated telemetry
    \citep[GASS;][]{2008IAUS..248..278M} to simulated DPAC data
    products \citep[GOG;][]{2014EAS....67..355A}, which were used both    internally and to support the astronomical community in preparing
    itself for the {\it Gaia} mission \citep{2012A&A...543A.100R,
      2014A&A...566A.119L}. These last simulations have been made
    available and will continue to be available alongside the released
    {\it Gaia} data.
  \item[\em Relativistic astrometric models.] The proper modelling and
    interpretation of astrometric data at the micro-arcsecond level
    accuracy aimed for by the {\it Gaia} mission requires the data
    processing to be fully compatible with general relativity (for
    example to accurately account for the effect of light bending by
    solar system objects on the apparent source directions). The
    relativistic model used for AGIS is described in
    \cite{2003AJ....125.1580K,2004PhRvD..69l4001K}, while the model
    employed for AVU-GSR is documented in
    \cite{2004ApJ...607..580D,2006ApJ...653.1552D}. The relativistic
    modelling of the observations also requires an up-to-date
    solar system ephemeris. For {\it Gaia} DR1, the DPAC employed the
    INPOP10e ephemeris \citep{2013arXiv1301.1510F}.
  \item[\em Auxiliary observations.] The {\it Gaia} photometric and
    spectroscopic data have to be tied to a physical flux and
    wavelength scale that is achieved through standard stars. In
    preparation of the {\it Gaia} mission, therefore, extensive
    observational campaigns to collect the relevant
    spectro-photometric data took place \citep{2012MNRAS.426.1767P} as
    well as monitoring campaigns to establish the photometric
    stability of candidate standard stars
    \citep{2015AN....336..515A}. For the RVS, a compilation of
    radial-velocity standards was undertaken
    \citep{2010A&A...524A..10C,2013A&A...552A..64S}, while for the
    solar system object processing, data on solar analogues had to be
    compiled to calibrate the interpretation of the {\it
      Gaia} photometry. A set of $\sim$30 {\it Gaia} FGK benchmark
    stars have been extensively studied to provide a reference for the
    astrophysical parameter inference system
    \citep{2014A&A...564A.133J,2014A&A...566A..98B,2015A&A...582A..49H,2015A&A...582A..81J,2016arXiv160508229H}. One
    of the largest auxiliary observation programmes is the GBOT effort
    already mentioned in
    Sect.~\ref{subsubsect:orbit_prediction,_monitoring,_and_control}.
  \item[\em Data publication.] All the results from the DPAC processing
    are published as {\it Gaia} data releases and this is a specific task
    within the consortium. It encompasses the extensive scientific
    validation of the data (using the combination of all DPAC
    results), the documentation of the data processing and the results
    thereof, the incorporation of the data into a publicly accessible
    archive, and the provision of various tools to enhance the
    interrogation and scientific exploitation of the {\it Gaia} data,
    including pre-computed cross-matches to other large surveys
    \citep{DPACP-17}. For {\it Gaia} DR1, the validation and publication of
    the data is summarised in \cite{DPACP-16,DPACP-19}.
\end{description}

\subsection{Organisation}\label{subsect:organisation}

The DPAC is a large entity in charge of a set of complex, large, and
interdependent data processing tasks. The DPAC is internally organised
into a set of relatively independently operating units, called
coordination units (CUs), which roughly correspond to the cyclic
data processing tasks outlined above, with some units also in charge of daily data processing. The actual processing takes place on
hardware spread around six data processing centres in Europe
(located at ESAC, Toulouse, Cambridge, Gen\`eve, Barcelona, and Torino),
which communicate through the main database located at ESAC
(hub-and-spokes topology). The main database also houses the DPAC end
data products from which the public data releases are produced. The
DPAC is managed at the top level by an executive body consisting of
the leaders of the coordination units and representatives from the
data processing centres. The Executive is supported in its tasks by
the DPAC project office (PO). More details on the DPAC organisational
structure can be found in
\cite{2008IAUS..248..224M,2009ASPC..411..470O,2010SPIE.7738E..10M}.

\section{End-of-mission scientific performance}\label{sect:scientific_performance}

The end-of-mission science performance of {\it Gaia}, based on CCD-level Monte Carlo simulations calibrated with measurements and then extrapolated to end-of-life conditions, has been published prior to launch in \cite{2012Ap&SS.341...31D} and was updated in \cite{2014EAS....67...23D} following the in-orbit commissioning phase of the mission. Here, we summarise the most up-to-date end-of-mission performance estimates based on in-orbit experience collected to date (mid 2016). An associated Python toolkit is available from \url{https://pypi.python.org/pypi/PyGaia/}. The actual scientific quality of {\it Gaia} Data Release~1 is described in \cite{DPACP-14} for the astrometry and in \cite{DPACP-11} for the photometry.

All performance estimates presented here include a 20\% scientific contingency margin to cover, among other things,
\begin{itemize}
\item scientific uncertainties and residual calibration errors in the on-ground data processing and analysis, for example uncertainties related to the spacecraft and solar system ephemeris, estimation errors in the sky background value that needs to be fed to the centroiding algorithm, the contribution to the astrometric error budget resulting from the mismatch between the actual and the calibrating point spread function, template mismatch in the spectroscopic cross-correlation, and residual errors in the derivation of the locations of the centroids of the reference spectral lines used for the spectroscopic wavelength calibration;
\item the fact that the sky does not contain, as assumed for the performance assessments, perfect stars but normal stars, which for example can be photometrically variable, have spectral peculiarities such as emission lines, have (unrecognised) companions, and can be located in crowded fields.
\end{itemize}
In the scientific performance assessments presented here, all known instrumental effects are included under the appropriate in-flight operating conditions. Because contamination (Fig.~\ref{fig:throughput}) is a small effect that is kept under control as necessitated by decontamination activities, and is not easy to model, it is not included and assumed to be covered by the 20\% science margin. All error sources are included as random variables with typical deviations (as opposed to best-case or worst-case deviations).

\subsection{Astrometry}\label{subsect:astrometry}

The astrometric science performance of the nominal, five-year mission is usually quantified by the end-of-mission parallax standard error $\sigma_\varpi$ in units of $\mu$as. This error, averaged over the sky and including 20\% science margin (Sect.~\ref{sect:scientific_performance}), depends to first order on the broadband $G$ magnitude of {\it Gaia},
\begin{equation}
  \sigma_\varpi [\mu{\rm as}] = c(V-I) \sqrt{-1.631 + 680.766 z_{12.09} + 32.732 z_{12.09}^2},\label{eq:astrometry_parallax}
\end{equation}
where
\begin{equation}
  z_{x}(G) = {\rm max}[10^{0.4 (x - 15)}, 10^{0.4 (G - 15)}],\label{eq:z}
\end{equation}
with $x$ denoting the bright-end, noise-floor magnitude, and
\begin{equation}
  c(V-I) = 0.986 + (1 - 0.986) (V-I).\label{eq:c}
\end{equation}
The quantity $c(V-I)$ represents a second-order $V-I$ colour term, which quantifies the widening of the point spread function for red(der) stars (e.g. $c(1) = 1$ and $c(3) = 1.03$). The model from Eq.~(\ref{eq:astrometry_parallax}) is valid for $3 \leq G < 20.7$~mag (see Sect.~\ref{subsect:bright-star_handling} for very bright stars with $G < 3$~mag). For stars that are brighter than $G \approx 12$~mag, shortened CCD integration times, through the use of six TDI gates (number 4 with 16 TDI lines and gate numbers 8--12 with 256--2900 TDI lines; Sect.~\ref{subsubsect:focal-plane_assembly}), are used to avoid saturation. Each gate effectively means a different geometric instrument and necessitates a dedicated geometric calibration with associated uncertainties. In addition, onboard magnitude-estimation errors result in a given bright star that is sometimes observed with the non-optimal TDI gate (Sect.~\ref{subsubsect:focal-plane_assembly}). The max function in Eq.~(\ref{eq:z}) hides these complications and simply returns a constant, bright-star parallax noise floor, at $\sigma_\varpi = 7~\mu$as, for stars with $3 \leq G \leq 12.09$~mag (assuming a G2V spectral type with $V-I = 0.75$~mag). Reaching this performance, however, will require full control of all error sources and fully (iteratively) calibrated instrument and attitude models, which can only reasonably be expected in the final data release. For stars at the very faint end (around $G \approx 20.5$~mag and fainter), the number of transits resulting in science data being available on ground may be reduced disproportionately as a result of onboard priority management and data deletion (Sects.~\ref{subsubsect:payload_data_handling_unit} and \ref{subsubsect:ground_stations}), onboard magnitude-estimation errors (Sect.~\ref{subsect:survey_coverage_and_completeness}), and finite onboard detection and confirmation probabilities (Sect.~\ref{subsubsect:video_processing_unit_and_algorithms}). Their standard errors can hence be larger than predicted through this model.

For sky-averaged position errors $\sigma_0$ [$\mu$as] at mid-epoch (i.e. the middle of the observation interval) and for (annual) proper-motion errors $\sigma_\mu$ [$\mu$as~yr$^{-1}$], the following relations can be used:
\begin{eqnarray}
\sigma_0 = 0.743 \sigma_\varpi;\nonumber\\
\sigma_{\alpha\ast} = 0.787 \sigma_\varpi;\nonumber\\
\sigma_\delta = 0.699 \sigma_\varpi;\nonumber\\
\sigma_\mu = 0.526 \sigma_\varpi;\nonumber\\
\sigma_{\mu_{\alpha\ast}} = 0.556 \sigma_\varpi;\nonumber\\
\sigma_{\mu_\delta} = 0.496 \sigma_\varpi,\label{eq:other_astrometric_params}
\end{eqnarray}
where the asterisk denotes true arcs on the sky (e.g. $\sigma_{\alpha\ast} = \sigma_{\alpha} \cos\delta$). The expected astrometric correlations between the five astrometric parameters were discussed in detail by \cite{2012A&A...543A..14H} and \cite{2012A&A...543A..15H}. The standard errors vary over the sky as a result of the scanning law (Sect.~\ref{subsect:scanning_law}). The main variation is with ecliptic latitude $\beta$. The mean, i.e. ecliptic longitude averaged, variations with $\beta$ are provided in Tab.~\ref{tab:sky_dependence}. The (approximate) ecliptic latitude can be calculated from the equatorial coordinates $(\alpha, \delta)$ or the galactic coordinates $(l, b)$ using\begin{eqnarray}
\sin\beta       &\approx&       0.9175 \sin\delta - 0.3978 \cos\delta \sin\alpha\nonumber\\
                &\approx&       0.4971 \sin b     + 0.8677 \cos b     \sin(l - 6\fdg{38}).
\end{eqnarray}

\begin{table*}[t]
  \caption{Numerical factor to be applied to the sky-averaged astrometric standard errors of Eqs.~(\ref{eq:astrometry_parallax}) and (\ref{eq:other_astrometric_params}) for the five astrometric parameters as a function of ecliptic latitude $\beta$, including the effect of the variation of the end-of-mission number of transits over the sky. The quantity $N_{\rm obs}$ in column~4 denotes the end-of-mission number of focal plane passages for AF, BP, and RP (both fields of view combined; recall that {\it Gaia} Data Release~1 is based on 14 months of data, corresponding on average to 16 field-of-view transits). For RVS, the number of focal plane transits is a factor $4 / 7 = 0.57$ smaller (Sect.~\ref{subsubsect:spectroscopic_instrument}). The transit numbers in column~4 are based on an assumed 6\% dead time (data loss). For the faintest objects ($G \gtrsim 20$~mag or $G_{\rm RVS} \gtrsim 14$~mag), the actual losses are larger (Sect.~\ref{subsubsect:ground_stations}).}\label{tab:sky_dependence}
  \begin{center}
    \begin{tabular}{ccccccccc}
      \hline\hline
$|\sin\beta|$ & $\beta_{\rm min}$ [$^{\circ}$] & $\beta_{\rm max}$ [$^{\circ}$] & $N_{\rm obs}$ & $\alpha\ast$ & $\delta$ & $\varpi$ & $\mu_{\alpha\ast}$ & $\mu_\delta$\\
      \hline
0.025   &0.0    &2.9    &61     &1.026  &0.756  &1.180  &0.725  &0.542\\
0.075   &2.9    &5.7    &61     &1.021  &0.757  &1.180  &0.722  &0.542\\
0.125   &5.7    &8.6    &62     &1.002  &0.751  &1.169  &0.710  &0.537\\
0.175   &8.6    &11.5   &62     &0.993  &0.752  &1.167  &0.703  &0.539\\
0.225   &11.5   &14.5   &63     &0.973  &0.751  &1.158  &0.689  &0.538\\
0.275   &14.5   &17.5   &65     &0.952  &0.742  &1.143  &0.673  &0.533\\
0.325   &17.5   &20.5   &66     &0.934  &0.740  &1.136  &0.662  &0.533\\
0.375   &20.5   &23.6   &68     &0.901  &0.730  &1.119  &0.640  &0.525\\
0.425   &23.6   &26.7   &71     &0.861  &0.718  &1.098  &0.614  &0.515\\
0.475   &26.7   &30.0   &75     &0.819  &0.705  &1.072  &0.584  &0.506\\
0.525   &30.0   &33.4   &80     &0.765  &0.691  &1.043  &0.548  &0.493\\
0.575   &33.4   &36.9   &87     &0.701  &0.673  &1.009  &0.500  &0.477\\
0.625   &36.9   &40.5   &98     &0.631  &0.650  &0.970  &0.541  &0.461\\
0.675   &40.5   &44.4   &122    &0.535  &0.621  &0.922  &0.381  &0.437\\
0.725   &44.4   &48.6   &144    &0.469  &0.607  &0.850  &0.327  &0.423\\
0.775   &48.6   &53.1   &106    &0.554  &0.636  &0.808  &0.386  &0.443\\
0.825   &53.1   &58.2   &93     &0.603  &0.654  &0.779  &0.422  &0.456\\
0.875   &58.2   &64.2   &85     &0.641  &0.669  &0.755  &0.447  &0.467\\
0.925   &64.2   &71.8   &80     &0.668  &0.680  &0.731  &0.466  &0.473\\
0.975   &71.8   &90.0   &75     &0.688  &0.706  &0.713  &0.481  &0.490\\
      \hline
Sky-average &0.0 &90.0  &81     &0.787  &0.699  &1.000  &0.556  &0.496\\
      \hline\hline
    \end{tabular}
  \end{center}
\end{table*}

The performance equations presented here refer to the standard errors, i.e. the precision of the astrometry. An assessment of (residual) systematic errors in the astrometry, linking to its accuracy and in particular to the parallax zero point, is much more difficult to provide. For astrometry, a potential contributor to systematic parallax errors are unmodelled, Sun-synchronous basic angle variations (Sect.~\ref{subsubsect:basic-angle_monitor}). The metrology data derived from the basic angle monitor should ultimately allow, after careful calibration \citep{DPACP-14}, the limitation of possible systematic effects in the final data release to $\mu$as levels.

\subsection{Photometry}\label{subsect:photometry}

For median straylight conditions over a spacecraft rotation period, the single field-of-view transit photometric standard error $\sigma_G$, in units of mag and including 20\% margin (Sect.~\ref{sect:scientific_performance}), of the $G$-band photometry is parametrised well by
\begin{equation}
\sigma_G {\rm [mag]} = 1.2~ 10^{-3} \left(0.04895 z_{12}^2 + 1.8633 z_{12} + 0.0001985\right)^{1/2},\label{eq:photometry_G}
\end{equation}
where $z_x(G)$ is defined in Eq.~(\ref{eq:z}). As for astrometry (Sect.~\ref{subsect:astrometry}), the bright-star errors were set to a constant noise floor. For the integrated BP and RP bands, a suitable parametrisation of the single field-of-view transit photometric standard errors $\sigma_{\rm BP/RP}$, in units of mag and including 20\% margin, for median straylight conditions depends on $G$ and on a $V-I$ colour term,
\begin{equation}
\sigma_{\rm BP/RP} ~{\rm [mag]} = 10^{-3} \left(10^{a_{\rm BP/RP}} z_{11}^2 + 10^{b_{\rm BP/RP}} z_{11} + 10^{c_{\rm BP/RP}}\right)^{1/2},\label{eq:photometry_XP}
\end{equation}
where
\begin{eqnarray}
  a_{\rm BP} &=& -0.000562 (V-I)^3 + 0.044390 (V-I)^2 +\nonumber\\
  && +0.355123 (V-I) + 1.043270;\nonumber\\
  b_{\rm BP} &=& -0.000400 (V-I)^3 + 0.018878 (V-I)^2 +\nonumber\\
  && +0.195768 (V-I) + 1.465592;\nonumber\\
  c_{\rm BP} &=& +0.000262 (V-I)^3 + 0.060769 (V-I)^2 +\nonumber\\
  && -0.205807 (V-I) - 1.866968;\nonumber\\
  a_{\rm RP} &=& -0.007597 (V-I)^3 + 0.114126 (V-I)^2 +\nonumber\\
  && -0.636628 (V-I) + 1.615927;\nonumber\\
  b_{\rm RP} &=& -0.003803 (V-I)^3 + 0.057112 (V-I)^2 +\nonumber\\
  && -0.318499 (V-I) + 1.783906;\nonumber\\
  c_{\rm RP} &=& -0.001923 (V-I)^3 + 0.027352 (V-I)^2 +\nonumber\\
  && -0.091569 (V-I) - 3.042268.
\end{eqnarray}
Resulting end-of-mission, median-straylight photometric errors can be estimated by division of the single field-of-view transit photometric standard errors from Eqs.~(\ref{eq:photometry_G}) and (\ref{eq:photometry_XP}) by the square root of the number of transits $N_{\rm obs}$, after taking an appropriate calibration error $\sigma_{\rm cal}$ (at field-of-view transit level) into account as follows:
\begin{eqnarray}
\sigma_{G,{\rm end-of-mission}} &=& 1.2 \sqrt{ {{ (\sigma_G/1.2)^2 + \sigma_{G,{\rm cal}}^2}\over{N_{\rm obs}}}  };\label{eq:photometry_G_eom}\\
\sigma_{\rm BP/RP,~end-of-mission} &=& 1.2 \sqrt{ {{ (\sigma_{\rm BP/RP}/1.2)^2 + \sigma_{\rm BP/RP,~cal}^2}\over{N_{\rm obs}}}  },\label{eq:photometry_XP_eom}
\end{eqnarray}
where $N_{\rm obs} = N_{\rm obs}(\beta)$ is given in Tab.~\ref{tab:sky_dependence} and the factors 1.2 refer to the 20\% science margin (Sect.~\ref{sect:scientific_performance}). A realistic estimate of the field-of-view transit level calibration error is $\sigma_{G,{\rm cal}} = 1$~milli-mag \citep{DPACP-11} and $\sigma_{\rm BP/RP,~cal} = 5$~milli-mag.

As described in \citeauthor{2013A&A...559A..74B} (\citeyear{2013A&A...559A..74B}; see also \citealt{2012MNRAS.426.2463L}), the BP/RP spectro-photometric data, sometimes in combination with the astrometric and spectroscopic data, allow one to classify objects and to estimate their astrophysical parameters. The accuracy of the estimation of the astrophysical parameters depends in general on $G$ and on the value of the astrophysical parameters themselves; in addition, the strong and ubiquitous degeneracy between effective temperature and extinction limits the accuracy with which either parameter can be estimated, in particular for faint stars \citep[e.g.][]{2011MNRAS.411..435B}. As an example, for FGKM stars at $G = 15$~mag with less than two magnitudes extinction, effective temperature $T_{\rm eff}$ can be estimated to 75--250~K, extinction to 0.06--0.15~mag, surface gravity $\log_{10}(g)$ to 0.2--0.5~dex, and metallicity [Fe/H] to 0.1--0.3~dex, where the ranges delimit optimistic and pessimistic estimates in view of template mismatch and calibration errors.

\subsection{Spectroscopy}\label{subsect:spectroscopy}

\begin{table*}[t]
  \caption{Parameters for the RVS performance model defined in Eq.~(\ref{eq:v_rad}). MP stands for metal poor (${\rm [Fe/H]} = -1.5$).}\label{tab:a_and_b}
  \begin{center}
    \begin{tabular}{cccccccccccc}
      \hline\hline
      SpT                    &   B0V & B5V   & A0V  & A5V  & F0V  & G0V & G5V     & K0V  & K1IIIMP & K4V  & K1III\\
      \hline
      $a$                    &  0.90 &  0.90 & 1.00 & 1.15 & 1.15 & 1.15 & 1.15  & 1.15 & 1.15    & 1.15 & 1.15\\
      $b$                    & 50.00 & 26.00 & 5.50 & 4.00 & 1.50 & 0.70 & 0.60  & 0.50 & 0.39    & 0.29 & 0.21\\
      $V - I$~[mag]          & -0.31 & -0.08 & 0.01 & 0.16 & 0.38 & 0.67 & 0.74 & 0.87 & 0.99    & 1.23 & 1.04\\
      $V - G_{\rm RVS}$~[mag] & -0.35 & -0.08 & 0.02 & 0.19 & 0.46 & 0.80 & 0.87 & 1.03 & 1.17    & 1.45 & 1.24\\
      \hline\hline
    \end{tabular}
  \end{center}
\end{table*}

Spectroscopy is being collected for a subset of the astrometric and photometric data to derive radial velocities and to perform stellar parametrisation. For the vast majority of stars, namely those that are faint, the individual transit spectra are too noisy to derive transit-level radial velocities. As a result, a single, end-of-mission composite spectrum is first reconstructed by co-adding all spectra collected during all RVS CCD crossings throughout the mission lifetime. A single, mission-averaged radial velocity is then extracted from this end-of-mission composite spectrum by cross-correlation with a synthetic template spectrum. The cross-correlation method finds the best match of the observed spectrum within a set of predefined synthetic spectra with different atmospheric parameters and subsequently assigns the astrophysical parameters of the best-fit template to the observed target. For the few million brightest targets, single field-of-view transit spectra will be used to derive associated epoch radial velocities. For this subset of objects also, the radial velocities of the components of (double-lined) spectroscopic binaries will be estimated using TODCOR \citep{1994ApJ...420..806Z}.

The radial-velocity performance is normally quantified through the sky-average, end-of-mission radial-velocity robust formal error, $\sigma_{\rm v,rad}$, in units of km~s$^{-1}$ and including 20\% margin (Sect.~\ref{sect:scientific_performance}). This error depends on spectral type and magnitude and, for $G_{\rm RVS}$ up to $\sim$16~mag, is suitably parameterised by
\begin{equation}
\sigma_{\rm v,rad}~ {\rm [km~s^{-1}]} = \sigma_{\rm floor} + b~ {\rm exp}(a [V - 12.7]),\label{eq:v_rad}
\end{equation}
where $a$ and $b$ are spectral-type dependent constants (Tab.~\ref{tab:a_and_b}) and $V$ denotes Johnson $V$ magnitude. The bright-star performance is limited by a noise floor $\sigma_{\rm floor}$. Recent investigations based on preliminary assessments of real data suggest that the bright-star noise floor will ultimately be at the level of 0.5~km~s$^{-1}$, and possibly better. Systematics in the radial velocities are expected to be kept under control to within a few 100~m~s$^{-1}$.

As described in \cite{2016A&A...585A..93R}, stellar parametrisation will be performed on the RVS spectra of individual stars with $G_{\rm RVS} \la 14.5$~mag. Stars with $G_{\rm RVS} \la 12.5$~mag are efficiently parametrised, including reliable estimations of the $\alpha$-element abundances with respect to iron. Typical internal errors for FGK metal-rich ($-0.5 \la {\rm [M/H]} \la 0.25$~dex) and intermediate-metallicity ($-1.25 \la {\rm [M/H]} \la -0.5$~dex) stars (dwarfs and giants) are around 40~K in $T_{\rm eff}$, 0.10~dex in $\log_{10}(g)$, 0.04~dex in [M/H], and 0.03~dex in [$\alpha$/Fe] at $G_{\rm RVS} = 10.3$~mag. These errors degrade to 155~K in $T_{\rm eff}$, 0.15~dex in $\log_{10}(g)$, 0.10~dex in [M/H], and 0.1~dex in [$\alpha$/Fe] at $G_{\rm RVS} \sim 12$~mag. Similar errors in $T_{\rm eff}$ and [M/H] are found for A-type stars, while the surface-gravity derivation is more precise (errors of 0.07 and 0.12 dex at $G_{\rm RVS} = 12.6$ and 13.4~mag, respectively). For the faintest stars, with $G_{\rm RVS} \ga 13$--14~mag, the input of effective temperature derived from the BP/RP spectro-photometry will allow the final, RVS-based parametrisation to be improved.

\subsection{Survey coverage and completeness}\label{subsect:survey_coverage_and_completeness}

The survey coverage of {\it Gaia} has some particular features:

{\bf Dense areas:} As explained in Sect.~\ref{subsubsect:video_processing_unit_and_algorithms}, the total number of samples that can be tracked simultaneously in the readout register of a CCD differs per instrument and is 20 in AF, 71 in BP and RP, and 72 in RVS. The associated maximum object densities (for the two superimposed viewing directions combined) depend on the along-scan window size and on the number of samples needed per object (plus the proximity-electronics settings). For the majority of (faint) objects, onboard across-scan binning of the window contents means that one object only requires one serial sample. In the absence of bright stars requiring 12 samples per object (10 for RVS), the maximum densities are $\sim$$1\,050\,000$~objects deg$^{-2}$ in the astrometric field (along-scan window size 12~pixels), $\sim$$750\,000$~objects deg$^{-2}$ for the BP and RP photometers (along-scan window size 60~pixels), and $\sim$$35\,000$~objects deg$^{-2}$ for the RVS spectrograph (along-scan window size 1296~pixels). The maximum density is proportional to the number of serial samples and is inversely proportional to the along-scan window length, which varies between the different instruments and, within a given instrument, varies with magnitude. When a bright star that needs to be windowed with full-pixel resolution enters the CCD, the above densities are temporarily reduced. For instance, in RVS, one bright star with $G_{\rm RVS} < 7$~mag consumes 10 serial samples, leaving 62 samples for faint stars, corresponding to $\sim$$30\,000$~objects deg$^{-2}$ (which is exceeded in $\sim$20\% of the sky). In AF, bright stars ($G < 13$~mag) are particularly detrimental because they have a longer window (18 along-scan pixels) and because each bright star consumes 12 of the 20 available serial samples (temporarily leaving only 8 samples for other stars, corresponding to $\sim$$420\,000$~objects deg$^{-2}$, which is exceeded in a few hundred square degrees on the sky). In case of a shortage of serial samples, object selection is based on object priority, with bright stars having a higher priority than faint stars (Sect.~\ref{subsubsect:video_processing_unit_and_algorithms}). Fortunately, the fact that each area on the sky is observed several dozen times over the course of the mission under different scanning angles means that there is no significant bias for faint sources close to bright sources in the final catalogue except for such objects receiving fewer transits. In (very) dense areas, the number of available transits for faint objects may be (greatly) reduced, up to the level of yielding a brighter completeness limit by up to several magnitudes (see also Sect.~\ref{subsect:dense-area_handling}).

{\bf Bright stars:} As already mentioned in Sect.~\ref{subsect:bright-star_handling}, the onboard detection efficiency at the bright end drops from $\sim$94\% at $G = 3$~mag to below 10\% for $G = 2$~mag and brighter \citep{2016arXiv160508347S}. Whereas special sky-mapper SIF images are acquired for the 230 very bright stars with $G < 3$~mag (Sect.~\ref{subsect:bright-star_handling}), in principle allowing the derivation of $G$-band photometry and astrometric information, a non-detection implies no BP/RP photometry and no RVS spectroscopy is collected; such data can only be acquired through a virtual object-based scheme, which is currently in preparation (Sect.~\ref{subsect:bright-star_handling}).

{\bf Close double stars:} As a result of the rectangular pixel size (Sect.~\ref{subsubsect:focal-plane_assembly}), the minimum separation to resolve a close, equal-brightness double star in the sky mapper is $0\farcs{23}$ in the along-scan and $0\farcs{70}$ in the across-scan direction, independent of the brightness of the primary \citep{2015A&A...576A..74D}; larger separations are required to resolve double stars with $\Delta G > 0$~mag. During the course of the mission, a given, close double star is observed many times with varying scanning angles such that it can be resolved on board in some transits and can stay unresolved in others. In the on-ground processing, however, the full resolution of the astrometric instrument, combined with the window size (at least 12 along-scan pixels of $0\farcs{06}$ each), allows one to systematically resolve double stars down to separations around $0\farcs{1}$. A special deblending treatment of close binaries will be performed in the BP/RP data processing.

{\bf Moving objects:} The majority of main-belt and near-Earth asteroids, at least up to speeds of 100~mas~s$^{-1}$, are properly detected \citep{2015A&A...576A..74D}. Moving objects may, after successful detection, leave the window (and even the CCD) at any moment during the focal plane transit because the window propagation assumes every object is a fixed star. In BP/RP, an additional window is assigned to bright objects ($G = 13$--18~mag) that have a large across-scan motion (Sect.~\ref{subsubsect:video_processing_unit_and_algorithms}).

{\bf Extended objects:} Unresolved, early-type elliptical galaxies and galaxy bulges will be mostly detected by {\it Gaia}, even with effective radii of several arcseconds, while late-type spiral galaxies, even those with weak bulges, will mostly remain undetected \citep{2015A&A...576A..74D}.

{\bf Faint stars:} The sky-mapper detection method is described in \cite{2015A&A...576A..74D}. In essence, the detection algorithm finds peaks in flat-fielded, local background-subtracted sky-mapper sample data and then accepts these as detections if their shape is consistent with that of a point source and their flux exceeds a certain, user-configurable threshold. This threshold has been set to $G = 20.7$~mag. The onboard magnitude estimation underlying the selection decision, however, has an error of $\sim$0.1~mag. Hence, the faint-end completeness in AF, BP, and RP is not sharp. In practice, the power-law slope of the object-detection-count histogram (in log space) as a function of $G$ magnitude already shows signs (as of mid 2016) of incompleteness starting just fainter than 20~mag, whereas object detections as faint as 21~mag are present as well. For RVS, the faint-end selection of targets is based on RP flux measurements. These measurements are essentially the sum of the flux of the RP samples corresponding to the RVS wavelength range, and are used as proxies for $G_{\rm RVS}$. The RVS faint completeness limit is hence not sharp either. In addition, the onboard software (Sect.~\ref{subsubsect:video_processing_unit_and_algorithms}) adapts the RVS threshold, through user-configurable look-up tables, to the instantaneous, straylight-dominated background level (Sect.~\ref{subsect:commissioning_and_performance_verification}), which means in practice that the faint limit varies between $\sim$15.5 and $\sim$16.2~mag over a spin period. At the end of the mission, however, taking the evolving scanning law into account (Sect.~\ref{subsect:scanning_law}), the effective faint limit will still be $G_{\rm RVS} \approx 16.2$~mag (or even a bit fainter, taking onboard RP flux-measurement errors into account).

\section{Summary}\label{sect:conclusions}

{\it Gaia} is the space-astrometry mission of the European Space Agency which, after successful commissioning, started scientific operations in mid-2014. The primary science goal of {\it Gaia} is to examine the kinematical, dynamical, and chemical structure and evolution of our Milky Way. In addition, the data of {\it Gaia} will have a strong impact on many other areas of astrophysical research, including stellar evolution and physics, star formation, stellar variability, the distance scale, multiple stars, exoplanets, solar system bodies, unresolved galaxies and quasars, and fundamental physics. With a focal plane containing more than 100 CCD detectors, {\it Gaia} surveys the heavens and repeatedly observes all objects down to $G \approx 20.7$~mag during its five-year nominal lifetime. The science data of {\it Gaia} comprise absolute astrometry (positions, proper motions, and parallaxes), broadband photometry in the unfiltered $G$ band, low-resolution blue and red (spectro-)photometry (BP and RP), and integrated $G_{\rm BP}$ and $G_{\rm RP}$ photometry for all objects. Medium-resolution spectroscopic data are collected for the brightest few hundred million sources down to $G_{\rm RVS} \approx 16.2$~mag. The concept and design of the spacecraft and the mission ultimately allows, after five years, stellar parallaxes (distances) to be measured with standard errors less than 10~$\mu$as for stars brighter than $G \approx 13$~mag, around 30~$\mu$as for stars around $G \approx 15$~mag, and around 600~$\mu$as around $G \approx 20$~mag. End-of-life photometric standard errors are in the milli-magnitude regime. The spectroscopic data allow the measurement of (mission-averaged) radial velocities with standard errors at the level of 1~km~s$^{-1}$ at $G_{\rm RVS} \approx 11$--$12$~mag and 15~km~s$^{-1}$ at $G_{\rm RVS} \approx 15$--$16$~mag, depending on spectral type. The {\it Gaia} Data Processing and Analysis Consortium (DPAC) is responsible for the processing and calibration of the {\it Gaia} data. The first intermediate release of {\it Gaia} data \citep{DPACP-8} comprises astrometry \citep{DPACP-14}, photometry \citep{DPACP-12}, and variability \citep{DPACP-15}; later releases will include BP/RP and RVS data. The validation of the data is described in \cite{DPACP-16} and the {\it Gaia} Archive is described in \cite{DPACP-19}.

\begin{acknowledgements}
This work has made use of results from the European Space Agency (ESA) space mission {\it Gaia}, the data from which were processed by the {\it Gaia} Data Processing and Analysis Consortium (DPAC). Funding for the DPAC has been provided by national institutions, in particular the institutions participating in the {\it Gaia} Multilateral Agreement. The {\it Gaia} mission website is \url{http://www.cosmos.esa.int/gaia}. The authors are current or past members of the ESA and Airbus DS {\it Gaia} mission teams and of the {\it Gaia} DPAC.
This work has financially been supported by:
the Algerian Centre de Recherche en Astronomie, Astrophysique et G\'{e}ophysique of Bouzareah Observatory;
the Austrian FWF Hertha Firnberg Programme through grants T359, P20046, and P23737;
the BELgian federal Science Policy Office (BELSPO) through various PROgramme de D\'eveloppement d'Exp\'eriences scientifiques (PRODEX) grants;
the Brazil-France exchange programmes FAPESP-COFECUB and CAPES-COFECUB;
the Chinese National Science Foundation through grant NSFC 11573054;
the Czech-Republic Ministry of Education, Youth, and Sports through grant LG 15010;
the Danish Ministry of Science;
the Estonian Ministry of Education and Research through grant IUT40-1;
the European Commission’s Sixth Framework Programme through the European Leadership in Space Astrometry (ELSA) Marie Curie Research Training Network (MRTN-CT-2006-033481), through Marie Curie project PIOF-GA-2009-255267 (SAS-RRL), and through a Marie Curie Transfer-of-Knowledge (ToK) fellowship (MTKD-CT-2004-014188); the European Commission's Seventh Framework Programme through grant FP7-606740 (FP7-SPACE-2013-1) for the {\it Gaia} European Network for Improved data User Services (GENIUS) and through grant 264895 for the {\it Gaia} Research for European Astronomy Training (GREAT-ITN) network;
the European Research Council (ERC) through grant 320360 and through the European Union’s Horizon 2020 research and innovation programme through grant agreement 670519 (Mixing and Angular Momentum tranSport of massIvE stars -- MAMSIE);
the European Science Foundation (ESF), in the framework of the {\it Gaia} Research for European Astronomy Training Research Network Programme (GREAT-ESF);
the European Space Agency in the framework of the {\it Gaia} project;
the European Space Agency Plan for European Cooperating States (PECS) programme through grants for Slovenia; the Czech Space Office through ESA PECS contract 98058;
the Academy of Finland; the Magnus Ehrnrooth Foundation;
the French Centre National de la Recherche Scientifique (CNRS) through action `D\'efi MASTODONS';
the French Centre National d'Etudes Spatiales (CNES);
the French L'Agence Nationale de la Recherche (ANR) investissements d'avenir Initiatives D’EXcellence (IDEX) programme PSL$\ast$ through grant ANR-10-IDEX-0001-02;
the R\'egion Aquitaine;
the Universit\'e de Bordeaux;
the French Utinam Institute of the Universit\'e de Franche-Comt\'e, supported by the R\'egion de Franche-Comt\'e and the Institut des Sciences de l'Univers (INSU);
the German Aerospace Agency (Deutsches Zentrum f\"{u}r Luft- und Raumfahrt e.V., DLR) through grants 50QG0501, 50QG0601, 50QG0602, 50QG0701, 50QG0901, 50QG1001, 50QG1101, 50QG140, 50QG1401, 50QG1402, and 50QG1404;
the Hungarian Academy of Sciences through Lend\"ulet Programme LP2014-17;
the Hungarian National Research, Development, and Innovation Office through grants NKFIH K-115709 and PD-116175;
the Israel Ministry of Science and Technology through grant 3-9082;
the Agenzia Spaziale Italiana (ASI) through grants I/037/08/0, I/058/10/0, 2014-025-R.0, and 2014-025-R.1.2015 to INAF and contracts I/008/10/0 and 2013/030/I.0 to ALTEC S.p.A.;
the Italian Istituto Nazionale di Astrofisica (INAF);
the Netherlands Organisation for Scientific Research (NWO) through grant NWO-M-614.061.414 and through a VICI grant to A.~Helmi;
the Netherlands Research School for Astronomy (NOVA);
the Polish National Science Centre through HARMONIA grant 2015/18/M/ST9/00544;
the Portugese Funda\c{c}\~ao para a Ci\^{e}ncia e a Tecnologia (FCT) through grants PTDC/CTE-SPA/118692/2010, PDCTE/CTE-AST/81711/2003, and SFRH/BPD/74697/2010; the Strategic Programmes PEst-OE/AMB/UI4006/2011 for SIM, UID/FIS/00099/2013 for CENTRA, and UID/EEA/00066/2013 for UNINOVA;
the Slovenian Research Agency;
the Spanish Ministry of Economy MINECO-FEDER through grants AyA2014-55216, AyA2011-24052, ESP2013-48318-C2-R, and ESP2014-55996-C2-R and MDM-2014-0369 of ICCUB (Unidad de Excelencia Mar\'{\i}a de Maeztu);
the Swedish National Space Board (SNSB/Rymdstyrelsen);
the Swiss State Secretariat for Education, Research, and Innovation through the ESA PRODEX programme, the Mesures d’Accompagnement, and the Activit\'es Nationales Compl\'ementaires;
the Swiss National Science Foundation, including an Early Postdoc.Mobility fellowship;
the United Kingdom Rutherford Appleton Laboratory;
the United Kingdom Science and Technology Facilities Council (STFC) through grants PP/C506756/1 and ST/I00047X/1; and
the United Kingdom Space Agency (UKSA) through grants ST/K000578/1 and ST/N000978/1.
The GBOT programme uses observations collected at (i) the European Organisation for Astronomical Research in the Southern Hemisphere with the VLT Survey Telescope (VST), under ESO programmes 092.B-0165, 093.B-0236, 094.B-0181, 095.B-0046, 096.B-0162, and 097.B-0304, (ii) the Liverpool Telescope, which is operated on the island of La Palma by Liverpool John Moores University in the Spanish Observatorio del Roque de los Muchachos of the Instituto de Astrof\'{\i}sica de Canarias with financial support from the United Kingdom Science and Technology Facilities Council, and (iii) telescopes of the Las Cumbres Observatory Global Telescope Network.
In addition to the authors of this paper, there are numerous people who have made essential contributions to {\it Gaia}, for instance those employed in the design, manufacturing, integration, and testing of the spacecraft and its modules, subsystems, and units. Many of those will remain unnamed yet spent countless hours, occasionally during nights, weekends, and public holidays, in cold offices and dark clean rooms. At the risk of being incomplete, we specifically name, in alphabetical order,
from Airbus DS (Toulouse):
Alexandre Affre,
Marie-Th\'er\`ese Aim\'e,
Audrey Albert,
Aur\'elien Albert-Aguilar,
Hania Arsalane,
Arnaud Aurousseau,
Denis Bassi,
Franck Bayle,
Pierre-Luc Bazin,
Emmanuelle Benninger,
Philippe Bertrand,
Jean-Bernard Biau,
Fran\c{c}ois Binter,
C\'edric Blanc,
Eric Blonde,
Patrick Bonzom,
Bernard Bories,
Jean-Jacques Bouisset,
Jo\"el Boyadjian,
Isabelle Brault,
Corinne Buge,
Bertrand Calvel,
Jean-Michel Camus,
France Canton,
Lionel Carminati,
Michel Carrie,
Didier Castel,
% Philippe Charvet, % author
% Fran\{c}ois Chassat, % author
Fabrice Cherouat,
Ludovic Chirouze,
Michel Choquet,
Claude Coatantiec,
Emmanuel Collados,
Philippe Corberand,
Christelle Dauga,
Robert Davancens,
Catherine Deblock,
Eric Decourbey,
Charles Dekhtiar,
Michel Delannoy,
Michel Delgado,
Damien Delmas,
Victor Depeyre,
Isabelle Desenclos,
Christian Dio,
Kevin Downes,
Marie-Ange Duro,
% Eric Ecale, % author
Omar Emam,
Elizabeth Estrada,
Coralie Falgayrac,
Benjamin Farcot,
Claude Faubert,
% Fr\'ed\'eric Faye, % author
S\'ebastien Finana,
Gr\'egory Flandin,
Loic Floury,
Gilles Fongy,
Michel Fruit,
Florence Fusero,
Christophe Gabilan,
J\'er\'emie Gaboriaud,
Cyril Gallard,
Damien Galy,
Benjamin Gandon,
Patrick Gareth,
Eric Gelis,
Andr\'e Gellon,
Laurent Georges,
Philippe-Marie Gomez,
Jos\'e Goncalves,
Fr\'ed\'eric Guedes,
Vincent Guillemier,
Thomas Guilpain,
St\'ephane Halbout,
Marie Hanne,
Gr\'egory Hazera,
Daniel Herbin,
Tommy Hercher,
Claude Hoarau le Papillon,
Matthias Holz,
Philippe Humbert,
Sophie Jallade,
Gr\'egory Jonniaux,
Fr\'ed\'eric Juillard,
Philippe Jung,
Charles Koeck,
Julien L'Hermitte,
R\'en\'e Laborde,
Anouk Laborie,
J\'er\^{o}me Lacoste-Barutel,
Baptiste Laynet,
Virginie Le Gall,
Marc Le Roy,
Christian Lebranchu,
Didier Lebreton,
Patrick Lelong,
Jean-Luc Leon,
Stephan Leppke,
Franck Levallois,
Philippe Lingot,
Laurant Lobo,
C\'eline Lopez,
Jean-Michel Loupias,
Carlos Luque,
S\'ebastien Maes,
Bruno Mamdy,
Denis Marchais,
Alexandre Marson,
% Benjamin Massart, % author
R\'emi Mauriac,
Philippe Mayo,
Caroline Meisse,
Herv\'e Mercereau,
Olivier Michel,
Florent Minaire,
Xavier Moisson,
% David Monteiro, % author
Denis Montperrus,
Boris Niel,
C\'edric Papot,
Jean-Fran\c{c}ois Pasquier,
Gareth Patrick,
Pascal Paulet,
Martin Peccia,
Sylvie Peden,
Sonia Penalva,
Michel Pendaries,
Philippe Peres,
Gr\'egory Personne,
Dominique Pierot,
Jean-Marc Pillot,
Lydie Pinel,
Fabien Piquemal,
% Vincent Poinsignon, % author
Maxime Pomelec,
Andr\'e Porras,
Pierre Pouny,
Severin Provost,
S\'ebastien Ramos,
Fabienne Raux,
Florian Reuscher,
Nicolas Riguet,
Mickael Roche,
Gilles Rougier,
Stephane Roy,
Jean-Paul Ruffie,
Fr\'ed\'eric Safa,
Claudie Serris,
Andr\'e Sobeczko,
Jean-Francois Soucaille,
Philippe Tatry,
Th\'eo Thomas,
Pierre Thoral,
Dominique Torcheux,
Vincent Tortel,
Stephane Touzeau,
Didier Trantoul,
Cyril V\'etel,
Jean-Axel Vatinel,
Jean-Paul Vormus,
Marc Zanoni,
from ESA:
Ricard Abello,
Ivan Aksenov,
Salim Ansari,
Philippe Armbruster,
Jean-Pierre Balley,
Rainer Bauske,
Thomas Beck,
% Gabriele Bellei, % author
Pier Mario Besso,
Carlos Bielsa,
Gerhard Billig,
Andreas Boosz,
Thierry Bru,
% Frank Budnik, % author
Joe Bush,
Marco Butkovic,
Jacques Cande\'e,
David Cano,
Carlos Casas,
Francesco Castellini,
David Chapmann,
Nebil Cinar,
Mark Clements,
Giovanni Colangelo,
Ana Colorado McEvoy,
Vincente Companys,
Federico Cordero,
Sylvain Damiani,
Paolo de Meo,
Fabio de Santis,
Fabienne Delhaise,
Gianpiero Di Girolamo,
% Federico di Marco, % author
Yannis Diamantidis,
John  Dodsworth,
Ernesto D\"olling,
Jane Douglas,
Jean Doutreleau,
Dominic Doyle,
Mark Drapes,
Frank Dreger,
Peter Droll,
Gerhard Drolshagen,
Bret Durrett,
Christina Eilers,
Yannick Enginger,
Alessandro Ercolani,
Robert Ernst,
% Maria Espina, % author
Hugh Evans,
Fabio Favata,
Stefano Ferreri,
Daniel Firre,
Michael Flegel,
Melanie Flentge,
Alan Flowers,
Jens Freih\"ofer,
C\'esar G\'omez Hern\'andez,
Juan Manuel Garcia,
Wahida Gasti,
Jos\'e Gavira,
Frank Geerling,
Franck Germes,
Gottlob Gienger,
B\'en\'edicte Girouart,
Bernard Godard,
Nick Godfrey,
Roy Gouka,
Cosimo Greco,
Robert Guilanya,
Kester Habermann,
Manfred Hadwiger,
Ian Harrison,
Angela Head,
Martin Hechler,
Kjeld Hjortnaes,
Jacolien Hoek,
Frank Hoffmann,
Justin Howard,
Arjan Hulsbosch,
Jos\'e Jim\'enez,
Simon Kellett,
Andrea Kerruish,
Kevin Kewin,
Oliver Kiddle,
Sabine Kielbassa,
Volker Kirschner,
% Arek Kowalczyk, % author
Holger Krag,
Benoi\^{\i}t Lain\'e,
Markus Landgraf,
Mathias Lauer,
Robert Launer,
Santiago Llorente,
% Alejandro Lopez-Lozano, % author
Guillermo Lorenzo,
% Tiago Loureiro, % author
James Madison,
% Jonas Marie, % author
Filip Marinic,
Arturo Mart\'{\i}n Polegre,
Ander Mart\'{\i}nez,
Marco Massaro,
% Ana Mestre, % author
Luca Michienzi,
% David Milligan, % author
Ali Mohammadzadeh,
Richard Morgan-Owen,
% Trevor Morley, % author
Prisca M\"uhlmann,
Michael M\"uller,
Pablo Munoz,
Petteri Nieminen,
Alfred Nillies,
Wilfried Nzoubou,
Alistair O'Connell,
Oscar Pace,
Mohini Parameswaran,
Ramon Pardo,
Taniya Parikh,
Panos Partheniou,
Dario Pellegrinetti,
Jos\'e-Louis Pellon-Bailon,
Michael Perryman,
Christian Philippe,
Alex Popescu,
% Florian Renk, % author
Alfonso Rivero,
Andrew Robson,
Gerd R\"ossling,
Martina Rossmann,
% Andreas Rudolph, % author
Markus R\"uckert,
Jamie Salt,
Giovanni Santin,
Rui Santos,
Stefano Scaglioni,
Melanie Schabe,
Dominic Sch\"afer,
Micha Schmidt,
Rudolf Schmidt,
Jean Sch\"utz,
Klaus-J\"urgen Schulz,
Julia Schwartz,
Andreas Scior,
J\"org Seifert,
Gunther Sessler,
Felicity Sheasby,
Heike Sillack,
Swamy Siram,
% Christopher Smith, % author
Claudio Sollazzo,
Steven Straw,
Mark Thompson,
Raffaele Tosellini,
Irren Tsu-Silva,
Livio Tucci,
Aileen Urwin,
Jean-Baptiste Valet,
Helma van de Kamp-Glasbergen,
Martin Vannier,
Kees van 't Klooster,
Enrico Vassallo,
David Verrier,
Sam Verstaen,
R\"udiger Vetter,
Jos\'e Villalvilla,
Raffaele Vitulli,
Mildred V\"ogele,
Sergio Volont\'e,
Catherine Watson,
Karsten Weber,
% Daniel Werner, % author
% Gary Whitehead, % author
Gavin Williams,
Alistair Winton,
Michael Witting,
Peter Wright,
Karlie Yeung,
Igor Zayer,
and from CERN:
Vincenzo~Innocente. We thank the referee, Joss Bland-Hawthorn, for constructive feedback that helped to clarify several items in the text.

\end{acknowledgements}

\bibliographystyle{aa} % aa.bst
\bibliography{29272} % 29272.bib

\clearpage
\onecolumn
\appendix

\section{Acronyms}\label{sect:acronyms}

The following table has been generated from the online {\it Gaia} acronym list.

\begin{longtable}{ll}
\caption{\label{tab:acronyms}Acronyms used in this paper.}\\
\hline\hline
Acronym & Description\\
\hline
\endfirsthead
\multicolumn{2}{c}%
{\tablename\ \thetable\ -- \textit{Continued from previous page.}}\\
\hline\hline
Acronym & Description\\
\hline
\endhead
\hline \multicolumn{2}{r}{\textit{Continued on next page.}}\\
\endfoot
\hline
\endlastfoot
AC   &   ACross-scan (direction) \\
ADC   &   Analogue-to-Digital Converter \\
AF   &   Astrometric Field \\
AGIS   &   Astrometric Global Iterative Solution \\
AL   &   ALong-scan (direction) \\
AOCS   &   Attitude and Orbit Control Subsystem \\
ASD   &   Auxiliary Science Data (VPU) \\
AVM   &   AVionics Model \\
BAM   &   Basic Angle Monitor \\
BP   &   Blue Photometer \\
CCD   &   Charge-Coupled Device (detector) \\
CDB   &   Configuration DataBase \\
CDU   &   Clock Distribution Unit \\
CPS   &   Chemical Propulsion Subsystem (orbit maintenance) \\
CTI   &   Charge-Transfer Inefficiency \\
CU   &   Coordination Unit (in DPAC) \\
DCS   &   De-compression and Calibration Services \\
$\Delta$-DOR   &   Delta Differential One-way Range \\
DIB   &   Diffuse Interstellar Band \\
DPAC   &   Data Processing and Analysis Consortium \\
DPC   &   Data Processing Centre \\
DR1   &   ({\it Gaia}) Data Release 1 (September 2016) \\
DS   &   (Airbus) Defence and Space \\
EAR   &   Event Anomaly Report \\
EIRP   &   Equivalent Isotropic Radiated Power \\
EO   &   Extended Object \\
EPSL   &   Ecliptic Poles Scanning Law \\
ESA   &   European Space Agency \\
ESAC   &   European Space Astronomy Centre (ESA) \\
ESO   & European Southern Observatory \\
ESOC   &   European Space Operations Centre (ESA) \\
ESTRACK   &   ESA Tracking Stations Network \\
E-SVM   &   Electrical SerVice Module \\
FL   &   First Look \\
GAIA   &   Global Astrometric Interferometer for Astrophysics (obsolete; now spelled as {\it Gaia}) \\
GBOT   &   Ground-Based Optical Tracking \\
GMSK   &   Gaussian Minimum Shift Keying \\
GOST   &   {\it Gaia} Observation Scheduling Tool \\
GSR   &   Global Sphere Reconstruction \\
ICRF   &   International Celestial Reference Frame \\
ICRS   &   International Celestial Reference System \\
IDT   &   Initial Data Treatment \\
IDU   &   Intermediate Data Update \\
IGSL   &   Initial {\it Gaia} Source List \\
INPOP   &   Int\'egrateur Num\'erique Plan\'etaire de l'Observatoire de Paris (ephemerides) \\
M2MM   &   M2 Mirror Mechanism (focus) \\
MDB   &   Main DataBase \\
MIT   &   MOC Interface Task \\
MMH   &   Mono-Methyl Hydrazine \\
MOC   &   Mission Operations Centre (ESOC) \\
MP   &   Metal-Poor (star) \\
MPS   &   Micro-Propulsion Subsystem (science mode) \\
M-SVM   &   Mechanical SerVice Module \\
NSS   &   Non-Single Star \\
NTO   &   (di-)Nitrogen TetrOxide \\
OBMT   &   OnBoard Mission Timeline \\
OBT   &   OnBoard Time (realised in CDU) \\
ODAS   &   One-Day Astrometric Solution \\
OGA-1   &   First On-Ground Attitude determination (in IDT) \\
OGA-2   &   Second (and improved) On-Ground Attitude determination (in ODAS/FL) \\
PAA   &   Phased-Array Antenna \\
PDHU   &   Payload Data-Handling Unit \\
PEM   &   Proximity-Electronics Module (CCD) \\
PLM   &   PayLoad Module \\
PO   &   (DPAC) Project Office \\
PSF   &   Point Spread Function \\
QSO   &   Quasi-Stellar Object \\
RAFS   &   Rubidium Atomic Frequency Standard (part of CDU) \\
RMS   &   Root-Mean-Square \\
RP   &   Red Photometer \\
RVS   &   Radial-Velocity Spectrometer \\
SEA   &   Source Environment Analysis \\
SED   &   Spectral Energy Distribution \\
SIF   &   Service Interface Function (VPU) \\
SM   &   Sky Mapper (detector) \\
SOC   &   Science Operations Centre (ESAC) \\
SP   &   Star Packet (VPU) \\
SSO   &   Solar System Object \\
TCB   &   Barycentric Coordinate Time \\
TDI   &   Time-Delayed Integration (CCD) \\
TMA   &   Three-Mirror Anastigmat (telescope) \\
TODCOR   &   TwO-Dimensional CORrelation \\
UTC   &   Coordinated Universal Time \\
VPA   &   Video Processing Algorithms (run on the VPU) \\
VPU   &   Video Processing Unit (runs the VPAs) \\
WFS   &   Wave-Front Sensor \\
XP   &   Shortcut for BP and/or RP (generic name for Blue and Red Photometer) \\
\end{longtable}

\end{document}